\documentclass[11pt,a4paper]{article}

\usepackage{tgheros}

\usepackage{amsfonts, amsmath, amssymb, amsthm, enumitem, float, mathrsfs, mathtools, multirow, nccmath, subfig, booktabs, caption}

\usepackage[mathcal]{euscript}
\usepackage[dvipsnames]{xcolor}

\usepackage{hyperref}
\hypersetup{colorlinks,
linkcolor={MidnightBlue},
citecolor={RoyalBlue},
urlcolor={Blue}}

\usepackage[title]{appendix}
\usepackage[nameinlink, noabbrev, capitalize]{cleveref}
\usepackage[hmargin=1.0in, vmargin=1.0in]{geometry}
\usepackage[round]{natbib}
\usepackage{setspace}

\def\be{\begin{equation}}
\def\ee{\end{equation}}
\def\bea{\begin{eqnarray}}
\def\eea{\end{eqnarray}}

\author{}
\title{}
 
\DeclareMathOperator*{\argmin}{\arg\!\min}

\DeclareMathOperator*{\diag}{\normalfont\textrm{diag}}

\DeclareMathOperator*{\VEC}{\normalfont\textrm{vec}}

\newtheorem{assumption}{Assumption}

\newtheorem{definition}{Definition}
\newtheorem{example}{Example}
\newtheorem{lemma}{Lemma}
\newtheorem{proposition}{Proposition}
\newtheorem{remark}{Remark}
\newtheorem{theorem}{Theorem}

\numberwithin{equation}{section}
\numberwithin{theorem}{section}
\numberwithin{lemma}{section}
\numberwithin{proposition}{section}

\newcommand\blfootnote[1]{
\begingroup
\renewcommand\thefootnote{}\footnote{#1}
\addtocounter{footnote}{-1}
\endgroup
}

\allowdisplaybreaks[4]

\begin{document}


\begin{titlepage}

\begin{center}

\begin{spacing}{1.5}
{\Large Time--Varying Quantile VAR Models: Estimation and Inference\blfootnote{
    
Gao and Peng would like to acknowledge the Australian Research Council Discovery Projects Program for its financial support under Grant Numbers: DP250100063.}}
\end{spacing}

\medskip

{\sc Jiti Gao$^{\dag}$, Fei Liu$^\sharp$ and Bin Peng$^{\dag}$}

\medskip

$^\dag$Monash University and $^\sharp$University of Bath

\bigskip

\today

\begin{abstract} 
We develop an econometric framework for impulse response analysis in time-varying quantile vector autoregressive (TV-QVAR) models. Unlike standard conditional-mean setting, TV-QVAR systems face a fundamental analytical challenge: neither time-varying parameters nor conditional quantiles propagate linearly through dynamic recursions. To overcome this, we establish a uniform Bahadur representation for local quantile VAR estimation, propose a smoothed check-function approach for tractable parameter inference, and introduce the best linear projection of the time-varying quantile impulse response (TV-QIR) function  to recover well-defined dynamic causal effects. 

Applying our methodology to U.S. industrial production growth and credit spreads from 1919 to 2025 uncovers pronounced structural shifts in tail interdependence, showing that downside asymmetry is concentrated at short horizons and diminishes as the forecast horizon lengthens.

\medskip
\medskip

\noindent\textit{Keywords}: Bahadur Representation; Growth-at-Risk; Impulse Response Analysis; Smoothing Quantile Loss Function; Stationary Approximation
\medskip

\noindent \textit{JEL Classification}: C14, C32, C65, O40

\medskip
\medskip
\medskip

\end{abstract}
\end{center}    
\end{titlepage}

\section{Introduction}\label{Sec.1}

Over the past two decades, two major methodological developments have reshaped empirical macroeconomics: time-varying parameter vector autoregression (TV-VAR) models, which capture evolving economic relationships, and quantile VAR (QVAR) models, which uncover asymmetric shock transmission and tail risks across the outcome distribution. While the TV-VAR literature (\citealp{Primiceri2005, Paul2020, Gao02012024, YGP2025}) and the emerging QVAR literature (e.g., \citealp{AdrianBoyarchenkoGiannone2019, CM2024, han2024estimation, Ruzicka2024, korobilis2025monitoring, wojciechowski2025identifying}) have each made substantial progress independently, they have evolved largely along separate tracks. To contextualize our contribution in what follows, we review theoretical and empirical developments in such key areas mentioned above that directly motivate the proposal and discussion of the theory, methodology and empirical application of this paper, before unifying these two strands into a single time-varying parameter quantile vector autoregression (TV-QVAR) framework.

VAR modelling, pioneered by \cite{Sims1980}, serves as the canonical framework for analyzing dynamic interdependencies among endogenous variables (\citealp{SW2001}). By accommodating contemporaneous feedback among macroeconomic variables, VAR models allow researchers to trace the transmission of structural shocks (such as monetary policy innovations) via impulse response analysis. While the VAR literature focuses predominantly on constant-parameter specifications, such static assumptions struggle to capture policy regime changes and evolving macroeconomic dynamics (\citealp{Primiceri2005, Paul2020}). To accommodate evolving structural relationships, parameter instability has been integrated into VAR systems. However, a key limitation of these existing TV-VAR studies is their exclusive focus on conditional-mean models, which fails to capture heterogeneous dynamics across different quantiles of the response distribution. In this paper, we incorporate TV parameters in the spirit of \cite{INOUE2024105726}, \cite{Gao02012024,GAO2025106035}, and \cite{maung2025nonparametric}, for example, while extending their conditional--mean settings to the conditional quantile framework.

In parallel, growing interest in tail risks and asymmetric transmission has motivated the development of quantile dynamic models (e.g., \citealp{CM2024, korobilis2025monitoring, wojciechowski2025identifying}). Within this literature, \citet{han2024estimation} trace how a current shock alters the expected future path of conditional quantiles, whereas \citet{Ruzicka2024} examines counterfactual shifts in the conditional quantiles of future outcomes, which admits a quantile local-projection representation under linearity. However, both frameworks are restricted to stationary environments. Meanwhile, \cite{CM2024} and \cite{korobilis2025monitoring} approach QVAR modelling from a Bayesian perspective. From a frequentist perspective, theoretical guarantees grounded in the Bahadur representation remain largely undeveloped in the QVAR literature.

This theoretical gap becomes even more acute when moving from one-step estimation to multi-step impulse response analysis in time-varying settings. In a conventional conditional-mean model, recursive iteration of the underlying data-generating process naturally yields a moving average representation, making impulse response analysis straightforward. In a quantile regression framework, by contrast, conditional quantiles do not propagate linearly under recursive forecasting; consequently, multi-step-ahead impulse responses cannot be derived by simply iterating the one-step-ahead quantile specification---such complexity that intensifies once time--varying parameters are introduced. 

To the best of our knowledge, how to develop quantile impulse response (QIR) analysis in a nonlinear and time-varying framework remains an open question, leaving its asymptotic inference largely unexplored. To bridge this gap, we introduce the best linear projection of the TV-QIR, and then establish a comprehensive toolkit for estimation and inference. Relative to existing approaches, our framework allows for asymptotically valid inference without requiring the multi-step conditional quantile response function to be correctly specified as linear.
\medskip

To better explain some challenges, we present a simple example here to show the primary obstacle by exploring an impulse response function involved in a TV quantile regression.

\begin{example}\label{def.rm2}  

    To focus on the main ideas, we consider a special scalar specification of the form:
    \begin{equation}\label{def.hstep0}
        y_{t} = y_{t-1} \beta_{t}(\eta_{t}),
    \end{equation}
    where $\eta_t \sim \text{i.i.d. } U(0,1)$ and $\beta_{t}(\cdot)$ is a TV coefficient function. Recursive substitution yields the $h$-step ahead representation
    \begin{equation}\label{def.hstep}
        y_{t} = y_{t-2} \beta_{t-1}(\eta_{t-1})\beta_{t}(\eta_{t}) = \cdots = y_{t-h} \prod_{\ell=0}^{h-1}\beta_{t-\ell}(\eta_{t-\ell}).
    \end{equation}
    Equation \eqref{def.hstep} exposes two fundamental obstacles to standard impulse response analysis. First, the time-varying coefficients $\{\beta_{t-\ell}(\cdot)\}_{\ell=0}^{h-1}$ compound multiplicatively across horizons rather than adding linearly. Second, because the quantile operator is nonlinear, the $\tau$-th conditional quantile of this product does not equal the product of the individual $\tau$-th quantiles $\prod_{\ell=0}^{h-1}\beta_{t-\ell}(\tau)$; instead, it depends jointly and nonlinearly on the entire sequence of innovations $\{\eta_{t},\ldots,\eta_{t-h+1}\}$ and varies intricately with the horizon $h$.
    
    In this context, how to formally define the QIR remains an open question in the literature, and the corresponding inference is unexplored. Even if we rewrite \eqref{def.hstep0} in an additive error form $y_{t} = y_{t-1} \beta_{t}(\tau) +e_{\tau,t}$ with $e_{\tau,t}\coloneqq  y_{t-1}(\beta_{t}(\eta_t) -\beta_{t}(\tau))$, the underlying challenge remains, mainly due to the fact that $\beta_t(\cdot)$ is now time--varying and completely nonlinear in both $t$ and $\tau$. Consequently, $y_t$ becomes highly nonstationary. 
       
\end{example}

Additionally, in the context of standard quantile regression, recent advancements have addressed the non--smoothness of the check function, which complicates both theoretical derivation and computational efficiency, particularly for large-scale datasets. While \cite{TWZ2022} and \cite{HE2023367} propose a convolution-type smoothed quantile loss as an alternative to the standard check function, our work approaches this smoothing technique from the perspective of generalized functions. Moreover, we establish a sharp approximation rate that remains independent of the specific models under study. Our knowledge also suggests that such result has not been formally presented elsewhere and is of independent interest to the broader field.

While the relevant brands of literature outlined above may account for one or more of the model estimation and inference issues we will address for the time--varying quantile VAR model setting, the novelty of this paper is addressing them all simultaneously to provide a unified framework and synthesize the existing paradigms by considering quantile regression for TV-VAR models and evaluating the impulse responses. To the best of our knowledge, meanwhile, rigorous discussions and results accommodating these features are not yet available in the literature. In what follows, we summarize our contributions below. 

\begin{enumerate}

    \item From a modelling standpoint, we propose a class of TV quantile VAR (TV-QVAR) models and then develop an estimation procedure incorporating a smoothed check function with local-linear kernel estimation. This framework not only extends the quantile autoregression of \citet{Koenker01092006} to a multivariate TV setting, but also generalizes the TV framework of \cite{Gao02012024} and \cite{INOUE2024105726} to the conditional quantile setting.
    
    \item From a theoretical perspective, as commented in Example \ref{def.rm2} and Appendix \ref{Appen.A1}, we face challenges to establish several fundamental results, which, to the best of our knowledge, are novel to the literature. These include, but are not limited to:
    
    \begin{itemize}
        \item establishing a Bahadur representation for the TV-QVAR regression, which facilitates the derivation of limiting distributions for the estimates of the coefficient functions;
        \item defining and analyzing the best linear projection of the TV-QIR, thereby overcoming the challenge that conditional quantiles do not propagate linearly through forward iteration of the underlying data--generating process; moreover, deriving the limiting distribution for the TV-QIR, which provides a rigorous foundation for inference;
        \item constructing a smoothed check function within the framework of generalized functions, and establishing a sharp rate of convergence between the standard check function and its smoothed counterpart, which is of independent interest.
    \end{itemize}
    
    \item Our finite--sample studies by simulated datasets show that the proposed model and estimation method works well numerically. 
    
    In an empirical study, our application to monthly U.S. industrial-production growth and the Baa-Aaa credit spread over 1919-2025 demonstrates the empirical relevance of the framework. The lower-tail TV-QVAR coefficient remains negative throughout the sample, while the upper-tail coefficient turns negative from the mid-1980s despite a positive constant-parameter estimate. The TV-QIR results further reveal stronger downside asymmetry at shorter horizons. These findings highlight important time--varying and horizon--dependent dynamics that cannot be captured by the time--invariant quantile models.
    
\end{enumerate} 

The rest of the paper is organized as follows. Section \ref{Sec.2} outlines the TV-VAR framework, identifies key methodological gaps in the existing literature, and establishes the corresponding theoretical results that resolve them. Section \ref{Sec.3} provides simulation results, Section \ref{Sec.4} examines the credit-growth application, and Section \ref{Sec.5} concludes. 

Appendix A contains supplementary results and implementation details. Specifically, Appendix \ref{Appen.A1} explains the inherent nonlinearity of the quantile framework, Appendix \ref{Appen.A2} presents additional empirical results omitted from the main text, Appendix \ref{Appen.A3} provides two preliminary lemmas that facilitate the establishment and derivation of the asymptotic distribution of the resulting VAR estimator in Section \ref{Sec.23}, and Appendix \ref{Appen.A4} discusses the numerical implementation of the estimation procedure. 

In the online supplement (Appendix B), Appendix \ref{Appen.B1} provides a jackknife method used for nonparametric bias correction, Appendix \ref{Appen.B2} introduces the notation and technical lemmas used throughout the mathematical proofs, which are provided in Appendices \ref{Appen.B3} and \ref{Appen.B4}.

Before proceeding further, we introduce some notations which will be repeatedly used in the paper. Vectors and matrices are always in bold font. For a real vector $\mathbf{u}=(u_1,\ldots, u_L)^\top$, let $\| u\|_\nu\coloneqq  (\sum_{i=1}^L |u_i|^\nu )^{1/\nu}$; for a random vector $\pmb{\xi}=(\xi_1,\ldots, \xi_L)^\top$, let $\|\pmb{\xi} \|_\nu \coloneqq  (\sum_{i=1}^L E|\xi_i|^\nu )^{1/\nu}$, where $\nu \ge 1$. For a matrix $\mathbf{A}$, we let $\| \mathbf{A}\|$, $\| \mathbf{A}\|_2$ and $\VEC(\mathbf{A})$ stand for its Frobenius norm,  Spectral norm and vectorization operation respectively. For two numbers $a$ and $b$, $a\lesssim b$ stands for $a =O(b)$, and accordingly, $a\asymp b$ stands for $a\lesssim b$ and $b\lesssim a$. For two random variables $x$ and $y$, $x\lesssim y$ stands for $x=O_P(y)$, and accordingly $x\asymp y$ represents $x\lesssim y$ and $y\lesssim x$. For any given smooth function, say $g(w)$, we let $g^{(\ell)}(w)$ define its $\ell^{th}$ derivative. For any given positive integer $L$, $[L] \coloneqq\{1,2,\ldots, L \}$. 

When carrying on quantile regression, we define the so-called check function as $$\rho_\tau(u) \coloneqq u(\tau- {\rm I}(u<0)),$$ where ${\rm I}(\cdot)$ is an indicator function. Accordingly, we let $\psi_\tau(u)$ be the subgradient of $\rho_\tau(u)$. To estimate TV parameters, we shall adopt the nonparametric kernel method. Throughout, $K(\cdot)$ stands for a kernel function, $\delta$ stands for a bandwidth, and $K_\delta(\cdot)\coloneqq K(\cdot/\delta)/\delta$. Accordingly, we define $v_j \coloneqq \int w^j K(w)\mathrm{d}w$ for $j\ge 0$. Apparently, $v_0=1$.

\section{The Setup, Theory and Methodology}\label{Sec.2}

Before presenting any theoretical results, we outline the data--generating process and identify the gaps in the existing literature.

Suppose that we have a set of $p\times 1$ observable variables:

\begin{eqnarray}\label{def.yts}
    \{\mathbf{y}_t = (y_{1,t},\ldots, y_{p, t})^\top\mid t=1-q, 2-q,\ldots,1,\ldots,T\},
\end{eqnarray}
where $q\ (\ge 0)$ is positive fixed integer accounting for the lag terms to be included in the model, and  the non-positive time indices are introduced purely for notational convenience in the subsequent regression analysis. Moreover, we suppose that the $\tau^{th}$ conditional quantile of each element in $\mathbf{y}_t$ (i.e., $y_{\ell, t}$) admits the following VAR($q$) representation:

\begin{eqnarray}\label{def.md1}
    \mathbf{y}_t = \sum_{j=1}^{q} \mathbf{A}_{\tau, j, t}\mathbf{y}_{t-j} + \mathbf{e}_{\tau, t},
\end{eqnarray}
where $\mathbf{A}_{\tau, j, t}$ is a $p\times p$ coefficient matrix and $\mathbf{e}_{\tau, t}=(e_{\tau,1,t},\ldots,e_{\tau,p,t})^\top$ is a $p\times1$ vector including random noises. 


For notational simplicity, we omit the intercept term in \eqref{def.md1}; including it introduces no fundamental technical challenges but unnecessarily complicates the notation. We shall also suppress $\tau$ wherever possible. Thus, the model \eqref{def.md1} is simplified as:

\begin{eqnarray}\label{def.md1.simple}
    \mathbf{y}_t = \sum_{j=1}^{q} \mathbf{A}_{j, t}\mathbf{y}_{t-j} + \mathbf{e}_{t},
\end{eqnarray} 
where $\mathbf{A}_{j, t} =(\mathbf{a}_{j1, t},\ldots,\mathbf{a}_{jp, t})^\top$ and $\mathbf{e}_{t} = (e_{1,t},\ldots, e_{p, t})^\top$.  Accordingly, for $\forall\ell \in [p]$, the $\tau^{th}$ conditional quantile of $y_{\ell, t}$ gives

\begin{eqnarray}\label{def.md2}
    Q_{\ell\mid \tau}(y_{\ell,t}\mid \mathbf{x}_{t-1})
    &=& \mathbf{x}_{t-1}^\top \pmb{\beta}_{\ell, t},
\end{eqnarray}
where $\mathbf{x}_{t-1} =\VEC (\mathbf{y}_{t-1}  ,\ldots, \mathbf{y}_{t-q} ) $ and $\pmb{\beta}_{\ell, t} =\VEC(\mathbf{a}_{1\ell, t} ,\ldots, \mathbf{a}_{q\ell, t} ) $. Obviously,   $\mathbf{x}_{t-1}$ and $\pmb{\beta}_{\ell, t}$ are $pq\times 1$ vectors. 

\begin{example}\label{Ex1}
From the modelling perspective, the above setting nests the following data--generating process:

\begin{eqnarray}\label{def.md3}
    y_{\ell, t} = \mathbf{x}_{t-1}^\top \pmb{\beta}_{\ell, t}(\eta_{\ell ,t})\quad\text{with}\quad \ell\in[p].
\end{eqnarray}
For \eqref{def.md3}, the literature  (e.g., \citealp{Koenker01092006}) normally requires that (1) for $\forall \ell\in [p]$, $\eta_{\ell,t}\sim U(0,1)$ is independent and identically distributed over $t$ and independent of $\mathcal F_{t-1}$ wherein $\mathcal F_{t-1}$ denotes the information set available up to time $t-1$; and (2) the right hand side of \eqref{def.md3} is monotonically increasing in $\eta_{\ell,t}$. In this case, simple algebra shows that $e_{\ell,t}$ of \eqref{def.md1.simple} takes the form of $e_{\ell,t} =\mathbf{x}_{t-1}^\top [\pmb{\beta}_{\ell,t}(\eta_{\ell,t}) -\pmb{\beta}_{\ell,t}(\tau)]$, and  \eqref{def.md2} holds. In Appendix \ref{Appen.A1}, we offer more details to further explain the inherent nonlinearity of the quantile framework using a process similar to \eqref{def.md3}.
\end{example}

We make a few comments below: the first one offers greater flexibility for practical implementation, while the second highlights the challenges of TV-QIR analysis.

\begin{remark}\label{def.rm1}
    It is possible to consider a more general condition by allowing the target quantile to vary across $\ell\in [p]$, such as  
    
    \begin{eqnarray*}
        Q_{\ell\mid \tau_\ell}(y_{\ell,t}\mid \mathbf{x}_{t-1})= \mathbf{x}_{t-1}^\top \pmb{\beta}_{\ell, t}\quad\text{for}\quad \forall\ell \in [p].
    \end{eqnarray*}
    However, this generalization provides no further substantive insights while unnecessarily encumbering the notation. Thus, we adopt a uniform $\tau$ across all $\ell \in [p]$ in \eqref{def.md2} solely for the sake of exposition. 
\end{remark}

In what follows, Section \ref{Sec.21} establishes a stationary approximation to the underlying data--generating process \eqref{def.md1}; Section \ref{Sec.22} analyzes the smoothing technique, yielding a theoretical result of independent interest to the broader literature; Section \ref{Sec.23} develops the estimation framework and derives its asymptotic properties; Section \ref{Sec.24} finally investigates QIR analysis. Although our ultimate goal is to infer impulse response in Section \ref{Sec.24}, one cannot recover the impulse response properly without understanding those fundamental results of Sections \ref{Sec.21}-\ref{Sec.23}.

\subsection{Stationary Approximation}\label{Sec.21}

Due to the recursive relationship of \eqref{def.md1} and the TV nature of the parameters, we firstly offer a stationary approximation. In order to do so, we introduce more notations, and present Assumption \ref{Assumption1}.

Note that \eqref{def.md1.simple} can be written as an AR(1) process:

\begin{eqnarray*}
    \mathbf{x}_t =
    \begin{pmatrix}
        \mathbf{A}_{[q-1], t}  & \mathbf{A}_{q, t} \\
        \mathbf{I}_{p(q-1)} & \mathbf{0}
    \end{pmatrix}
    \mathbf{x}_{t-1}+
    \begin{pmatrix}
        \mathbf{e}_{t}\\
        \mathbf{0} 
    \end{pmatrix}\eqqcolon \tilde{\mathbf{A}}_{t} \mathbf{x}_{t-1} +\tilde{\mathbf{e}}_{t},
\end{eqnarray*}
where $\mathbf{A}_{[q-1], t} \coloneqq (\mathbf{A}_{1,t},\ldots, \mathbf{A}_{q-1,t})$, and $ \tilde{\mathbf{A}}_{t}$ is the companion matrix. Under mild conditions, for $\forall \tau\in (0,1)$, $\mathbf{y}_t$ admits an MA($\infty$) process with a drifting term:

\begin{eqnarray}\label{def.yt.ma}
    \mathbf{y}_{t} &=& \sum_{\ell=0}^\infty
    \mathbf{c}_{\ell,t}\mathbf{e}_{t-\ell}
    =\bar{\mathbf v}_t+
    \sum_{\ell=0}^\infty\mathbf c_{\ell,t}\bar{\mathbf e}_{t-\ell},
\end{eqnarray}
where $\mathbf{c}_{\ell,t}\coloneqq \mathbf{s}\mathbf{C}_{\ell,t}\mathbf{s}^\top$, $\mathbf{s}\coloneqq(\mathbf{I}_p,\mathbf{0}_{p\times p(q-1)})$, $\mathbf{C}_{\ell,t}\coloneqq\prod_{j=0}^{\ell-1}\tilde{\mathbf A}_{t-j}$ for $\ell\ge1$, and $\mathbf{C}_{0,t}\coloneqq\mathbf I_{pq}$, $\bar{\mathbf e}_t\coloneqq\mathbf e_t-E[\mathbf e_t]$, and $\bar{\mathbf v}_t\coloneqq\sum_{\ell=0}^{\infty} \mathbf c_{\ell,t}E[\mathbf e_{t-\ell}]$.

Throughout the rest of this paper, we consider estimation and inference for rescaled time, i.e., $\gamma_t\coloneqq t/T$. It means heuristically that with increasing $T$ more and more data of each local structure are observed (\citealp{Dahlhaus}). Accordingly, for $\forall j\in [q]$, we specify

\begin{eqnarray}\label{def.Ajt}
    \mathbf{A}_{j,t}\eqqcolon\mathbf{A}_j(\gamma_t)\quad\text{and}\quad \mathbf{A}_{j, t} \equiv \mathbf{A}_{j, 0}\quad\text{for}\quad t\le 0.
\end{eqnarray}
Thus, for $\forall\gamma\in [0,1]$, we have 

\begin{eqnarray*}
    \mathbf{c}_{\ell}(\gamma) =\mathbf{s}\tilde{\mathbf{A}}^\ell (\gamma) \mathbf{s}^\top \quad \text{and}\quad\tilde{\mathbf{A}}(\gamma)=\begin{pmatrix}
        \mathbf{A}_{[q-1]}(\gamma)  & \mathbf{A}_{q}(\gamma) \\
        \mathbf{I}_{p(q-1)} & \mathbf{0}
    \end{pmatrix},
\end{eqnarray*}
where $\mathbf{A}_{[q-1]}(\gamma) \coloneqq (\mathbf{A}_{1}(\gamma),\ldots, \mathbf{A}_{q-1}(\gamma))$.

Let $\pmb\psi_t \coloneqq \left(\psi_\tau(e_{1,t}),\ldots, \psi_\tau(e_{p,t})\right)^\top$, $w_{t,\gamma}\coloneqq(\gamma_t-\gamma)/\delta$, $\mathbf W_{t,\gamma} \coloneqq\VEC\!\left( \mathbf x_{t-1},w_{t,\gamma}\mathbf x_{t-1} \right)$, and $k_{\delta,t}^{(r)}(\gamma) \coloneqq w_{t,\gamma}^r K_\delta(\gamma_t-\gamma)$ for $r\in\{0,1,2,3,4\}$. Finally, we provide the following assumption to regulate the variables involved in \eqref{def.yt.ma}.

\begin{assumption}\label{Assumption1}
\item 

\begin{enumerate}[leftmargin=*]

\item

\begin{enumerate}[leftmargin=*]
\item   For $\forall \tau\in (0, 1)$,  \eqref{def.md2} holds. Additionally,  for some $\nu\ge4$, $\max_{t} \|\mathbf e_t\|_\nu \le C.$

 \item  For $\forall \ell \in [p]$, let
$f_{\ell,t}(w\mid\mathcal F_{t-1})$ be the conditional density of
$e_{\ell,t}$.  The followings  hold almost surely
\begin{eqnarray*}
 &&E\!\left[\psi_\tau(e_{\ell,t})\mid\mathcal F_{t-1}\right]=0,
 \quad
 \max_{t\in[T],\ell\in[p]}\sup_{w\in \mathbb R}
 |f_{\ell,t}(w\mid\mathcal F_{t-1})|
 \le C,
 \nonumber\\
 &&\max_{t\in[T],\ell\in[p]}\sup_{|w|\le\varepsilon_0}
 \left|
 f_{\ell,t}(w\mid\mathcal F_{t-1})
 -f_{\ell,t}(0\mid\mathcal F_{t-1})
 \right|
 \le C|w|,
\end{eqnarray*}
for some $\varepsilon_0>0$.
\end{enumerate}

\item \begin{enumerate}[leftmargin=*]
    \item Let \eqref{def.Ajt} hold, and $0<\sup_{\gamma\in [0,1]}\rho(\tilde{\mathbf{A}}(\gamma))\le \rho_y<1$, where $\rho(\cdot)$ stands for the Spectral radius, and the components of $\mathbf{A}_j(\gamma)$ are twice continuously differentiable on $[0,1]$. 
     
    \item For $r\in\{0,1,2,3,4\}$, there exist deterministic and continuous matrices
$\{\mathbf D_{\ell,\gamma}^Q\}_{\ell\in[p]}$, such that
\begin{eqnarray*}
 &&\max_{\ell\in[p]}
 \sup_{\gamma\in[\delta,1-\delta]}
 \left\|
 \frac1T\sum_{t=1}^{T}
 f_{\ell,t}(0\mid\mathcal F_{t-1})
 \mathbf x_{t-1}\mathbf x_{t-1}^{\top}
 k_{\delta,t}^{(r)}(\gamma)
 -v_r\mathbf D_{\ell,\gamma}^Q
 \right\|
 =o_P(1),
\end{eqnarray*}
and $\inf_{\gamma\in[0,1]}
 \lambda_{\min} (
 \diag\{\mathbf D_{1,\gamma}^{Q},\ldots,
 \mathbf D_{p,\gamma}^{Q}\}
 )>0.$

\item For $\forall \gamma \in (0,1)$, there exists a deterministic positive-semidefinite matrix
$\mathbf M_\gamma^Q$ such that
\begin{eqnarray*}
\frac1{T}\sum_{t=1}^{T}
\left[
E [
\pmb\psi_t\pmb\psi_t^\top
\mid \mathcal F_{t-1}
]
\otimes
\{
\mathbf W_{t,\gamma} \mathbf W_{t,\gamma}^\top
\}
\right] \frac{1}{\delta} \, K^2(w_{t,\gamma})
&\to_P&
\diag\{a_1\mathbf M_\gamma^Q,
a_2\mathbf M_\gamma^Q\},
\end{eqnarray*}
where $a_1=\int K^2(w)\mathrm dw$ and
$a_2=\int w^2K^2(w)\mathrm dw$.

\end{enumerate}
\end{enumerate}
\end{assumption}

Assumption \ref{Assumption1} serves a dual purpose: first, it ensures the existence of a stationary approximation for $\mathbf{y}_t$ at each time point $t$; second, it guarantees the feasibility of establishing a Bahadur representation.

The conditional quantile restriction in Assumption \ref{Assumption1} is in the same spirit as Assumption A.1 in \cite{Koenker01092006}, but we do not require identical distribution due to the TV nature.  More importantly, the restriction is formulated as a  martingale difference condition rather than temporal independence, which therefore accommodates Example \ref{Ex1}.  The uniform moment and conditional density bounds regulate the heterogeneity associated with the error components.

Assumption \ref{Assumption1}.2 infers that all absolute eigenvalues of $\tilde{\mathbf{A}}_{t}$ are less than 1. In other words, $|\mathbf{I}_p -\mathbf{A}_{1,t}z-\ldots - \mathbf{A}_{q,t}z^q  |\ne 0$ for $|z|\le 1$ uniformly in $t$. As a result, it ensures that $\mathbf{y}_t$ will not have any nonstationary or explosive behavior under the TV framework. See \citet[p. 260]{Lut2005} for more detailed discussion.

The remaining high-level conditions in Assumption \ref{Assumption1} ensure the required sample convergence. They can be verified under standard weak dependence frameworks (e.g., the functional dependence conditions in Assumption \ref{Assumption4}.2) alongside the stated moment and density regularity conditions. For now, we leave the dependence structure unspecified to preserve generality.

With these conditions in hand, we present the first result of this paper.

\begin{proposition}\label{prop.1}
Let Assumptions \ref{Assumption1}.1.(a) and \ref{Assumption1}.2.(a) hold. There exists a time series $\{\mathbf{y}_t^* = \sum_{\ell=0}^\infty \mathbf{c}_{\ell}(\gamma_t) \mathbf{e}_{t-\ell} \mid t\in [T] \}$ such that  $\max_{t\in [T]}\|\mathbf{y}_t - \mathbf{y}_t^*\|_\nu=O(1/T)$, wherein $\nu$ is specified in Assumption \ref{Assumption1}.1.(a), 
\end{proposition}

By Proposition \ref{prop.1}, $\mathbf{y}_t^*$ serves as a stationary approximation of $\mathbf{y}_t$ for all $t$. The moments of $\mathbf{y}_t^*$ are functions of $\gamma_t$  only, and evolve smoothly over time, ensuring that a well-defined limiting Bahadur representation is attainable.

\subsection{Smoothing the Check Function}\label{Sec.22}

Having characterized the properties of $\{\mathbf{y}_t\}$, we present another result before moving on to the estimation procedure. To the best of our knowledge, the following result has not been formally presented elsewhere and is of independent interest to the broader field.

Building on \cite{Horowitz1998}, a growing body of quantile regression literature has advocated for the use of smoothed check functions (e.g., \citealp{HE2023367}). In this work, we also construct a smoothed check function within the framework of generalized functions, and establish a sharp rate of convergence between the standard check function and its smoothed counterpart. 

Formally, define for $\forall m>0$
 
\begin{eqnarray*}
\rho_{\tau,m}(u)=\int \rho_\tau(x)\phi_m(x-u)\mathrm{d}x\quad\text{with}\quad \phi_m(x)=m\cdot\phi(mx),
\end{eqnarray*}
where $\phi(\cdot)$ is a smooth function satisfying Assumption \ref{Assumption2} to be specified below. The function sequence $\{\rho_{\tau,m}(u)\}$ is called a regular sequence of $\rho_\tau(u)$ if $m$ takes positive integers only (\citealp{kanwal1983}), while in this article we allow $m$ to take any value no less than 1. This will become clearer after presenting Lemma \ref{LM1} below. 

To investigate $\rho_{\tau,m}(u)$, we impose the following conditions.

\begin{assumption}\label{Assumption2}
Suppose that $\phi(\cdot)$ is nonnegative, symmetric on $[-1,1]$, at least second order continuously differentiable, $\lim_{u\to \pm 1}\phi^{(\ell)}(u) =0$ for $\ell\in \{ 0,1\}$, and $\int_{-1}^1 \phi(u)dv=1$. 
\end{assumption}

Assumption \ref{Assumption2} imposes a set of standard conditions on the user-defined function $\phi(\cdot)$. For example, we can let

\begin{eqnarray*} 
\phi(u)=C\exp (-1/(1-u^2) )   \text{ for }|u|\le 1,\quad\text{and}\quad \phi(u)=0 \text{ elsewhere},
\end{eqnarray*}
where $C$ is chosen\footnote{Our numerical implementation suggests $C\approx 0.444$, which is carried out throughout all numerical studies of the paper. 
} to ensure $\int_{-1}^1 \phi(u)\mathrm{d}v=1$. 

\begin{lemma}\label{LM1}
Under Assumption \ref{Assumption2}, for $\forall m\in [1,\infty)$, 

\begin{enumerate}[leftmargin=*]
    \item $\rho_{\tau,m}(u)$ is convex;
    
    \item $0\le \rho_{\tau,m}(u)-\rho_\tau(u)\le \frac{1}{2m}$ for $|u|\le m^{-1}$, and $\rho_{\tau,m}(u)-\rho_\tau(u) =0$ elsewhere.
\end{enumerate}
\end{lemma}

\begin{figure}[htb!]
    \centering
    \includegraphics[scale=0.6]{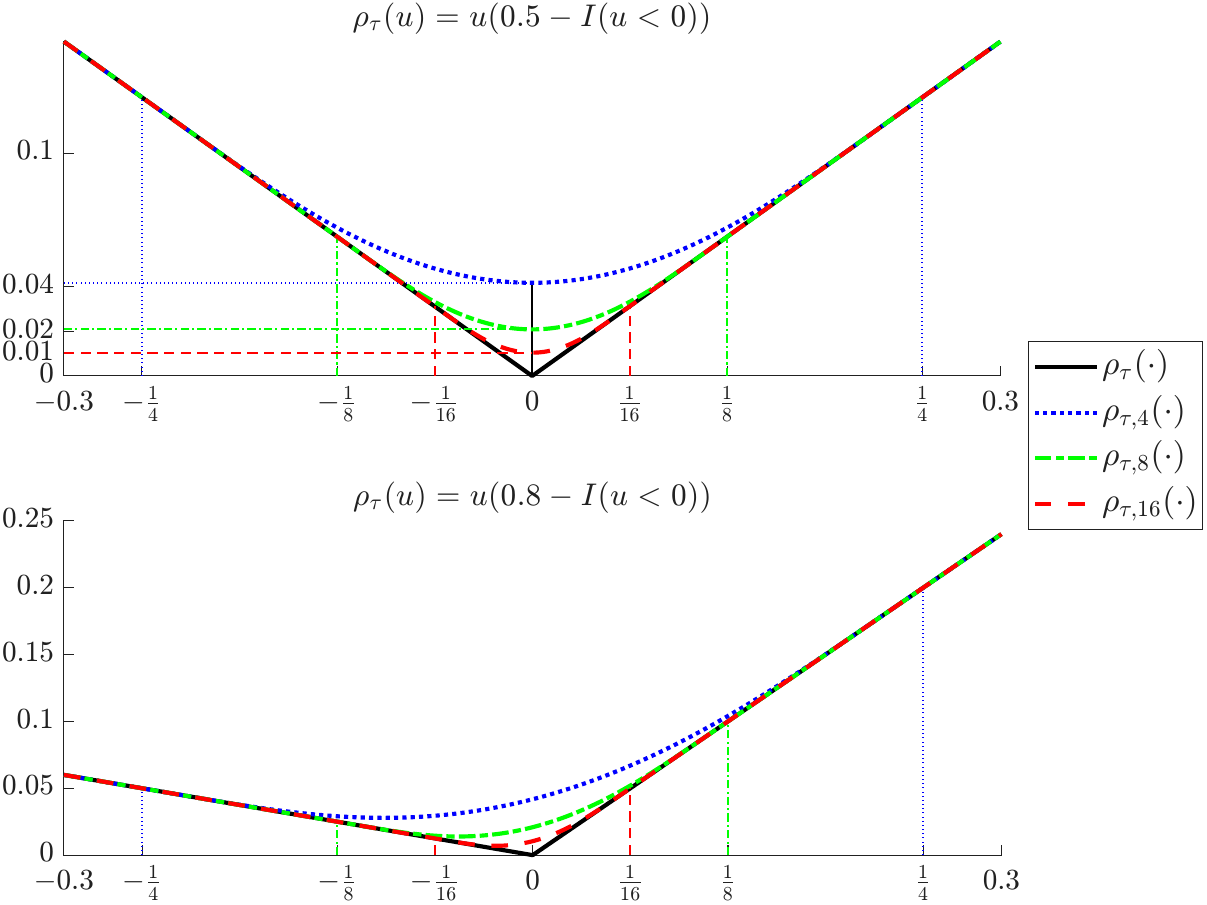}  
    \caption{Plots of $\rho_{\tau,m}(u)$ for $\tau =0.5, 0.8$}\label{Fig1}  
\end{figure}

Lemma \ref{LM1} demonstrates that our approximation not only smooths the check function, but also ensures that the approximation error can be made arbitrarily small. To illustrate this property, Figure \ref{Fig1} plots $\rho_{\tau,m}(u)$ for $\tau \in \{0.5, 0.8\}$  and $m \in \{4, 8, 16\}$. Clearly, $\rho_{\tau,m}(u)$ is smooth for all $m$. Figure \ref{Fig1} also highlights several key values, confirming that the analytical bounds established in Lemma \ref{LM1} hold numerically. With this result in hand, we are now ready to define the objective function and proceed with estimation.

\subsection{Estimation}\label{Sec.23}

By Assumption \ref{Assumption1}, we may write $\pmb{\beta}_{\ell,t} = \pmb{\beta}_\ell(\gamma_t)$ whenever necessary. For subsequent analysis, our goal is to recover 

\begin{eqnarray}\label{def.beta0}
    \pmb{\beta}_{\gamma}^0 \coloneqq (\pmb{\beta}_1(\gamma),\ldots, \pmb{\beta}_p(\gamma))\quad\text{for}\quad \forall\gamma\in (0,1),
\end{eqnarray}
of which the element-wise first and second derivatives will be employed in a few results below, and are defined as follows:

\begin{eqnarray}\label{def.beta12}
    \pmb{\beta}_{\gamma}^{j} \coloneqq (\pmb{\beta}_1^{(j)}(\gamma),\ldots, \pmb{\beta}_p^{(j)}(\gamma))\quad\text{for}\quad \forall j\in \{1,2\}.
\end{eqnarray}

As we have TV coefficients involved, we employ local-linear regression to conduct the estimation, and require the kernel function and the bandwidth to satisfy the following conditions.

\begin{assumption}\label{Assumption3}
\item 
\begin{enumerate}[leftmargin=*]
\item The kernel function $K(w) \ (\ge 0)$ is a symmetric Lipschitz-continuous function supported on $[-1,1]$, and $\int_{-1}^1K(w)\mathrm{d}w=1$.  
\item  The bandwidth satisfies $\delta\asymp T^{-\vartheta}$ with $\vartheta\ge1/5$ and $\frac{\log T}{T^{1-2/\nu}\delta} \rightarrow0$, where $\nu$ has been defined in Assumption \ref{Assumption1}.1.(a).

\end{enumerate}
\end{assumption}

In Assumption \ref{Assumption3}, the first condition is standard. The second condition ensures that $T\delta\to \infty$ and is used to control the tail behavior when deriving  the uniform bound. It is automatically satisfied whenever $\vartheta<1-2/\nu$.  In the literature, one would expect (i) $\delta\asymp T^{-1/5}$ in order to reach the optimal convergence rate, or (ii) $T\delta^5\to0$ in order to remove the nonparametric bias via an under-smoothing condition. The current Assumption \ref{Assumption3} allows for both cases.

We then present the following objective function:

\begin{eqnarray}\label{def.obj1}
    Q(\mathbf{B}^0, \mathbf{B}^1) =\frac{1}{T}\sum_{t=1}^{T}\sum_{\ell=1}^p\rho_{\tau, m} (y_{\ell,t} - \mathbf{x}_{t-1}^\top \mathbf{b}_{\ell}^{0} -\mathbf{z}_{t-1}^\top \mathbf{b}_{\ell}^1 ) K_\delta(\gamma_t -\gamma) ,
\end{eqnarray}
where $\mathbf{z}_{t-1}\coloneqq (\gamma_t-\gamma)\mathbf{x}_{t-1}$, $\mathbf{B}^0 \coloneqq (\mathbf{b}_{1}^0,\ldots, \mathbf{b}_{p}^0)$ and $\mathbf{B}^1 \coloneqq (\mathbf{b}_{1}^1,\ldots, \mathbf{b}_{p}^1)$ are two generic matrices, and $\mathbf{b}_{\ell}^0$ and $\mathbf{b}_{\ell}^1$ are respectively the counterparts of $\pmb{\beta}_{\ell}(\cdot)$ and $\pmb{\beta}_{\ell}^{(1)}(\cdot)$. Reiterating Remark \ref{def.rm1}, \eqref{def.obj1} in fact considers the estimation of each $\ell$ separately, thereby confirming that the remark holds. It aligns well with the simplest OLS regression for VAR models (e.g., \citealp[Chapter 3]{Lut2005}).

The estimates are thus obtained by 

\begin{eqnarray}\label{def.est}
    (\widehat{\mathbf{B}}^0, \widehat{\mathbf{B}}^1) = \argmin_{\mathbf{B}^0, \mathbf{B}^1} Q(\mathbf{B}^0, \mathbf{B}^1),
\end{eqnarray}
where $\widehat{\mathbf{B}}^{j} =(\widehat{\mathbf{b}}_1^{j},\ldots, \widehat{\mathbf{b}}_p^{j})$ with $j\in \{0, 1\}$.

To investigate \eqref{def.est}, we present the following theorem, wherein we firstly define a few variables and then give a Bahadur representation associated with the objective function \eqref{def.obj1}.

\begin{theorem}[Bahadur Representation]\label{TM.1} 
Let

\begin{eqnarray*}
    && \mathbb{B} \coloneqq \left\{ (\mathbf{B}^0, \mathbf{B}^1) \mid
    \|\Delta \mathbf{B}_\delta\| =o(1)\right\},\quad
    \Delta \mathbf{B}_\delta \coloneqq
    \mathbf B_\delta-\pmb\beta_{\gamma,\delta},\quad \mathbf B_\delta\coloneqq
    \VEC(\mathbf{B}^0, \delta\mathbf{B}^1),\notag\\
    &&\pmb\beta_{\gamma,\delta}\coloneqq
    \VEC(\pmb{\beta}_\gamma^0 ,\delta\pmb{\beta}_\gamma^1),\quad \pmb{\xi}_{\gamma,\delta}
    \coloneqq\VEC(\tfrac{\delta^2}{2}\pmb{\beta}_\gamma^2,
    \mathbf{0}_{pq\times p}),\quad 
    \pmb{\Sigma}_{\mathbf f,\gamma}
    \coloneqq\diag\{\mathbf D_{1,\gamma}^Q,\ldots,
    \mathbf D_{p,\gamma}^Q\},
\end{eqnarray*}
where $\mathbf D_{\ell,\gamma}^Q$ is defined in Assumption \ref{Assumption1}. Under Assumptions
\ref{Assumption1}-\ref{Assumption3} and $m\to\infty$,

\begin{eqnarray*}
&&
\sup_{(\mathbf{B}^0, \mathbf{B}^1)\in \mathbb{B}}
\frac{\left|
Q(\mathbf{B}^0, \mathbf{B}^1)-Q_{\mathbf e}
+Q_1(\mathbf{B}^0, \mathbf{B}^1)
-Q_2(\mathbf{B}^0, \mathbf{B}^1)
\right|}{
\|\mathbf B_\delta-\pmb\beta_{\gamma,\delta}\|^2
+\delta^4+\frac{\log T}{T\delta}}
=o_P(1)
\end{eqnarray*}
uniformly in $\gamma\in [\delta, 1-\delta]$, where

\begin{eqnarray*}
&& Q_{\mathbf{e}} = \frac{1}{T}\sum_{t=1}^{T}\sum_{\ell=1}^p \rho_{\tau, m}\left(e_{\ell,t} \right)  \cdot K_\delta(\gamma_t -\gamma)  ,\notag \\
&&Q_1(\mathbf{B}^0, \mathbf{B}^1) = -\frac{1}{T}\sum_{t=1}^{T}\sum_{\ell=1}^p \rho_{\tau, m}^{(1)} (e_{\ell,t}  ) \cdot  (\mathbf{x}_{t-1}^\top\pmb{\beta}_{\ell, t} -  \mathbf{x}_{t-1}^\top \mathbf{b}_{\ell}^0 -\mathbf{z}_{t-1}^\top \mathbf{b}_{\ell}^1) \cdot K_\delta(\gamma_t -\gamma),\notag\\
&&Q_2(\mathbf{B}^0, \mathbf{B}^1)
= \frac{1}{2}\int_{-1}^{1}
(\Delta \mathbf{B}_\delta-w^2\pmb{\xi}_{\gamma,\delta})^\top
\diag\{\pmb\Sigma_{\mathbf f,\gamma},
w^2\pmb\Sigma_{\mathbf f,\gamma}\}
(\Delta \mathbf{B}_\delta-w^2\pmb{\xi}_{\gamma,\delta})
K(w)\mathrm{d}w .
\end{eqnarray*}
\end{theorem}

Theorem \ref{TM.1} establishes a Bahadur representation for the objective function \eqref{def.obj1} that incorporates both the smoothed check function and the nonparametric kernel function. Consequently, it does not only capture the TV nature of the model, but also accounts for  the smoothing technique introduced in Section \ref{Sec.22}. Furthermore, this result demonstrates that the quantile regression is approximately equivalent to minimizing a quadratic form, thereby facilitating the analysis of the estimates  in \eqref{def.est}. 

Note also that the term $Q_2(\mathbf{B}^0, \mathbf{B}^1)$ can be further expanded as follows:

\begin{eqnarray}\label{def.Q2}
    Q_2(\mathbf{B}^0, \mathbf{B}^1)
    &=&
    \frac{1}{2}\int_{-1}^{1}
    \VEC(\mathbf{B}^0-\pmb{\beta}_\gamma^0
    -\tfrac{\delta^2w^2}{2}\pmb{\beta}_\gamma^2)^\top
    \pmb\Sigma_{\mathbf f,\gamma}
    \VEC(\mathbf{B}^0-\pmb{\beta}_\gamma^0
    -\tfrac{\delta^2w^2}{2}\pmb{\beta}_\gamma^2)
    K(w)\mathrm{d}w \notag\\
    &&
    +\frac{v_2}{2}
    \VEC(\delta\mathbf B^1-\delta\pmb{\beta}_\gamma^1)^\top
    \pmb\Sigma_{\mathbf f,\gamma}
    \VEC(\delta\mathbf B^1-\delta\pmb{\beta}_\gamma^1).
\end{eqnarray}

The right-hand side of \eqref{def.Q2} is convenient as it completely separates $\mathbf{B}^0$ and $\mathbf{B}^1$ in the quadratic form, thereby facilitating the derivation of the asymptotic distribution of $\widehat{\mathbf{B}}^0$ later on.

With Theorem \ref{TM.1} in hand, we can now establish the convergence rates for the estimators in \eqref{def.est} and derive the asymptotic distribution. To maintain the flow of the main text, two preliminary lemmas concerning the convergence rates are deferred to Appendix \ref{Appen.A3} of the main text. Building on them, we can then establish the resulting asymptotic distribution in the following theorem.

\begin{theorem}\label{TM.2} 
Let Assumptions \ref{Assumption1}-\ref{Assumption3} hold, and $\sqrt{T\delta}/m\to 0$. For $\forall \gamma\in (0,1)$,

\begin{eqnarray*}
    \sqrt{T\delta}\VEC\Big(\widehat{\mathbf{B}}^0 -\pmb{\beta}_{\gamma}^0-\frac{v_2\delta^2}{2}\pmb{\beta}_\gamma^2 \Big)\to_D N(\mathbf{0}, \mathbf{S}\pmb{\Omega}_\gamma \mathbf{S}^\top),
\end{eqnarray*}
where $\pmb\Omega_\gamma\coloneqq
 \diag\{\pmb\Sigma_{\mathbf f,\gamma},
 v_2\pmb\Sigma_{\mathbf f,\gamma}\}^{-1}
 \diag\{a_1\mathbf M_\gamma^Q,a_2\mathbf M_\gamma^Q\}
 \diag\{\pmb\Sigma_{\mathbf f,\gamma},
 v_2\pmb\Sigma_{\mathbf f,\gamma}\}^{-1},$  
$\mathbf M_\gamma^Q$ is defined in Assumption \ref{Assumption1}, and $\mathbf{S}=(\mathbf{I}_{p^2q},\mathbf{0}_{p^2q} )$ is a selection matrix.
\end{theorem}

For each $\ell\in[p]$, it is easy to show that $E [|\psi_\tau(e_{\ell,t})|^2\mid\mathcal F_{t-1}]=\tau(1-\tau).$ If, in addition, the components are conditionally independent over $\ell$ given $\mathcal F_{t-1}$, then $E [\pmb\psi_t\pmb\psi_t^\top
 \mid\mathcal F_{t-1}  ] = \tau(1-\tau)\mathbf I_p.$ Consequently, for some deterministic matrix $\pmb\Sigma_\gamma$, it is easy to obtain that
\begin{eqnarray*}
 \frac1{T\delta}\sum_{t=1}^{T}
 \mathbf W_{t,\gamma}\mathbf W_{t,\gamma}^\top
 K^2(w_{t,\gamma})
 &\to_P&
 \diag\{
 a_1\pmb\Sigma_\gamma,
 a_2\pmb\Sigma_\gamma
 \},
\end{eqnarray*}
then the restriction presented in Assumption \ref{Assumption1} can be further simplified as $\mathbf M_\gamma^Q=\tau(1-\tau)( \mathbf I_p\otimes\pmb\Sigma_\gamma ).$ This gives a special case for the covariance structure.  See \citet[Theorem 2]{Koenker01092006} for the scalar case. In general, $\mathbf M_\gamma^Q$ is retained because the components over $\ell$ may be contemporaneously dependent and their conditional second moments may depend on the lags involved. 

\medskip

Before moving on to impulse response analysis, we select the number of lags. For simplicity, we consider the selection on a set of grid points, $\mathbb{U} \coloneqq \{\tilde{\gamma}_1,\ldots, \tilde{\gamma}_L\}$, in which $L$ is finite and $\tilde{\gamma}_\ell\in (0,1)$ for all $\ell \in [L]$. Accordingly, we define an information criterion as follows:

\begin{eqnarray}\label{def.IC}
    \text{IC}(\tilde{q}) = \frac{1}{L}\sum_{\ell=1}^LQ_{\tilde{\gamma}_\ell \mid \tilde{q}}(\widehat{\mathbf{B}}_{ \mid \tilde{q}}^0, \widehat{\mathbf{B}}_{ \mid \tilde{q}}^1) + \omega_T\cdot \tilde{q},
\end{eqnarray}
where $Q_{\tilde{\gamma}_\ell \mid \tilde{q}}(\cdot, \cdot)$ corresponds to $Q(\cdot, \cdot)$ of \eqref{def.obj1} using $\tilde{q}$ lags and evaluated at $\gamma =\tilde{\gamma}_\ell$, and $(\widehat{\mathbf{B}}_{\mid \tilde{q}}^0, \widehat{\mathbf{B}}_{\mid \tilde{q}}^1)$ stand for the estimates obtained via minimizing $Q_{\tilde{\gamma}_\ell\mid \tilde{q}}( \mathbf{B}_{ \mid \tilde{q}}^0, \mathbf{B}_{ \mid \tilde{q}}^1)$. It should be understood that the dimensions of $(\mathbf{B}_{\mid \tilde{q}}^0, \mathbf{B}_{\mid \tilde{q}}^1)$ vary with respect to $\tilde{q}$.  

Accordingly, we select the optimal number of lags as follows:

\begin{eqnarray}\label{def.estq}
    \widehat{q} =\argmin_{0\le \tilde{q}\le q^*} \text{IC}(\tilde{q}),
\end{eqnarray}
where $q^*$ is a sufficiently large constant. 

\begin{theorem}\label{TM.3} 
    Under Assumptions \ref{Assumption1}-\ref{Assumption3}, suppose further that  (1) $\sqrt{T\delta}/m\to 0$, $\omega_T\to 0$, and $T\delta \omega_T\to \infty$, and (2) there is at least one $\ell\in [L]$ such that $\|\mathbf{A}_q(\tilde\gamma_\ell) \|\ne 0$. Then $\Pr(\widehat{q} =q)\to 1$.
\end{theorem}

Having estimated the key unknown quantities, including the lag order of model \eqref{def.md1}, we turn to conduct impulse response analysis in Section \ref{Sec.24}. However, the existing literature relies almost exclusively on a conditional-mean framework. In what follows, we propose to estimate impulse responses within a quantile regression setting, an approach for which, to the best of our knowledge, theoretical justification has yet to be established.

\subsection{Impulse Response Analysis}\label{Sec.24}

In this section, we define and estimate the impulse responses based on the TV quantile autoregression framework. Recall that the model \eqref{def.md1.simple} characterizes the one-step ahead conditional $\tau$-quantile of each component $y_{\ell,t}$ given the past information $\mathbf{x}_{t-1}$. However, multi-step ahead responses do not follow by iterating it, as the conditional quantile does not propagate linearly through the recursion (see Appendix \ref{Appen.A1} for details). Thereby, the response has to be estimated directly at each horizon. Throughout this section, we continue to suppress \(\tau\) in notation whenever no confusion arises.

To facilitate causal interpretation and establish the weak-dependence conditions required for the QIR analysis, we impose a recursive contemporaneous restriction together with a functional-dependence representation for the innovations. In particular, the conditional quantile restriction $E[\psi_\tau(e_{\ell,t})\mid\mathcal F_{t-1}]=0$ in Assumption \ref{Assumption1} does not generally imply an analogous restriction for multi-step QIR regression innovations, since conditional quantiles do not satisfy the law of iterated expectations. Furthermore, overlapping forecast horizons can induce serial dependence in these innovations, so that martingale-difference arguments is generally inapplicable. We therefore adopt the functional-dependence framework of \citet{Wu2005} to establish the weak-dependence properties required for QIR asymptotics.

\begin{assumption}\label{Assumption4}

\item 
\begin{enumerate}[leftmargin=*]
\item  The variables $(y_{1,t}, \ldots, y_{p,t})$ are ordered such that, for each
$j \in [p]$ and every $\tau \in (0,1)$,
\begin{eqnarray*}
  Q_\tau\!\left(y_{j,t} \,\middle|\,
      y_{1,t},\ldots,y_{j-1,t},y_{j+1,t},,\ldots ,y_{p,t},\,\mathbf{x}_{t-1}\right)= Q_\tau\!\left(y_{j,t} \,\middle|\, y_{1,t},\ldots,y_{j-1,t},\,\mathbf{x}_{t-1}\right).
\end{eqnarray*}
\item  Let $\{\pmb\eta_t\}$ be an i.i.d. auxiliary sequence and $\mathcal E_t=(\ldots,\pmb\eta_{t-1},\pmb\eta_t)$, and let the model innovation admit the representation $\mathbf e_t=G_e(\gamma_t,\mathcal E_t)$, where $\sup_{\gamma\in[0,1]}\|G_e(\gamma,\mathcal E_t)\|_\nu\le C$ and $\|G_e(\gamma,\mathcal E_t)
 -G_e(\gamma',\mathcal E_t)\|_\nu
 \le C|\gamma-\gamma'|.$ Let $\mathcal E_t^{*(k)}$ be obtained from $\mathcal E_t$ by replacing
$\pmb\eta_{t-k}$ with an independent copy and leaving all other coordinates
unchanged.  Then, for $k\ge0$,
\begin{eqnarray*}
 \mathfrak d_{\nu,e}(k)
 &\coloneqq&
 \sup_{\gamma\in[0,1]}
 \|G_e(\gamma,\mathcal E_t)
 -G_e(\gamma,\mathcal E_t^{*(k)})\|_\nu
 \le C\rho_e^k.
\end{eqnarray*}
 
\end{enumerate}    

\end{assumption}

Assumption \ref{Assumption4}.1 requires that the contemporaneous values $y_{j+1,t}, \ldots, y_{p,t}$ carry no additional information for the $\tau$-quantile of $y_{j,t}$ beyond $(y_{1,t},\ldots,y_{j-1,t}, \mathbf{x}_{t-1})$, which is the quantile analogue of Cholesky  identification in structural VARs \citep[e.g.,][]{Sims1980}. Assumption \ref{Assumption4}.2 adopts the functional dependence framework of \citet{Wu2005}, which nests Example \ref{Ex1} for instance as a special case (see Appendix \ref{Appen.A1} for detailed justification). Closely related constructions have been used for the univariate TV quantile regression setting \citep[e.g.,][]{ZhouWu2009}.  

For each pair $(i,j)\in[p]^2$, the response we seek measures how far the $\tau^{th}$ conditional quantile of the future outcome $y_{i,t+h}$ moves when the shock variable $y_{j,t}$ is raised by one unit, holding fixed the contemporaneous identification controls and the history. Writing $\mathcal C_t^{[j]} \coloneqq \{y_{1,t},\ldots,y_{j-1,t},\,\mathbf{x}_{t-1}\}$ for the conditioning set held fixed, this counterfactual quantile response is

\begin{eqnarray}\label{def.qcf}
    \Delta^{(h)}_{ij}(\gamma_t;s)
    \;\coloneqq\;
    Q_\tau\!\bigl(y_{i,t+h}\,\big|\,y_{j,t}=s+1,\;\mathcal C_t^{[j]}\bigr)
    - Q_\tau\!\bigl(y_{i,t+h}\,\big|\,y_{j,t}=s,\;\mathcal C_t^{[j]}\bigr).
\end{eqnarray}
This response cannot be recovered by iterating the one-step ahead quantile VAR forward as the quantile operator is not additive.  Additionally, the $h$-step ahead conditional quantile is nonlinear in the conditioning variables, so \eqref{def.qcf} varies with the evaluation point $s$ and is not a single coefficient.

To obtain the best linear summary of \eqref{def.qcf}, we introduce some additional notation:

\begin{eqnarray*}
    \mathbf x_t^{[j]} \coloneqq (y_{1,t},\ldots,y_{j,t},\;\mathbf{x}_{t-1}^\top )^\top \in \mathbb{R}^{j+pq},
\end{eqnarray*}
and, for $\forall \gamma\in[0,1]$, let

\begin{eqnarray*}
    &&\tilde{\mathbf e}_t(\gamma) \coloneqq G_e(\gamma,\mathcal E_t),\quad \tilde{\mathbf y}_t(\gamma) \coloneqq \sum_{\ell=0}^{\infty} \mathbf c_\ell(\gamma)\tilde{\mathbf e}_{t-\ell}(\gamma),\quad\tilde{\mathbf x}_{t-1}(\gamma)\coloneqq\VEC (\tilde{\mathbf y}_{t-1}(\gamma),\ldots,\tilde{\mathbf y}_{t-q}(\gamma) ),\notag \\
    &&\tilde{\mathbf x}_t^{[j]}(\gamma)
    \coloneqq
     (
    \tilde y_{1,t}(\gamma),\ldots,
    \tilde y_{j,t}(\gamma),
    \tilde{\mathbf x}_{t-1}(\gamma)^\top
     )^\top.
\end{eqnarray*}

Recall that Proposition \ref{prop.1} introduces the approximation process \(\mathbf y_t^*\). We now construct a further approximation, \(\tilde{\mathbf y}_t(\gamma)\), by holding the time index \(\gamma_t\) fixed at \(\gamma\) in the innovation generating process. For each fixed \(\gamma\), the resulting process is stationary and serves as a theoretical device for the QIR asymptotic analysis. In particular, it provides a basis for defining the population impulse response function, establishing the CLT for the QIR estimator, and characterizing the associated long-run covariance. Lemma \ref{Appendix.B4} links the two approximating processes and establishes bounds on the corresponding approximation errors.
 
Formally, we introduce the following definition for the TV-QIR function.

\begin{definition}\label{def.tvqirf}
 We define the  best linear projection of the time-varying quantile impulse response  (TV-QIR) of variable $y_{i,t+h}$ to a unit shock to variable $y_{j,t}$ as
\begin{eqnarray*}
    \Psi^{(h)}_{ij}(\gamma_t)
    \;\coloneqq\;
    \pmb{\iota}_j^\top\pmb{\alpha}^{(h)}_{ij}(\gamma_t),
\end{eqnarray*}
where $\pmb{\iota}_j \in \mathbb{R}^{j+pq}$ is the $j^{th}$ standard selection vector identifying the coefficient on $y_{j,t}$, and $\pmb{\alpha}^{(h)}_{ij}(\gamma)$ is the smooth coefficient function of $\gamma$ that solves

\begin{eqnarray}\label{def.alpha.proj}
    \pmb{\alpha}^{(h)}_{ij}(\gamma)
    &\coloneqq&
    \argmin_{\mathbf{a}\in\mathbb{R}^{j+pq}}
    E [
    \rho_\tau (
    \tilde y_{i,t+h}(\gamma)
    -\tilde{\mathbf x}_t^{[j]}(\gamma)^\top\mathbf a
     ) ].
\end{eqnarray}
\end{definition}

Relative to the literature of quantile impulse responses \citep[e,g,][]{Ruzicka2024,han2024estimation}, our theory explicitly targets the best linear quantile projection, allowing the multi-step conditional quantile to be nonlinear. Consequently, the validity of inference for this projection parameter does not rely on correct linear specification of the multi-step conditional quantile. The counterfactual response \eqref{def.qcf} coincides with $\Psi^{(h)}_{ij}( \gamma)$ for evaluation point $s$  under the additional  restriction

\begin{eqnarray*}
    Q_\tau  ( \tilde y_{i,t+h}(\gamma) \mid \tilde{\mathbf x}_t^{[j]}(\gamma)) =\tilde{\mathbf x}_t^{[j]}(\gamma)^\top \pmb\alpha^{(h)}_{ij}(\gamma).
\end{eqnarray*}

In the general case, $\Psi^{(h)}_{ij}( \gamma)$ remains the best linear $\tau$-quantile approximation in the sense of \cite{angrist2006quantile}. 

The coefficient $\pmb{\alpha}^{(h)}_{ij}(\gamma)$ is then estimated by local-linear quantile regression using the smoothed check function $\rho_{\tau,m}$ of Section \ref{Sec.22}. With the slope regressor $\mathbf z_t^{[j]}\coloneqq (\gamma_t-\gamma)\mathbf x_t^{[j]}$, the  objective function at horizon $h$ is

\begin{eqnarray}\label{def.lp.obj}
    Q^{(h)}_{ij}(\mathbf{a}^0, \mathbf{a}^1)
    = \frac{1}{T-h}\sum_{t=1}^{T-h}
      \rho_{\tau,m} (
        y_{i,t+h}
        - \mathbf x_t^{[j]\top}\mathbf{a}^0
        - \mathbf z_t^{[j]\top}\mathbf{a}^1
      )
      K_\delta(\gamma_t - \gamma),
\end{eqnarray}
and the estimators are obtained as
\begin{eqnarray*}
    (\widehat{\mathbf{a}}^{(h,0)}_{ij}(\gamma),\;
           \widehat{\mathbf{a}}^{(h,1)}_{ij}(\gamma))
    = \argmin_{\mathbf{a}^0,\,\mathbf{a}^1}
       Q^{(h)}_{ij}(\mathbf{a}^0, \mathbf{a}^1),
    \qquad
    \widehat{\Psi}^{(h)}_{ij}(\gamma)
    = \pmb{\iota}_j^\top\widehat{\mathbf{a}}^{(h,0)}_{ij}(\gamma).
\end{eqnarray*}

The QIR objective function \eqref{def.lp.obj} has a structure similar to the quantile VAR objective function \eqref{def.obj1}. Under additional assumptions, we can show that the asymptotic properties of $\widehat{\Psi}^{(h)}_{ij}(\gamma)$ parallel to those of the quantile VAR estimates.

To proceed, define the filtration generated by the auxiliary innovations by $\mathcal H_t\coloneqq\sigma(\pmb\eta_s:s\le t)$ and write $d_{\ell,t}(w;\gamma)\coloneqq f_{\tilde e_{\ell,t}(\gamma)\mid\mathcal H_{t-1}}
(w\mid\mathcal H_{t-1})$.  Let $f^{(h)}_{ij,t,T}(\cdot\mid\mathbf x_t^{[j]})$ and $f^{(h)}_{ij}(\cdot\mid\mathbf x;\gamma)$ be the conditional densities of the actual and stationary-approximation outcomes, respectively. Define

\begin{eqnarray*}
g^{(h)}_{ij,t,T}(w\mid\mathbf x_t^{[j]})
&\coloneqq&
f^{(h)}_{ij,t,T} (
\mathbf x_t^{[j]\top}\pmb\alpha^{(h)}_{ij}( \gamma_t)+w
\mid \mathbf x_t^{[j]} ),\notag\\
\tilde g^{(h)}_{ij}(w\mid\mathbf x;\gamma)
&\coloneqq&f^{(h)}_{ij} (
\mathbf x^\top\pmb\alpha^{(h)}_{ij}( \gamma)+w
\mid \mathbf x;\gamma ).
\end{eqnarray*}

For the long-run variance, we define the stationary approximation projection residual and score by

\begin{eqnarray*}
 \tilde u^{(h)}_{i,t}(\gamma)
 \coloneqq
 \tilde y_{i,t+h}(\gamma)
 -\tilde{\mathbf x}_t^{[j]}(\gamma)^\top
 \pmb\alpha^{(h)}_{ij}( \gamma),\quad
 \tilde{\mathbf s}^{(h)}_{ij,t}(\gamma)
 &\coloneqq&
 \psi_\tau (\tilde u^{(h)}_{i,t}(\gamma) )
 \tilde{\mathbf x}_t^{[j]}(\gamma).
\end{eqnarray*}

\begin{assumption}\label{Assumption5}

\item
\begin{enumerate}[leftmargin=*] 
\item For $i\in[p]$, suppose $\max_{t\in[T]}\sup_{\gamma\in[0,1]}\sup_{w\in\mathbb R}
 d_{i,t}(w;\gamma)
 \le C$ almost surely.  Moreover, for some $\varepsilon_0>0$, $|w|\le\varepsilon_0$,

\begin{eqnarray*}
 &&\max_{t\in [T]} |
 g^{(h)}_{ij,t,T}(w\mid\mathbf x_t^{[j]})
 -g^{(h)}_{ij,t,T}(0\mid\mathbf x_t^{[j]})|
 \le C|w|.
\end{eqnarray*}
almost surely.

\item The fitted stationary residual-density ordinate
$\tilde g^{(h)}_{ij}(0\mid\mathbf x;\gamma')$ is Lipschitz continuous uniformly over
$\gamma'$ in a neighborhood of $\gamma\in(0,1)$ and
$\mathbf x\in\mathbb R^{j+pq}$.  In addition,
\begin{eqnarray*}
 \max_{
 |\gamma_t-\gamma|\le\delta}
  \|
 g^{(h)}_{ij,t,T}(0\mid\mathbf x_t^{[j]})
 -\tilde g^{(h)}_{ij} (
 0\mid\tilde{\mathbf x}_t^{[j]}(\gamma_t);\gamma_t
  )
 \|_2
 &=&o(1).
\end{eqnarray*}

\item Define $\mathbf D^{(h)}_{ij}(\gamma) \coloneqq E [ \tilde g^{(h)}_{ij} ( 0\mid\tilde{\mathbf x}_t^{[j]}(\gamma);\gamma ) \tilde{\mathbf x}_t^{[j]}(\gamma) \tilde{\mathbf x}_t^{[j]}(\gamma)^\top ]$ and assume 
$\lambda_{\min} (\mathbf D^{(h)}_{ij}(\gamma) )>0$. Additionally, let $\mathbf{V}^{(h)}_{ij}( \gamma)=\sum_{k=-\infty}^{\infty} E [ \tilde{\mathbf s}^{(h)}_{ij,t}(\gamma)  \tilde{\mathbf s}^{(h)}_{ij,t+k}(\gamma)^\top ]$, where $\mathbf{V}^{(h)}_{ij}( \gamma)$ is positive definite.

\item $\pmb\alpha^{(h)}_{ij}( \gamma)$ is twice continuously differentiable in a neighbourhood of $\gamma$.
\end{enumerate}
\end{assumption}

The regularity conditions on the density function in Assumption \ref{Assumption5} are standard in local quantile asymptotic properties.  Analogous boundedness and Lipschitz conditions on conditional and stationary approximation densities are used by \citet{ZhouWu2009}, while \citet{Koenker01092006} impose conditional density regularity to obtain the local-quadratic expansion for quantile autoregression.  Our theory only requires this condition at the fitted point and is therefore weaker than a uniform approximation of the full conditional density.

\begin{theorem}\label{TM.4}
Under Assumptions \ref{Assumption1}-\ref{Assumption5}, and
$\sqrt{T\delta}/m\to 0$, for each pair $(i,j)$ and $\forall\gamma\in(0,1)$,
\begin{eqnarray*}
    \sqrt{T\delta}\left(
      \widehat{\Psi}^{(h)}_{ij}( \gamma)
      - \Psi^{(h)}_{ij}( \gamma)
      - \frac{v_2\delta^2}{2}\Psi^{(h,2)}_{ij}( \gamma)
    \right) \to_D
    N (0,\;\sigma^{(h)2}_{ij}( \gamma) ),
\end{eqnarray*}
where $\Psi^{(h,2)}_{ij}( \gamma)=\pmb{\iota}_j^\top\pmb{\alpha}^{(h,2)}_{ij}( \gamma)$ and $\sigma^{(h)2}_{ij}( \gamma) = a_1\pmb{\iota}_j^\top\big(\mathbf{D}^{(h)}_{ij}(\gamma)\big)^{-1} \mathbf{V}^{(h)}_{ij}( \gamma)\big(\mathbf{D}^{(h)}_{ij}(\gamma)\big)^{-1}\pmb{\iota}_j$,
in which $\pmb{\alpha}^{(h,2)}_{ij}( \gamma)$ denotes the second derivative of $\pmb{\alpha}^{(h)}_{ij}( \gamma)$ with respect to $\gamma$.
\end{theorem}

\medskip

For feasible point-wise inference, we introduce the heteroskedasticity and autocorrelation consistent (HAC) estimator for the asymptotic covariance. Let $\mathcal L(\cdot)$, $b_T$ and $\ell_{H,T}$ be the kernel function, bandwidth and truncation parameter, respectively, used in the construction of HAC estimator, and set $\mathcal L_{b_T}(u)\coloneqq b_T^{-1}\mathcal L(u/b_T)$. Define the  residual  

\begin{eqnarray*}
    \widehat u^{(h)}_{ij,t}(\gamma) \coloneqq y_{i,t+h}-\mathbf x_t^{[j]\top} \widehat{\mathbf a}^{(h,0)}_{ij}(\gamma),
\end{eqnarray*}
and the sample scores:

\begin{eqnarray*}
    \widehat{\mathbf s}^{(h)}_{ij,t}(\gamma) \coloneqq
 \psi_\tau (\widehat u^{(h)}_{ij,t}(\gamma) )
 \mathbf x_t^{[j]}\quad\text{and}\quad \overline{\mathbf s}^{(h)}_{ij}(\gamma)
 \coloneqq \left(\sum_{t=1}^{T-h}K_{\delta}(\gamma_t-\gamma)\right)^{-1}\sum_{t=1}^{T-h}K_{\delta}(\gamma_t-\gamma)\widehat{\mathbf s}^{(h)}_{ij,t}(\gamma).
\end{eqnarray*}

We introduce the following estimators for $\mathbf{D}^{(h)}_{ij}(\gamma)$ and $\mathbf{V}^{(h)}_{ij}(\gamma)$:

\begin{eqnarray*}
 \widehat{\mathbf D}^{(h)}_{ij}(\gamma)
 &\coloneqq&
 \frac1{T-h}\sum_{t=1}^{T-h}
 \mathcal L_{b_T}\!\left(\widehat u^{(h)}_{ij,t}(\gamma)\right)
 \mathbf x_t^{[j]}\mathbf x_t^{[j]\top}
 K_\delta(\gamma_t-\gamma),\notag \\
  \widehat{\mathbf\Omega}^{(h)}_{ij}( \gamma)
 &\coloneqq&
 \widehat{\mathbf\Gamma}^{(h)}_{ij,0}(\gamma)
 +\sum_{k=1}^{\ell_{H,T}}
 \left(1-\frac{k}{\ell_{H,T}+1}\right)
 \left\{
 \widehat{\mathbf\Gamma}^{(h)}_{ij,k}(\gamma)
 +\widehat{\mathbf\Gamma}^{(h)}_{ij,k}(\gamma)^{\top}
 \right\}.\notag
\end{eqnarray*}
For $0\le k\le\ell_{H,T}$,
\begin{eqnarray*}
 \widehat{\mathbf\Gamma}^{(h)}_{ij,k}(\gamma)
 &\coloneqq&
 \frac1{(T-h)\delta}
 \sum_{t=1}^{T-h-k}
 K(w_{t,\gamma})K(w_{t+k,\gamma})
 \left\{\widehat{\mathbf s}^{(h)}_{ij,t}(\gamma)
 -\overline{\mathbf s}^{(h)}_{ij}(\gamma)\right\}
 \left\{\widehat{\mathbf s}^{(h)}_{ij,t+k}(\gamma)
 -\overline{\mathbf s}^{(h)}_{ij}(\gamma)\right\}^{\top}.
\end{eqnarray*}
Using $\widehat{\mathbf D}^{(h)}_{ij}(\gamma)$ and $\widehat{\mathbf\Gamma}^{(h)}_{ij,k}(\gamma)$, we can construct the HAC estimator for $\sigma^{(h)2}_{ij}(\gamma)$:
\begin{eqnarray*}
 \widehat\sigma^{(h)2}_{ij}(\gamma)
 &\coloneqq&
 \pmb\iota_j^\top
 \widehat{\mathbf D}^{(h)}_{ij}(\gamma)^{-1}
 \widehat{\mathbf\Omega}^{(h)}_{ij}(\gamma)
 \widehat{\mathbf D}^{(h)}_{ij}(\gamma)^{-1}
 \pmb\iota_j.
\end{eqnarray*}

Under a set of conventional regularity conditions on the kernel function, bandwidth, and truncation parameter, we can establish the consistency of $\widehat\sigma^{(h)2}_{ij}(\gamma)$. Specifically, we introduce these conditions in the following assumption.

\begin{assumption}\label{Assumption6}
\item
\begin{enumerate}[leftmargin=*]
\item The kernel $\mathcal L(w)$ is  Lipschitz continuous  on  the support $[-1,1]$.
\item The  bandwidth $b_T$ and the truncation order $\ell_{H,T}$ satisfy $b_T\to0$, $\ell_{H,T}\to\infty$, and 
\begin{eqnarray*}
 &&
 \frac{\log^4(1/b_T)}{T\delta b_T}\to0,\quad
 \frac{\delta^2+(T\delta)^{-1/2}}{b_T^2}\to0,\quad
 \ell_{H,T}\left\{\delta+(T\delta)^{-1/2}\right\}^{1/5}\to0.
\end{eqnarray*}
\end{enumerate}
\end{assumption}
Assumption \ref{Assumption6} is used only for feasible covariance estimation and can be easily justified. For example, the conditions can be satisfied by setting $\delta^2=O((T\delta)^{-1/2})$, $b_T=(T\delta)^{-1/8}$ and $\ell_{H,T}=\lceil c_\ell\log(T\delta)\rceil$ with the constant $c_\ell>0$.

The consistency of the HAC estimator is established in the following lemma. 
\begin{theorem}\label{TM.5}
Under Assumptions \ref{Assumption1}--\ref{Assumption6}, and
$\sqrt{T\delta}/m\to 0$, for each pair $(i,j)$ and $\forall \gamma\in(0,1)$,
\begin{eqnarray*}
 \widehat\sigma^{(h)2}_{ij}(\gamma)
 &\to_P&
 \sigma^{(h)2}_{ij}(\gamma).
\end{eqnarray*}
\end{theorem}

In addition to the consistent estimation of the asymptotic covariance, we can use jackknife method to correct the leading $O(\delta^2)$ bias in the CLT. The details are provided in Appendix \ref{Appen.B1}.

\section{Simulation}\label{Sec.3}

In this section, we conduct simulation to examine the theoretical findings in Section \ref{Sec.2}. Before proceeding, we fix a few quantities for simplicity.

We adopt the Epanechnikov kernel for $K(\cdot)$ defined in Assumption \ref{Assumption3}. As mentioned in Section \ref{Sec.2}, the optimal $\delta$ should be proportional to $T^{-1/5}$. To check the robustness on the choices of $\delta$ and $m$, we consider the following combination: $\delta_j =c_jT^{-1/5}$ and $m_j =(T\delta_j)^{3/4}$, where $c_1 =1.5$, $c_2 =1.75$, $c_3 =2$ and $c_4 =2.25$.  By doing so, it is easy to check that $\sqrt{T\delta}/m\to 0$. Additionally, we always let $\omega_T =\sqrt{\tau(1-\tau)\log T}/(T\delta)$ in order to select the optimal lag. Some additional details regarding numerical implementation are provided in Appendix \ref{Appen.A4} for the sake of space.  

Given that \eqref{def.est} has no closed-form solution, we evaluate $\gamma$ and $\tau$ over the following finite sets to alleviate the computational burden:

\begin{eqnarray*}
\gamma \in \mathbb{D}_\gamma = \{ 0.2, 0.4, 0.6, 0.8\}\quad \text{and}\quad \tau\in \mathbb{D}_\tau =\{0.1,0.3,0.5, 0.7,0.9 \}.
\end{eqnarray*}
For each data--generating process (DGP) below, we perform $R = 400$ replications.

To carry on dynamic quantile regression, a fundamental requirement is for $\forall \ell \in [p]$
 
\begin{eqnarray}\label{def.sim1}
    \Pr(e_{\ell,t}\le 0\mid \mathcal{F}_{t-1}) =\tau,
\end{eqnarray}
where $\mathbf{e}_t =(e_{1,t},\ldots, e_{p,t})^\top$. We consider two DGPs below to account for \eqref{def.sim1}.

\medskip

\textbf{DGP 1} --- We start with a simple case, which is also a special scenario of Example \ref{Ex1}: 

\begin{eqnarray*}
    \mathbf{y}_t = \pmb{\alpha}_t(\mathbf{u}_t) + \mathbf{A}_{1,t}\mathbf{y}_{t-1},
\end{eqnarray*}
where $\pmb{\alpha}_t(\mathbf{u}_t) = (\alpha_{1,t}(u_{1, t}),\ldots, \alpha_{p,t}(u_{p, t}))^\top$,  $u_{\ell, t}\sim U(0,1)$, and $\mathbf{A}_{1,t} =(\mathbf{A}_{1 1,t},\ldots, \mathbf{A}_{1p,t})^\top$. Accordingly, for $\forall\ell\in [p]$ and $\forall \tau\in(0,1)$, we can rewrite the model as

\begin{eqnarray*}
    y_{\ell, t} = \alpha_{\ell,t}(\tau) + \mathbf{y}_{t-1}^\top \mathbf{A}_{1\ell,t} +e_{\ell, t},
\end{eqnarray*}
where $e_{\ell,t}\coloneqq \alpha_{\ell,t}(u_\ell) -\alpha_{\ell,t}(\tau)$. Thus, simple algebra shows that

\begin{eqnarray*}
    \Pr(e_{\ell,t}\le 0) =\Pr( \alpha_{\ell,t}(u_\ell) \le \alpha_{\ell,t}(\tau)) = \tau,
\end{eqnarray*}
which is true provided $\alpha_{\ell,t}(u_{\ell, t})$ is monotonically increasing in $u_{\ell, t}$.

That said, we let 

\begin{eqnarray*}
    \pmb{\alpha}_t(\mathbf{u}_t) =\begin{pmatrix}
        u_{1,t} - 2+\exp(\gamma_t) \\
        u_{2,t} - 2+\exp(\gamma_t)
    \end{pmatrix}\quad\text{and}\quad\mathbf{A}_{1,t} =\begin{pmatrix}
        \frac{1}{2}\exp(\gamma_t-1) & 0 \\
        0.3\cos(\gamma_t \pi)& \frac{1}{2}\exp(\gamma_t-1)
    \end{pmatrix}.
\end{eqnarray*}
Based on the above design, it is obvious that $q=1$ and $p=2$. 

Firstly, we assume that $p$ is known, and examine the estimation defined in \eqref{def.est}. As the above DGP includes the intercept term, we modify the objective function defined in \eqref{def.obj1} in an obvious minor to accommodate this change. For $\forall \tau\in \mathbb{D}_\tau$, we estimate 

\begin{eqnarray*}
    \pmb{\alpha} (\gamma\mid \tau) = \exp(\gamma) +\tau -2 \quad\text{and}\quad\mathbf{A}_{1}(\gamma\mid \tau) =\begin{pmatrix}
        \frac{1}{2}\exp(\gamma -1) & 0 \\
        0.3\cos(\gamma \pi)& \frac{1}{2}\exp(\gamma -1)
    \end{pmatrix}.
\end{eqnarray*}
After $R$ replications, we report the following measures

\begin{eqnarray*}
    \text{RMSE}_{\pmb{\alpha}} &=&\left\{\frac{1}{R \cdot \sharp \mathbb{D}_\tau\cdot \sharp \mathbb{D}_\gamma}\sum_{i=1}^R  \sum_{\tau \in \mathbb{D}_\tau} \sum_{\gamma\in \mathbb{D}_\gamma} \| \widehat{\pmb{\alpha}}_i(\gamma\mid \tau)-\pmb{\alpha}(\gamma\mid \tau)\|^2 \right\}^{1/2}, \notag \\
    \text{RMSE}_{\mathbf{A}} &=&\left\{\frac{1}{R \cdot \sharp \mathbb{D}_\tau\cdot \sharp \mathbb{D}_\gamma}\sum_{i=1}^R  \sum_{\tau \in \mathbb{D}_\tau} \sum_{\gamma\in \mathbb{D}_\gamma} \| \widehat{\mathbf{A}}_{1,i}(\gamma\mid \tau)-\mathbf{A}_1(\gamma\mid \tau)\|^2 \right\}^{1/2} ,
\end{eqnarray*}
where $\widehat{\pmb{\alpha}}_i(\gamma\mid \tau)$ and $\widehat{\mathbf{A}}_{1,i}(\gamma\mid \tau)$ stand for the estimates of $\pmb{\alpha}(\gamma\mid \tau)$ and $\mathbf{A}_1(\gamma\mid \tau)$ respectively at the $i^{th}$ replication.

Also, we examine the selection of the lag terms via \eqref{def.estq}, and let $\mathbb{U} \equiv\mathbb{D}_\gamma$, where $\mathbb{U}$ is defined in \eqref{def.IC}. We define the following measure:

\begin{eqnarray*}
    \text{S}_{q}= \frac{1}{R \cdot \sharp \mathbb{D}_\tau}\sum_{i=1}^R  \sum_{\tau \in \mathbb{D}_\tau}  I(\widehat{q}_{i\mid\tau}=q),
\end{eqnarray*}
where $\widehat{q}_{i\mid\tau}$ stands for the value $\widehat{q}$ obtained at the $i^{th}$ replication for $\forall \tau\in \mathbb{D}_\tau$.

We summarize the results in Table \ref{Tb.1} below. As expected, RMSEs decrease as $T$ goes up. Also, the selection of the lag terms is accurate in general. Moreover, the results do not vary with respect to $c_j$ very much showing robustness on the choices of $\delta$ and $m$.

\textbf{DGP 2} --- The DGP 1 serves as a simple extension of \cite{Koenker01092006}, wherein there is no correlation between $e_{1,t}$ and $e_{2,t}$. Therefore, the connection between the elements of $\mathbf{y}_t$ is fully captured by the coefficient matrix, i.e., $\mathbf{A}_{1,t}$. In what follows, we consider a more complex setting, so that the connection between the elements of $\mathbf{y}_t$ is also captured by the TV dependence among the elements of the error term.

Specifically, we consider the structure: $\mathbf{e}_t = \pmb{\pi}_t \pmb{\epsilon}_t-\pmb{\theta}_t$, where $\pmb{\epsilon}_t\sim N(\pmb{\mu}_\epsilon, \mathbf{I}_p)$ is a $p\times 1$ random vector, $\pmb{\pi}_t =(\pmb{\pi}_{1,t},\ldots, \pmb{\pi}_{p,t})^\top$ is a $p\times p$ deterministic matrix, and $\pmb{\theta}_t =(\theta_{1,t},\ldots, \theta_{p,t})^\top$ is a $p\times 1$ deterministic vector. For $\forall \ell \in [p]$, \eqref{def.sim1} is equivalent to 

\begin{eqnarray}\label{def.sim2}
    \Pr(\pmb{\pi}_{\ell,t}^\top \pmb{\epsilon}_t -\theta_{\ell,t}\le 0)=\tau.
\end{eqnarray}

To ensure \eqref{def.sim2}, we now provide some direct calculation to demonstrate the possible choices of $\pmb{\pi}_t$ and $\pmb{\theta}_t$. Simple algebra shows that

\begin{eqnarray*}
    \|\pmb{\pi}_{\ell,t}\|_2^{-1}\pmb{\pi}_{\ell,t}^\top \pmb{\epsilon}_t -\|\pmb{\pi}_{\ell,t}\|_2^{-1}\pmb{\pi}_{\ell,t}^\top \pmb{\mu}_\epsilon -c_\tau\sim N(-c_\tau, 1),
\end{eqnarray*}
where $c_\tau =F^{-1}(\tau)$, and $F(\cdot)$ is the cumulative distribution function of $N(0,1)$. For $\forall \ell\in [p]$, we thus require that

\begin{eqnarray*}
    \|\pmb{\pi}_{\ell,t}\|_2^{-1} \theta_{\ell,t}  = \|\pmb{\pi}_{\ell,t}\|_2^{-1}\pmb{\pi}_{\ell,t}^\top \pmb{\mu}_\epsilon + c_\tau .
\end{eqnarray*}
Consequently, we obtain that $\pmb{\pi}_t\pmb{\mu}_\epsilon-\pmb{\theta}_t= -c_\tau \pmb{\Pi}_t\mathbf{1}_p,$ where $\pmb{\Pi}_t =\diag\{ \|\pmb{\pi}_{1,t}\|_2 ,\ldots, \|\pmb{\pi}_{p,t}\|_2\}$.

We can now calculate the quantities $E[\mathbf{e}_t]$ and $E[\mathbf{e}_t\mathbf{e}_t^\top]$:

\begin{eqnarray*}
    E[\mathbf{e}_t] = - c_\tau \pmb{\Pi}_t\mathbf{1}_p,
\end{eqnarray*}
and

\begin{eqnarray*}
    E[\mathbf{e}_t \mathbf{e}_t^\top] =\pmb{\pi}_t\pmb{\pi}_t^\top +(\pmb{\pi}_t\pmb{\mu}_\epsilon-\pmb{\theta}_t)(\pmb{\pi}_t\pmb{\mu}_\epsilon-\pmb{\theta}_t)^\top =\pmb{\pi}_t\pmb{\pi}_t^\top +c_\tau^2\pmb{\Pi}_t\mathbf{1}_p\mathbf{1}_p^\top \pmb{\Pi}_t.
\end{eqnarray*}
If we require further that  $\|\pmb{\pi}_{\ell,t}\|_2=1$, $\pmb{\pi}_t\pmb{\mu}_\epsilon-\pmb{\theta}_t=- c_\tau \pmb{\Pi}_t\mathbf{1}_p$ reduces to $\pmb{\pi}_t\pmb{\mu}_\epsilon-\pmb{\theta}_t= -c_\tau \mathbf{1}_p.$ We obtain the simplified moments as follows:

\begin{eqnarray*}
    E[\mathbf{e}_t]=-c_\tau \mathbf{1}_p\quad\text{and}\quad E[\mathbf{e}_t\mathbf{e}_t^\top]= \pmb{\pi}_t \pmb{\pi}_t^\top + c_\tau^2 \mathbf{1}_p \mathbf{1}_p^\top .
\end{eqnarray*}
We would like to emphasize that the restriction $\|\pmb{\pi}_{\ell,t}\|_2=1$ is simply for the ease of computation, and it is unnecessary in practice.

Based on the above calculation, we let $p=2$, and  for $t\ge 1$

\begin{eqnarray*}
     \pmb{\epsilon}_t \sim N(\mathbf{1}_2, \mathbf{I}_2)\quad\text{and}\quad \pmb{\pi}_t =\pmb{\pi} (\gamma_t) =\begin{pmatrix}
        1 & 0 \\
        \sin(\gamma_t\pi) & \cos(\gamma_t\pi)
    \end{pmatrix}.
\end{eqnarray*}
Correspondingly, $\pmb{\theta}_t=\pmb{\theta}(\gamma_t)= (1, \sin(\gamma_t\pi)+\cos(\gamma_t\pi))^\top+c_\tau \mathbf{1}_p$. Additionally, we let for $\forall t\le 0$

\begin{eqnarray*}
    E[\mathbf{e}_t]=E[\mathbf{e}_0] =-c_\tau \mathbf{1}_p\quad\text{and}\quad  E[\mathbf{e}_t\mathbf{e}_t^\top]=E[\mathbf{e}_0\mathbf{e}_0^\top] =\mathbf{I}_2+c_\tau^2 \mathbf{1}_p \mathbf{1}_p^\top.
\end{eqnarray*}

By doing do, the TV dependence of the VAR system is fully reflected in the DGP. Finally, we let 

\begin{eqnarray*}
    \mathbf{y}_t = \mathbf{A}_{1,t}\mathbf{y}_{t-1} + \mathbf{e}_t\quad\text{and}\quad \mathbf{A}_{1,t} =\begin{pmatrix}
        \frac{2}{3}\exp(\gamma_t-1) & 0 \\
        0.3\cos(\gamma_t \pi)& \frac{2}{3}\exp(\gamma_t-1)
    \end{pmatrix}.
\end{eqnarray*}
To sum up, we now have $p=2$ and $q=1$. For this DGP, we use the measure $\text{RMSE}_{\mathbf{A}}$ and $\text{S}_q$ only.

The finding in Table \ref{Tb.1} is very similar to those from DGP 1. $\text{RMSE}_{\pmb{\alpha}}$ decreases as $T$ goes up. Also, the selection of the lag terms is accurate in general. Again, the results do not vary with respect to $c_j$ very much showing robustness on the choices of $\delta$ and $m$.

\begin{table}[htb!]
    \centering\small
\caption{Results of DGP 1 and DGP 2}\label{Tb.1} 
\begin{tabular}{llccccccccc}
    \hline\hline
     &      & \multicolumn{4}{c}{$\text{RMSE}_{\pmb{\alpha}}$}        & \multicolumn{4}{c}{$\text{RMSE}_{\mathbf{A}}$}        & \multirow{2}{*}{$S_q$} \\
     & T    & $\delta_1$    & $\delta_2$    & $\delta_3$    & $\delta_4$    & $\delta_1$    & $\delta_2$    & $\delta_3$    & $\delta_4$    &                     \\
DGP 1 & 500  & 0.195 & 0.192 & 0.190 & 0.191 & 0.156 & 0.155 & 0.155 & 0.156 & 1.000               \\
     & 750  & 0.164 & 0.161 & 0.160 & 0.161 & 0.129 & 0.128 & 0.128 & 0.129 & 1.000               \\
     & 1000 & 0.140 & 0.138 & 0.138 & 0.140 & 0.114 & 0.113 & 0.114 & 0.115 & 1.000               \\
     &      &       &       &       &       &       &       &       &       &                     \\
DGP 2 & 500  &       &       &       &       & 0.192 & 0.178 & 0.169 & 0.166 & 0.953               \\
     & 750  &       &       &       &       & 0.162 & 0.146 & 0.136 & 0.129 & 0.975               \\
     & 1000 &       &       &       &       & 0.152 & 0.138 & 0.127 & 0.121 & 0.975               \\
     \hline\hline
\end{tabular}
\end{table}

\medskip

Next, we evaluate QIR using the same DGPs and innovation draws as above.  So we set $q=1$ and $h\in\{1,2,4,8\}$.  Simple algebra gives the true value of the response function $\pmb\Psi^{(h)}(\gamma) = \mathbf A_1(\gamma)^h\mathbf P(\gamma)$, where
\begin{eqnarray*}
 \mathbf P(\gamma)
 &=& \mathbf I_2 \text{ in DGP 1},\, \text{and}\,\,
 \mathbf P(\gamma)=
 \begin{pmatrix}1&0\\ \sin(\pi\gamma)&1\end{pmatrix}
\text{ in DGP 2}.
\end{eqnarray*}
To evaluate the estimation accuracy, we use $\delta=2T^{-1/5}$ and $m=\lfloor(T\delta)^{3/4}\rfloor$.   Table \ref{tab:qlp-point-main} reports RMSE, the average of absolute bias, and the average of SD for each horizon.

\begin{table}[htb!]
\centering\small
\caption{Estimation Accuracy of Quantile Impulse Responses}
\label{tab:qlp-point-main}
\resizebox{0.95\textwidth}{!}{%
\begin{tabular}{llcccccccccccc}
\hline\hline
 &  & \multicolumn{4}{c}{RMSE} & \multicolumn{4}{c}{Absolute $$bias$$} & \multicolumn{4}{c}{SD} \\
 & $T/h$ & $1$ & $2$ & $4$ & $8$ & $1$ & $2$ & $4$ & $8$ & $1$ & $2$ & $4$ & $8$ \\
DGP 1 & 500 & 0.086 & 0.089 & 0.091 & 0.092 & 0.013 & 0.015 & 0.013 & 0.014 & 0.081 & 0.086 & 0.088 & 0.089 \\
 & 750 & 0.070 & 0.073 & 0.075 & 0.074 & 0.013 & 0.012 & 0.012 & 0.012 & 0.066 & 0.071 & 0.072 & 0.072 \\
 & 1000 & 0.062 & 0.065 & 0.065 & 0.066 & 0.011 & 0.012 & 0.010 & 0.011 & 0.058 & 0.062 & 0.063 & 0.063 \\
 & & & & & & & & & & & & & \\
DGP 2 & 500 & 0.106 & 0.115 & 0.115 & 0.115 & 0.020 & 0.023 & 0.016 & 0.010 & 0.097 & 0.108 & 0.110 & 0.111 \\
 & 750 & 0.091 & 0.097 & 0.098 & 0.097 & 0.020 & 0.019 & 0.015 & 0.010 & 0.082 & 0.090 & 0.093 & 0.093 \\
 & 1000 & 0.081 & 0.086 & 0.088 & 0.087 & 0.018 & 0.020 & 0.016 & 0.010 & 0.073 & 0.079 & 0.083 & 0.083 \\
\hline\hline
\end{tabular}}
\end{table}

RMSE and empirical standard deviation decline with $T$ at every horizon. Averaging across the four horizons, RMSE decreases from $0.090$ at $T=500$ to $0.065$ at $T=1000$ under DGP~1, and from $0.113$ to $0.086$ under DGP~2.

For inference, we assess the performance of the HAC estimator for QIR asymptotic covariance. To eliminate the local-linear regression bias,  we use $\delta=cT^{-1/4}$ with $c\in\{0.75,1,1.25\}$.   Table \ref{tab:qlp-inference-main} reports the coverage rate of the 95\% confidence interval which is averaged over all  evaluation points.

\begin{table}[htb!]
\centering\small
\caption{95\% Coverage of Quantile Impulse Responses}
\label{tab:qlp-inference-main}
\resizebox{0.95\textwidth}{!}{%
\begin{tabular}{llcccccccccccc}
\hline\hline
 &  & \multicolumn{4}{c}{$c=0.75$} & \multicolumn{4}{c}{$c=1$} & \multicolumn{4}{c}{$c=1.25$} \\
 & $T/h$ & $1$ & $2$ & $4$ & $8$ & $1$ & $2$ & $4$ & $8$ & $1$ & $2$ & $4$ & $8$ \\
DGP 1 & 500 & 0.926 & 0.918 & 0.917 & 0.909 & 0.934 & 0.929 & 0.926 & 0.921 & 0.937 & 0.934 & 0.933 & 0.928 \\
 & 750 & 0.936 & 0.930 & 0.928 & 0.923 & 0.943 & 0.938 & 0.936 & 0.931 & 0.946 & 0.941 & 0.938 & 0.937 \\
 & 1000 & 0.941 & 0.939 & 0.936 & 0.931 & 0.946 & 0.941 & 0.939 & 0.937 & 0.948 & 0.941 & 0.944 & 0.940 \\
 & & & & & & & & & & & & & \\
DGP 2 & 500 & 0.896 & 0.890 & 0.895 & 0.887 & 0.907 & 0.904 & 0.909 & 0.907 & 0.913 & 0.910 & 0.917 & 0.913 \\
 & 750 & 0.906 & 0.904 & 0.908 & 0.902 & 0.914 & 0.915 & 0.918 & 0.913 & 0.920 & 0.918 & 0.920 & 0.919 \\
 & 1000 & 0.913 & 0.917 & 0.914 & 0.909 & 0.917 & 0.922 & 0.920 & 0.918 & 0.922 & 0.927 & 0.926 & 0.925 \\
\hline\hline
\end{tabular}}
\end{table}

With $c=1$, average coverage across horizons increases from $0.928$ at $T=500$ to $0.941$ at $T=1000$ under DGP~1, and from $0.907$ to $0.919$ under DGP~2. Averaging across all sample sizes and horizons, the coverage rates for $c=0.75$, $1$, and $1.25$ are $0.928$, $0.935$, and $0.939$ under DGP~1, and $0.903$, $0.914$, and $0.919$ under DGP~2, respectively. Thus, coverage generally improves with both the sample size and  and the inference bandwidth, which coincides with the theoretical results.

\section{Empirical Study}\label{Sec.4}

Understanding how tighter credit conditions relate to downside risks to economic growth is important for assessing macroeconomic vulnerability. The Growth-at-Risk (GaR) analysis of \citet{AdrianBoyarchenkoGiannone2019} shows that tighter financial conditions are associated with pronounced declines in the lower quantiles of future growth, while the upper quantiles remain comparatively stable. Although existing GaR regressions allow these relationships to differ across quantiles and forecast horizons, their coefficients are typically held fixed over time. This restriction may obscure changes in the sensitivity of downside growth to credit conditions and the extent of asymmetry across the growth distribution.

Our TV-QVAR framework and TV-QIR relax this restriction by allowing quantile-specific coefficients to evolve smoothly over time. We apply these methods to monthly U.S. industrial-production growth and the Moody's Baa-Aaa corporate bond spread from February 1919 to December 2025. The TV-QVAR characterizes one-month-ahead conditional quantile dynamics, while the TV-QIR estimates growth responses at different horizons (e.g., three, six, nine, and twelve months). Together, they allow us to assess whether the association between credit conditions and future growth remains concentrated in the lower tail and how this pattern evolves over time and across horizons.

\subsection{A Time-Varying Quantile GaR Model}\label{Sec.41}
We begin with a financing channel linking credit conditions to future growth. For simplicity, we consider only the first lag in this derivation. In estimation, we assess the lag length using the information criterion developed in Section~\ref{Sec.23}. Let \(J_t\) denote financing-sensitive investment and other expenditures planned for period \(t\) using information available at \(t-1\), \(g_t\) economic growth, and \(f_t\) the corporate credit spread. Planned expenditure satisfies
\begin{equation}\label{emp.investment}
    J_t=j_0(\gamma_t)+j_g(\gamma_t)g_{t-1} 
    -j_f(\gamma_t)f_{t-1},
\end{equation}
where \(\gamma_t=t/T\), and \(j_0(\gamma)\), \(j_g(\gamma)\), and \(j_f(\gamma)\) vary smoothly over time. When \(j_f(\gamma)>0\), a wider spread raises the cost of external finance and reduces expenditure planned for period \(t\).

The resulting contraction need not affect all parts of the growth distribution equally. Financial accelerator mechanisms may amplify its effect on downside growth when borrower balance sheets deteriorate \citep{BernankeGertlerGilchrist1999}. We therefore specify
\begin{equation}\label{emp.growth}
    Q_\tau(g_t\mid\mathcal F_{t-1})
    =a_0(\gamma_t,\tau)+a_g(\gamma_t,\tau)g_{t-1}
    +a_J(\gamma_t,\tau)J_t,
\end{equation}
where \(a_J(\gamma,\tau)\) captures the transmission of expenditure to the conditional \(\tau\)-quantile of growth. The dependence on \((\gamma,\tau)\) allows both persistence and expenditure transmission to vary over time and across the growth distribution.

Substituting \eqref{emp.investment} into \eqref{emp.growth} yields
\begin{equation}\label{emp.reduced}
    Q_\tau(g_t\mid\mathcal F_{t-1})
    =\rho_0(\gamma_t,\tau)+\rho_g(\gamma_t,\tau)g_{t-1}
    +\rho_f(\gamma_t,\tau)f_{t-1},
\end{equation}
where $\rho_0(\gamma,\tau) =a_0(\gamma,\tau)+a_J(\gamma,\tau)j_0(\gamma)$, $\rho_g(\gamma,\tau) =a_g(\gamma,\tau)+a_J(\gamma,\tau)j_g(\gamma)$, and $\rho_f(\gamma,\tau) =-a_J(\gamma,\tau)j_f(\gamma)$. When \(a_J(\gamma,\tau)>0\), a wider spread lowers the corresponding growth quantile, with stronger downside transmission implying a more negative \(\rho_f(\gamma,\tau)\) at lower quantiles. Equation \eqref{emp.reduced} is a TV extension of the GaR quantile predictive regression of \citet{AdrianBoyarchenkoGiannone2019}. We embed it in a first-order TV-QVAR for \(\mathbf{y}_t=(g_t,f_t)^\top\), corresponding to the general specification in Section~\ref{Sec.2} with \(p=2\) and \(q=1\). We report \(\rho_f(\gamma_t,\tau)\sigma_f\), where \(\sigma_f\) is the full-sample standard deviation of the spread, so that the estimates represent the change in annualized growth associated with a one-standard-deviation widening of the credit spread.

The TV-QVAR characterizes one-month-ahead conditional dynamics. To examine transmission over longer horizons, we take \(t\) as the forecast origin and define annualized average industrial production growth over the next \(h\) months as $g_{t,t+h}^{(h)} =\frac{1}{h}\sum_{s=1}^{h}g_{t+s}$. For \(h\in\{3,6,9,12\}\), we estimate the TV-QIR
\begin{align}\label{emp.qlp}
    Q_\tau (g_{t,t+h}^{(h)}\mid\mathcal{F}_t )
    ={}&\alpha_h(\gamma_t,\tau)
    +\phi_{g,h}(\gamma_t,\tau)g_t
    +\beta_{f,h}(\gamma_t,\tau)f_t \notag\\
    &+\phi_{g,-1,h}(\gamma_t,\tau)g_{t-1}
    +\beta_{f,-1,h}(\gamma_t,\tau)f_{t-1}.
\end{align}
In estimation, we report \(\beta_{f,h}(\gamma_t,\tau)\sigma_f\), the TV-QIR of future growth to a one-standard-deviation financial shock under the recursive ordering \((g_t,f_t)^\top\).

\subsection{Variables and Data}\label{Sec.42}

We measure monthly growth at an annualized rate as \(g_t=1200\Delta\log(IP_t)\), where \(IP_t\) is the Federal Reserve's Industrial Production: Total Index (INDPRO). Financial conditions are measured by the spread between Moody's seasoned Baa and Aaa corporate bond yields, \(f_t=\text{BAA}_t-\text{AAA}_t\). A wider spread indicates higher relative financing costs for lower-rated corporations and hence tighter credit conditions. The three underlying series are obtained from the Federal Reserve Bank of St. Louis FRED database.\footnote{The series identifiers are \href{https://fred.stlouisfed.org/series/INDPRO}{INDPRO}, \href{https://fred.stlouisfed.org/series/BAA}{BAA}, and \href{https://fred.stlouisfed.org/series/AAA}{AAA}.} The monthly sample contains 1,283 observations from February 1919 to December 2025. Table~\ref{tab:gar-summary} reports summary statistics. ADF tests with an intercept reject the unit-root null for monthly growth, the credit spread in levels, and all forward-growth measures.

\begin{table}[htb!]
    \centering
    \caption{Summary statistics and unit-root tests}
    \label{tab:gar-summary}
    \small
    \begin{tabular*}{0.9\textwidth}{@{\extracolsep{\fill}}lrrrrr@{}}
        \toprule
        Variable & No. of Obs & Mean & S.D. & $t$-statistic & $p$-value \\
        \midrule
        Monthly industrial-production growth & 1,283 & 2.84 & 22.86 & $-10.54$ & 0.000 \\
        Baa--Aaa credit spread                & 1,283 & 1.16 &  0.68 &  $-3.46$ & 0.009 \\
        Three-month forward growth            & 1,280 & 2.90 & 17.54 &  $-6.88$ & 0.000 \\
        Six-month forward growth              & 1,277 & 2.88 & 13.62 &  $-6.03$ & 0.000 \\
        Nine-month forward growth             & 1,274 & 2.85 & 11.62 &  $-7.17$ & 0.000 \\
        Twelve-month forward growth           & 1,271 & 2.84 & 10.59 &  $-5.40$ & 0.000 \\
        \bottomrule
    \end{tabular*}
    \vspace{0.3em}

    \begin{minipage}{0.9\textwidth}
    \footnotesize\textit{Notes:} The sample runs from February 1919 through December 2025. The credit spread is measured in percentage points. The ADF regressions include an intercept, with lag lengths selected by the AIC from a maximum of 24 monthly lags.
    \end{minipage}
\end{table}

We select the lag order using the information criterion defined in \eqref{def.IC}, evaluated at $\gamma\in\{0.2,0.4,0.6,0.8\}$, with penalty $\omega_T=1.5\widehat s_{\max}\frac{\sqrt{\tau(1-\tau)\log T}}{T\delta}$, where $\widehat s_{\max}=\max\{\widehat\sigma_g,\widehat\sigma_f\}$ and $\widehat\sigma_g$ and $\widehat\sigma_f$ denote the sample standard deviations of growth and the credit spread, respectively. Table~\ref{tab:gar-lag-ic} reports the criterion values for lag orders ranging from one to ten. The criterion selects two lags at $\tau=0.1$ and $0.5$, but one lag at $\tau=0.9$. To maintain a parsimonious specification common to all quantiles, we adopt one lag for the baseline analysis and consider a two-lag QIR specification as a robustness check.

\begin{table}[htb!]
    \centering
    \caption{Information criterion for lag selection}
    \label{tab:gar-lag-ic}
    \footnotesize
    \setlength{\tabcolsep}{3pt}
    \begin{tabular*}{0.9\textwidth}{@{\extracolsep{\fill}}l*{10}{r}@{}}
        \toprule
        & \multicolumn{10}{c}{Lag order $q$} \\
        \cmidrule(l){2-11}
        $\tau$ & 1 & 2 & 3 & 4 & 5 & 6 & 7 & 8 & 9 & 10 \\
        \midrule
        $0.1$ & 2.4026 & \textbf{2.3683} & 2.3820 & 2.4056 & 2.4363 & 2.4557 & 2.4935 & 2.5016 & 2.5458 & 2.5850 \\
        $0.5$ & 4.7926 & \textbf{4.7152} & 4.7663 & 4.8303 & 4.8950 & 4.9568 & 5.0025 & 5.0745 & 5.1491 & 5.2181 \\
        $0.9$ & \textbf{2.4709} & 2.4712 & 2.4989 & 2.5321 & 2.5749 & 2.6093 & 2.6278 & 2.6605 & 2.6656 & 2.6423 \\
        \bottomrule
    \end{tabular*}
\end{table}

\subsection{Estimation Results}\label{Sec.43}

Table~\ref{tab:gar-constant} reports time-constant estimates of \eqref{emp.reduced} and \eqref{emp.qlp}, using the same variables, controls, quantiles, and inference procedures. The quantile VAR spread coefficient is negative at the lower quantile ($-7.45$) and positive at the upper quantile ($10.16$). The QIR point estimates are negative at all four horizons and all three quantiles, although several confidence intervals for the tail responses include zero.

We then estimate the TV-QVAR models at \(\tau\in\{0.1,0.5,0.9\}\). Point estimates are obtained using the local-linear smoothed quantile estimator developed in Section~\ref{Sec.23}, with \(\delta=2.5T^{-1/4}\) and \(m=0.5(T\delta)^{3/4}\). The variance nuisance functions are estimated using local-constant smoothing with the same time bandwidth \(\delta\). The quantile VAR inference uses a lag-zero sandwich estimator, while the QIR inference uses a Bartlett HAC kernel with truncation lag $\ell_h=\max\!\{h-1,\lceil\log\big((T-h)\delta\big)\rceil\}$. 

\begin{table}[htb!]
    \centering
    \caption{Time-constant QVAR and QIR estimates}
    \label{tab:gar-constant}
    \small
    \begin{tabular*}{0.88\textwidth}{@{\extracolsep{\fill}}lccc@{}}
        \toprule
        & $\tau=0.1$ & $\tau=0.5$ & $\tau=0.9$ \\
        \midrule
        QVAR & $-7.45$ & $-0.57$ & $10.16$ \\
                    & [$-9.98,-4.93$] & [$-1.85,0.70$] & [$6.78,13.53$] \\
        \addlinespace
        QIR, $h=3$ & $-11.32$ & $-10.90$ & $-6.97$ \\
                    & [$-15.05,-7.58$] & [$-14.07,-7.74$] & [$-34.58,20.63$] \\
        \addlinespace
        QIR, $h=6$ & $-4.91$ & $-4.15$ & $-2.32$ \\
                    & [$-9.22,-0.60$] & [$-10.52,2.21$] & [$-25.71,21.07$] \\
        \addlinespace
        QIR, $h=9$ & $-9.56$ & $-4.19$ & $-3.38$ \\
                    & [$-37.24,18.11$] & [$-8.33,-0.06$] & [$-13.11,6.35$] \\
        \addlinespace
        QIR, $h=12$ & $-13.93$ & $-5.35$ & $-7.59$ \\
                    & [$-30.39,2.52$] & [$-10.27,-0.44$] & [$-17.96,2.79$] \\
        \bottomrule
    \end{tabular*}
    \vspace{0.3em}

    \begin{minipage}{0.88\textwidth}
    \footnotesize\textit{Notes:} Each entry is the change in annualized industrial-production growth associated with a one-standard-deviation increase in the Baa--Aaa spread; point-wise 95\% confidence intervals are shown in brackets.
    \end{minipage}
\end{table}


\begin{figure}[t]
    \centering 
    \includegraphics[width=\textwidth]{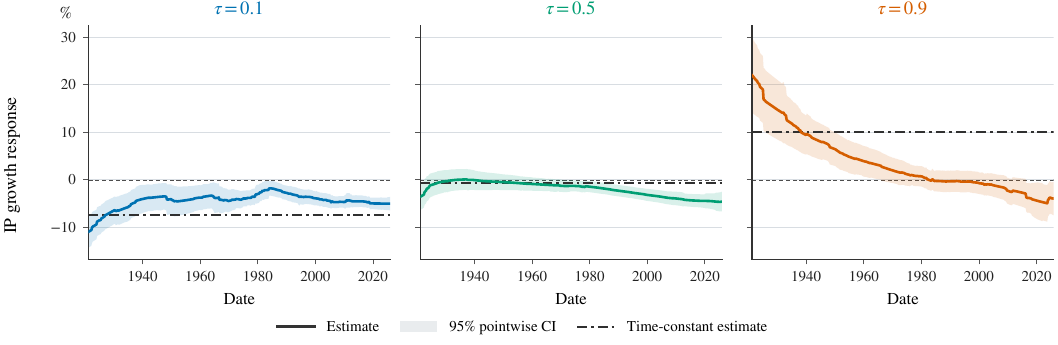}
    \caption{TV-QVAR estimates of the lagged credit-spread coefficient}
    \label{fig:gar-tvqvar-spread}
    \vspace{0.3em}
    \begin{minipage}{0.9\textwidth}
    \footnotesize\textit{Notes:} {\footnotesize The coefficient is scaled to a one-standard-deviation increase in the Baa--Aaa corporate bond spread; the vertical axis is measured in annualized percentage points. Solid lines are TV estimates, shaded regions are  95\% confidence intervals, and horizontal dash-dotted lines are time-constant QVAR estimates}.
    \end{minipage}
\end{figure}

\begin{figure}[t]
    \centering
    \includegraphics[width=\textwidth]{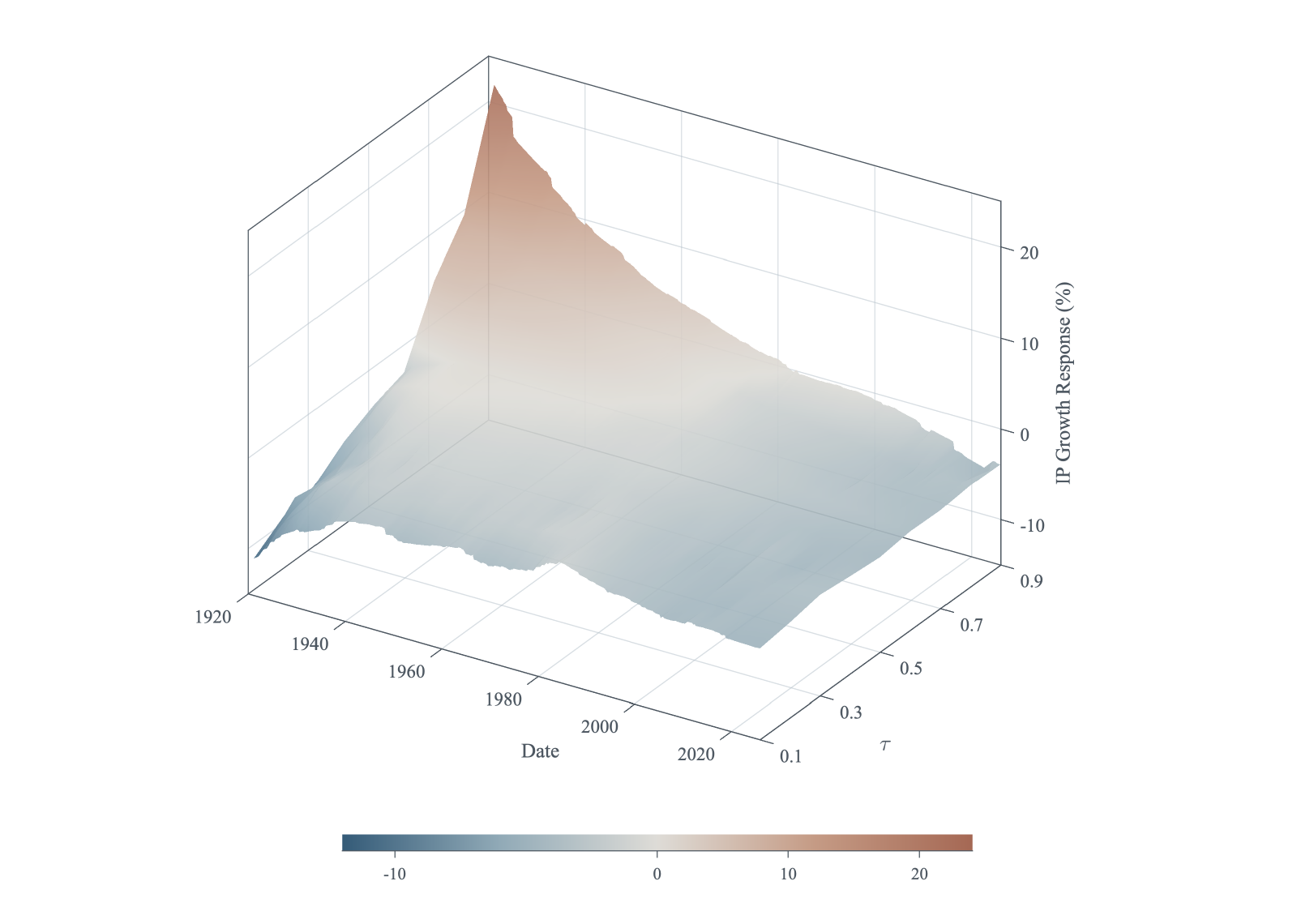}
    \caption{TV-QVAR credit-spread coefficients across time and quantiles}
    \label{fig:gar-tvqvar-surface}
    \vspace{0.2em}
    \begin{minipage}{0.8\textwidth}
    \footnotesize\textit{Notes:}The coefficient is scaled to a one-standard-deviation increase in the Baa--Aaa corporate bond spread; the vertical axis is measured in annualized percentage points. The surface is formed by estimates at $\tau=0.1,0.2,\ldots,0.9$  joined by linear interpolation.
    \end{minipage}
\end{figure}

\begin{figure}[p]
    \centering 
    \includegraphics[width=\textwidth]{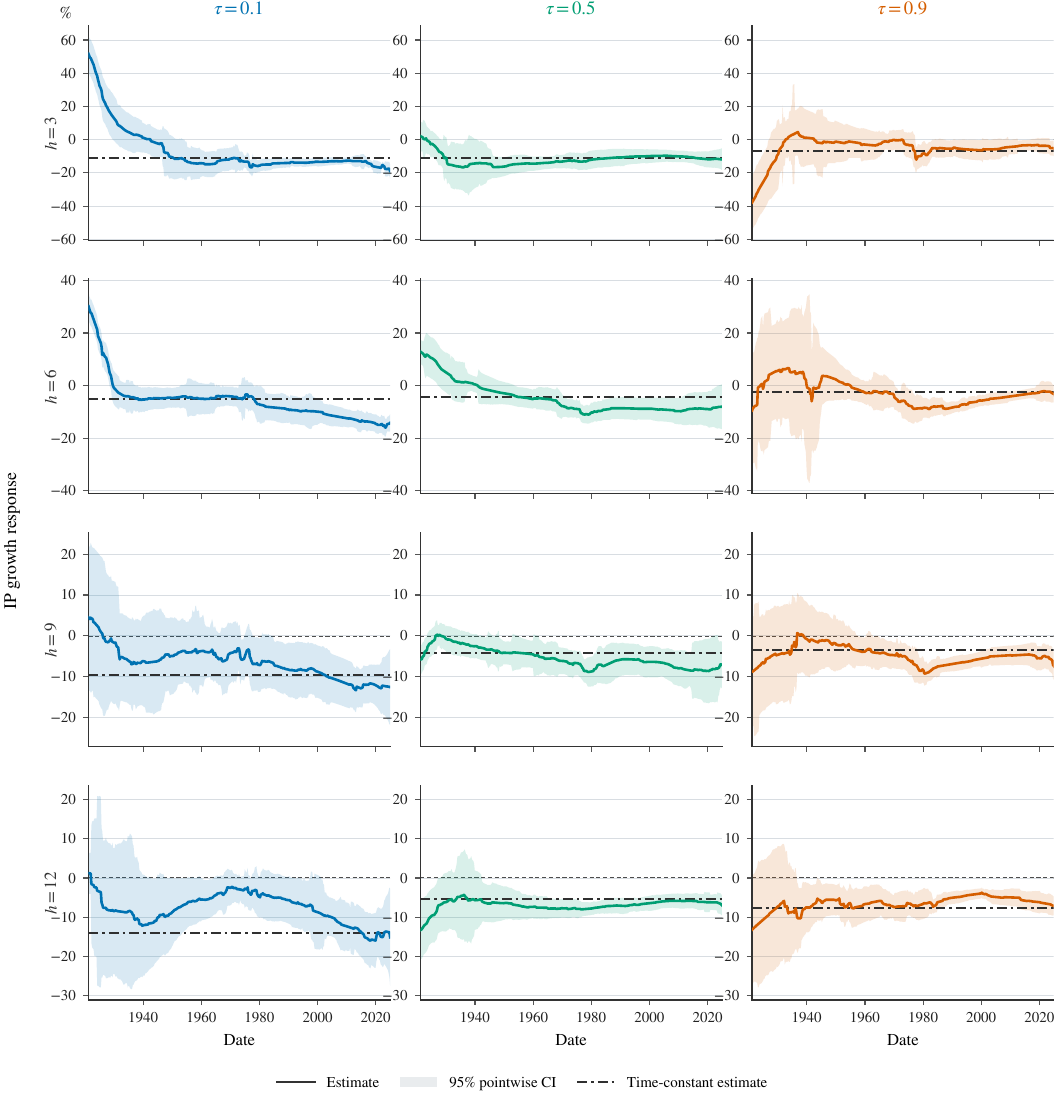}
     \caption{TV-QIR Estimates of the Credit-Spread Response}
    \label{fig:gar-tvqlp-spread}
    \vspace{0.2em}
    \begin{minipage}{0.8\textwidth}
    \footnotesize\textit{Notes:}  Each coefficient is scaled to a one-standard-deviation increase in the spread; the vertical axis is measured in annualized percentage points. Solid lines are TV estimates, shaded regions are point-wise 95\% confidence intervals, and horizontal dash-dotted lines are time-constant QIR estimates.
    \end{minipage}
\end{figure}

Figure~\ref{fig:gar-tvqvar-spread} reports the TV-QVAR coefficient on the lagged credit spread. The lower-tail estimate is negative throughout the sample and remains below the median and upper-tail estimates. The upper-tail coefficient is positive during part of the early sample and remains negative from the mid-1980s onward. The time-constant upper-tail estimate in Table~\ref{tab:gar-constant} shown by the horizontal line is positive. The lower-tail coefficient is especially negative in the 1920s and 1930s. \citet{Bernanke1983} documents how disruptions to credit intermediation deepened the contraction in the early 1930s. In later decades, negative coefficients also appear at the median and upper quantile, linking tighter credit conditions to weaker growth outside the lower tail. In their historical cross-country data, \citet{JordaSchularickTaylor2017} find that real and financial activity are more closely correlated in economies with higher leverage. To examine time variation across a broader range of quantiles, we further estimate the TV-QVAR model at $\tau=0.2,\ldots,0.8$. Figure~\ref{fig:gar-tvqvar-surface} presents the resulting coefficient surface, which reveals pronounced heterogeneity across quantiles and substantial variation over time. The coefficient for credit-spread  is generally more negative at lower quantiles, confirming greater downside sensitivity to credit conditions. The differences between lower- and upper-quantile coefficients are particularly pronounced in the early decades but narrow over time as the upper-quantile coefficients decline and eventually turn negative.

Figure~\ref{fig:gar-tvqlp-spread} reports the TV-QIR responses. At three months, the lower-tail response is more negative than the upper-tail response over much of the sample. The medians are $-12.99$, $-11.83$, and $-3.66$ percentage points at $\tau=0.1$, $0.5$, and $0.9$, respectively. The confidence bands are wider than in the TV-QVAR, especially in the early decades. At six months, most responses are negative, but their ordering across quantiles changes over time and the confidence bands overlap more frequently. At nine and twelve months, the responses are predominantly negative and closer across the three quantiles. Their point-wise confidence bands overlap over much of the sample.

We conduct two robustness checks. First, we add the second lags of growth and the credit spread to the TV-QIR regressions. Figure~\ref{fig:gar-tvqlp-two-lags} in Appendix~\ref{Appen.A2} compares these estimates with the one-lag baseline. At three months, the lower-tail response remains more negative than the upper-tail response over much of the sample. At nine and twelve months, the responses remain predominantly negative, with smaller differences across quantiles. Second, we replace average growth over the forecast horizon with monthly growth at its endpoint. Appendix~\ref{Appen.A2} compares these estimates with the baseline TV-QIR paths. The stronger downside response at three months remains evident after 1973, and the twelve-month responses are predominantly negative across quantiles. The six- and nine-month responses are more sensitive to the choice of growth measure.

\section{Conclusion}\label{Sec.5}

{\small Time-constant quantile models capture distributional asymmetry but cannot account for changes in quantile-specific dynamics over time. We develop a TV-QVAR and TV-QIR framework to address this limitation. The TV-QIR approach complements the one-step ahead TV-QVAR model  by directly estimating multi-horizon quantile responses, which generally cannot be obtained by iterating conditional quantiles from the TV-QVAR model. We establish asymptotic results for smoothed local-linear estimation and point-wise inference, including a consistent HAC covariance estimator that accommodates dependence in the QIR scores. Simulations demonstrate improved estimation precision and confidence interval coverage as the sample size increases.

Our application to U.S. credit conditions and economic growth highlights the importance of time variation in characterizing distributional asymmetry. The lower-tail TV-QVAR coefficient remains negative throughout the sample, while the upper-tail coefficient turns negative from the mid-1980s onward, despite a positive constant-parameter estimate. This suggests that the negative association between credit spreads and future growth extends beyond the lower tail in the later sample period. The TV-QIR results further reveal a horizon-dependent pattern, with pronounced downside asymmetry at shorter horizons that diminishes over longer horizons}.

\begin{appendix}

\section*{Appendix A}    

\renewcommand{\thelemma}{A\arabic{lemma}}
\setcounter{lemma}{0} 
\renewcommand{\theproposition}{A\arabic{proposition}}
\setcounter{proposition}{0} 
\renewcommand{\theassumption}{A\arabic{assumption}}
\setcounter{assumption}{0} 
\renewcommand{\thesubsection}{A\arabic{subsection}}
\setcounter{subsection}{0} 
\renewcommand{\theequation}{A.\arabic{equation}}
\setcounter{equation}{0} 
\renewcommand{\thefigure}{A.\arabic{figure}}
\setcounter{figure}{0}
\renewcommand{\thetable}{A.\arabic{table}}
\setcounter{table}{0}
  
In this Appendix, Appendix \ref{Appen.A1} explains the inherent nonlinearity of the quantile framework; Appendix \ref{Appen.A2} presents additional empirical results omitted from the main text; Appendix \ref{Appen.A3} provides two preliminary lemmas to facilitate the development of the asymptotic distribution of VAR estimation in Section \ref{Sec.23}; Appendix \ref{Appen.A4} discusses the numerical implementation of the estimation procedure.

\subsection{Nonlinearity of Quantile Framework}\label{Appen.A1}

In this section, we demonstrate the importance of accounting for the inherent nonlinearity of the quantile framework and provide a formal justification for Assumption \ref{Assumption4}, which nests stationary quantile autoregression (QAR) processes  studied by \citet{Koenker01092006} as a special case.

Without loss of generality, consider  the following QAR(1) process

\begin{eqnarray*}
 y_t&=&\theta_0(\eta_t)+\theta_1(\eta_t)y_{t-1},
\end{eqnarray*}
where $\eta_t$ follows $\mathrm{i.i.d.}\ U(0,1)$. We can then write
\begin{eqnarray*}
 y_t&=&\sum_{j=0}^{\infty}
 \left\{\prod_{i=0}^{j-1}\theta_1(\eta_{t-i})\right\}
 \theta_0(\eta_{t-j})
 \eqqcolon H(\mathcal E_t),
\end{eqnarray*}
where $\mathcal E_t=(\ldots,\eta_{t-1},\eta_t)$.

Under their monotonicity condition, the $\tau$-quantile residual is
\begin{eqnarray*}
    e_{\tau,t}=\theta_0(\eta_t)-\theta_0(\tau) +\{\theta_1(\eta_t)-\theta_1(\tau)\}y_{t-1}
    =\theta_0(\eta_t)-\theta_0(\tau) +\{\theta_1(\eta_t)-\theta_1(\tau)\}H(\mathcal E_{t-1})
    \eqqcolon G_e(\gamma,\mathcal E_t)
\end{eqnarray*}

Suppose that $\|\theta_0(\eta_t)\|_\nu<\infty$ and there exists $\rho_e\in(0,1)$ such that $\|\theta_1(\eta_t)\|_\nu<\rho_e$, By Minkowski's inequality,
\begin{eqnarray*}
 \|y_t\|_\nu
 &\le&\sum_{j=0}^{\infty}
 \|\theta_1(\eta_t)\|_\nu^j\|\theta_0(\eta_t)\|_\nu
 \le\frac{\|\theta_0(\eta_t)\|_\nu}{1-\rho_e}<\infty.
\end{eqnarray*}
Additionally, since $\eta_t$ is independent of $y_{t-1}$,
\begin{eqnarray*}
 \|e_{\tau,t}\|_\nu
 &\le&\|\theta_0(\eta_t)-\theta_0(\tau)\|_\nu
 +\|\theta_1(\eta_t)-\theta_1(\tau)\|_\nu\|y_{t-1}\|_\nu
 <\infty.
\end{eqnarray*}
For fixed $t$ and $k\ge0$, we replace $\eta_{t-k}$ with an independent copy $\eta_{t-k}'$, leaving all other innovations unchanged, and denote the resulting variables by $y_s^{*(k)}$ and $e_{\tau,s}^{*(k)}$. The two processes agree before $t-k$, so

\begin{eqnarray}\label{eq.just.1}
 y_{t-k}-y_{t-k}^{*(k)}
 &=&\theta_0(\eta_{t-k})-\theta_0(\eta_{t-k}')
 +\{\theta_1(\eta_{t-k})-\theta_1(\eta_{t-k}')\}y_{t-k-1}.
\end{eqnarray}
The original and copied innovations are independent of $y_{t-k-1}$. Hence
\begin{eqnarray*}
 \|y_{t-k}-y_{t-k}^{*(k)}\|_\nu
 &\le&2\|\theta_0(\eta_t)\|_\nu
 +2\|\theta_1(\eta_t)\|_\nu\|y_{t-1}\|_\nu
 \le C.
\end{eqnarray*}
We can then iterate the representation in \eqref{eq.just.1} and obtain 
\begin{eqnarray*}
 y_{t-1}-y_{t-1}^{*(k)}
 &=&\left\{\prod_{s=t-k+1}^{t-1}\theta_1(\eta_s)\right\}
 \{y_{t-k}-y_{t-k}^{*(k)}\}.
\end{eqnarray*}
It follows that
\begin{eqnarray*}
 \|y_{t-1}-y_{t-1}^{*(k)}\|_\nu
 &=&\|\theta_1(\eta_t)\|_\nu^{k-1}
 \|y_{t-k}-y_{t-k}^{*(k)}\|_\nu
 \le C\rho_e^{k-1}.
\end{eqnarray*}
Since $\eta_t$ is unchanged and independent of past processes,
\begin{eqnarray*}
 \| G_e(\gamma,\mathcal E_t)-G_e(\gamma,\mathcal E_t^{*(k)})\|_\nu
&=&
\|\{\theta_1(\eta_t)-\theta_1(\tau)\}
 \{y_{t-1}-y_{t-1}^{*(k)}\}\|_\nu
\nonumber\\
 &=&\|\theta_1(\eta_t)-\theta_1(\tau)\|_\nu
 \|y_{t-1}-y_{t-1}^{*(k)}\|_\nu
 \le C\rho_e^k.
\end{eqnarray*}
This verifies Assumption \ref{Assumption4}.2.

\subsection{Additional Empirical Results}\label{Appen.A2}

\begin{figure}[htb!]
    \centering 
    \includegraphics[width=\textwidth]{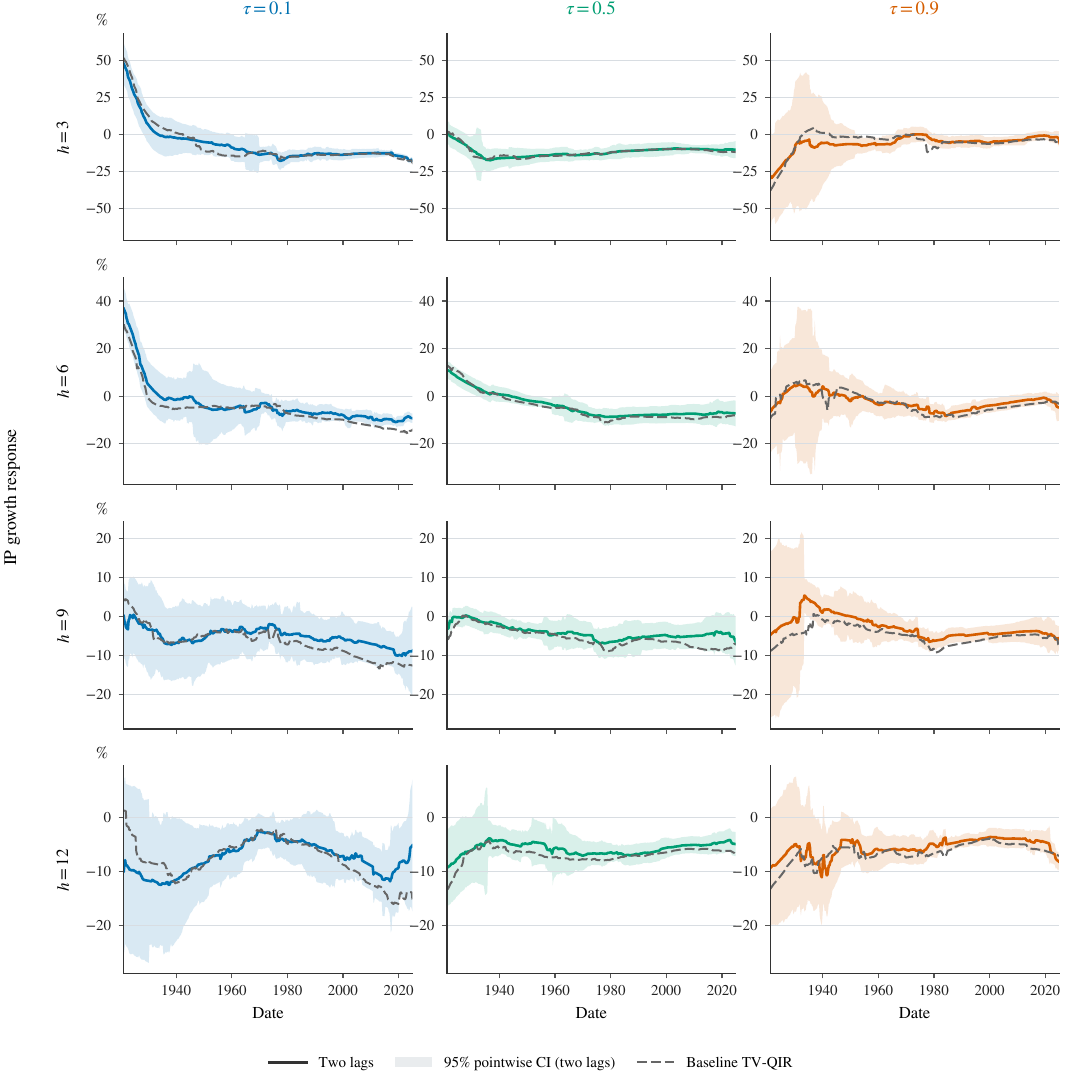}
    \caption{Quantile Impulse Responses with Two Lags}
     \label{fig:gar-tvqlp-two-lags}
    \vspace{0.3em}
    \begin{minipage}{0.98\textwidth}
    \footnotesize\textit{Notes:} The two-lag QIR adds $g_{t-2}$ and $f_{t-2}$ to the baseline regressors in \eqref{emp.qlp}. Solid lines are responses to a one-standard-deviation increase in the spread, shaded regions are point-wise 95\% confidence intervals, and gray dashed lines show the baseline TV-QIR estimates from Figure~\ref{fig:gar-tvqlp-spread}. The vertical axis is measured in annualized percentage points.
    \end{minipage}
\end{figure}

\paragraph{Lag Specification} The information criterion in Table~\ref{tab:gar-lag-ic} selects two lags for the lower and median quantiles. We add $g_{t-2}$ and $f_{t-2}$ to each baseline QIR regression, using the same estimation settings. Figure~\ref{fig:gar-tvqlp-two-lags} reports the two-lag responses and their point-wise confidence intervals; gray dashed lines show the baseline paths.

\begin{figure}[htb!]
    \centering
    \includegraphics[width=\textwidth]{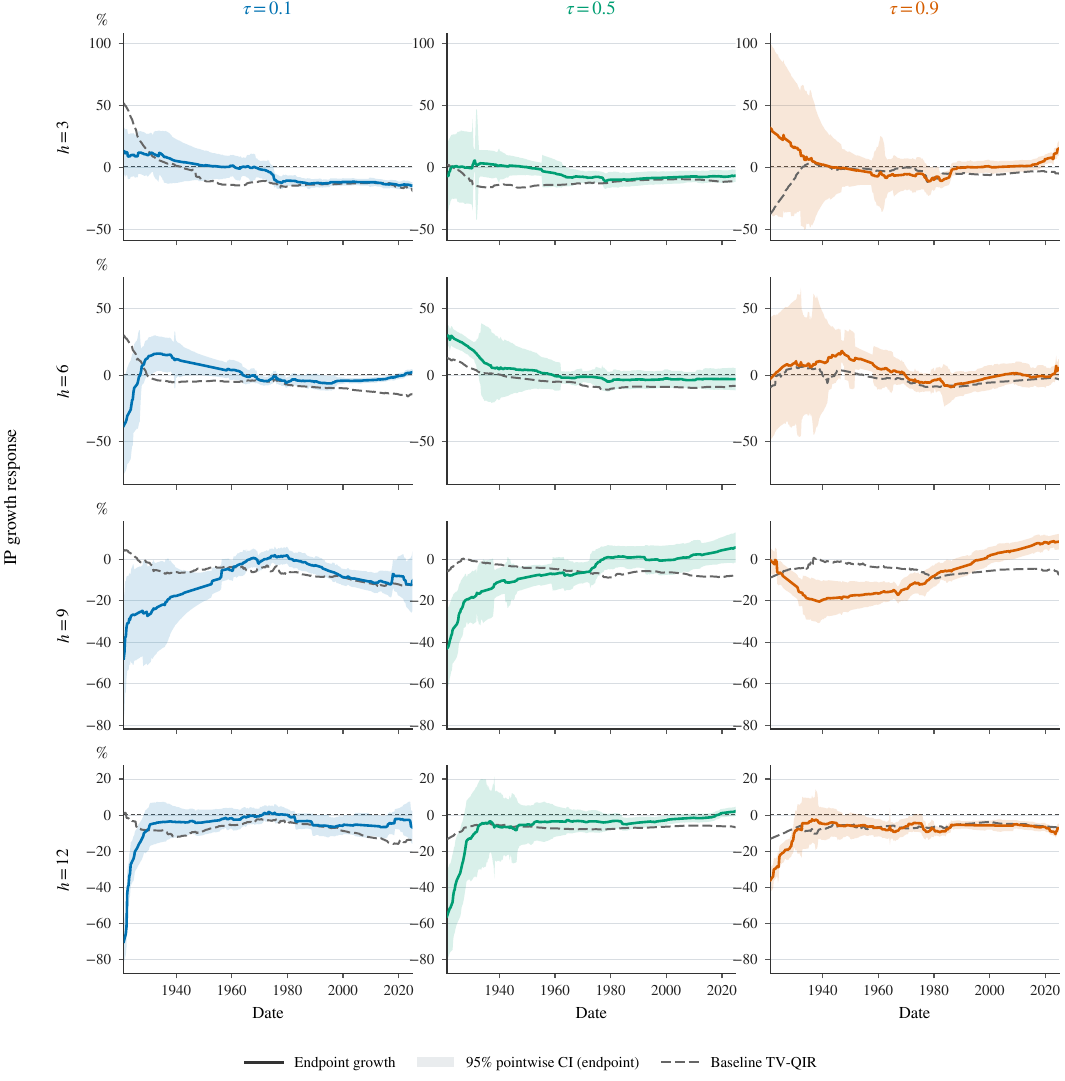}
    \caption{Quantile Impulse Responses for Endpoint Industrial-Production Growth}
    \label{fig:gar-tvqlp-endpoint}
    \vspace{0.3em}
    \begin{minipage}{0.98\textwidth}
    \footnotesize\textit{Notes:} The dependent variable is annualized monthly industrial-production growth in month $t+h$. Solid lines are responses to a one-standard-deviation increase in the spread, shaded regions are point-wise 95\% confidence intervals, and gray dashed lines show the baseline TV-QIR estimates from Figure~\ref{fig:gar-tvqlp-spread}. The vertical axis is measured in annualized percentage points.
    \end{minipage}
\end{figure}

\paragraph{Endpoint Growth as the QIR Outcome} The baseline TV-QIR measures average annualized growth over the forecast horizon. As a robustness check, we instead consider annualized monthly growth at the horizon endpoint,

\begin{equation}\label{emp.endpointgrowth}
    g_{t+h}=1200\left(\log IP_{t+h}-\log IP_{t+h-1}\right), 
    \qquad h\in\{3,6,9,12\}.
\end{equation}
The forecast origin is $t$, and the coefficient functions are indexed by $\gamma_t$ as in \eqref{emp.qlp}. We use the baseline regressors and estimation settings. Figure~\ref{fig:gar-tvqlp-endpoint} compares the endpoint-growth responses with the baseline paths, shown as gray dashed lines.

\subsection{Preliminary Lemmas for VAR Estimation}\label{Appen.A3}

Firstly, we present the rates of convergence of $(\widehat{\mathbf{B}}^0, \widehat{\mathbf{B}}^1)$.

\begin{lemma}\label{Appendix.A1} 
Let Assumptions \ref{Assumption1}-\ref{Assumption3} hold and $\sqrt{T\delta}/m\to 0$. Then

\begin{enumerate}[leftmargin=*]
    \item $\sup_{\gamma\in [\delta, 1-\delta]}\|\VEC(\widehat{\mathbf{B}}^0, \delta\widehat{\mathbf{B}}^1) -\VEC(\pmb{\beta}_\gamma^0 ,\delta\pmb{\beta}_\gamma^1)\|=O_P(\frac{\sqrt{\log T}}{\sqrt{T\delta}})$,
    \item for $\forall\gamma\in (0, 1)$, $\|\VEC(\widehat{\mathbf{B}}^0, \delta\widehat{\mathbf{B}}^1) -\VEC(\pmb{\beta}_\gamma^0 ,\delta\pmb{\beta}_\gamma^1)\|=O_P(\frac{1}{\sqrt{T\delta}}).$
\end{enumerate}
\end{lemma} 

The rates involved Lemma \ref{Appendix.A1} are both optimal from the nonparametric regression standpoint.

To derive asymptotic distribution, we define a new estimator to bridge $(\widehat{\mathbf{B}}^0,  \widehat{\mathbf{B}}^1)$ and $(\pmb{\beta}_\gamma^0 , \pmb{\beta}_\gamma^1)$. Specifically, we denote it by $(\tilde{\pmb{\beta}}_\gamma^0, \tilde{\pmb{\beta}}_\gamma^1)$:

\begin{eqnarray}\label{def.beta.tilde}
    \VEC(\tilde{\pmb{\beta}}_\gamma^0,
    \delta\tilde{\pmb{\beta}}_\gamma^1)
    &\coloneqq&
    \VEC(\pmb{\beta}_\gamma^0,
    \delta\pmb{\beta}_\gamma^1)
    +v_2\pmb{\xi}_{\gamma,\delta}
    +\diag\{\pmb\Sigma_{\mathbf f,\gamma},
    v_2\pmb\Sigma_{\mathbf f,\gamma}\}^{-1}
    \mathbf J_\gamma
\end{eqnarray}
and $\mathbf{J}_\gamma \coloneqq  \frac{1}{T}\sum_{t=1}^{T}\pmb{\rho}_{m,t}^{(1)}\otimes \VEC(\mathbf{x}_{t-1}, \delta^{-1}\mathbf{z}_{t-1})\cdot K_\delta(\gamma_t -\gamma)$, where the dimensions of $\tilde{\pmb{\beta}}_\gamma^0$ and $\tilde{\pmb{\beta}}_\gamma^1$ are comfortable with $\pmb{\beta}_\gamma^0$ and $\pmb{\beta}_\gamma^1$ respectively, and $\pmb{\rho}_{m,t}^{(1)}\coloneqq (\rho_{\tau, m}^{(1)} (e_{1,t}  ),\ldots, \rho_{\tau, m}^{(1)} (e_{p,t}  ))^\top$.  

For the newly defined estimator, we are able to establish the following results.

\begin{lemma}\label{Appendix.A2}
Let Assumptions \ref{Assumption1}-\ref{Assumption3} hold, and $\sqrt{T\delta}/m\to 0$. 

\begin{enumerate}[leftmargin=*]
\item \begin{enumerate}
\item $\sup_{\gamma\in [\delta, 1-\delta]} \| \VEC(\tilde{\pmb{\beta}}_\gamma^0,\delta \tilde{\pmb{\beta}}_\gamma^1) -\VEC(\pmb{\beta}_\gamma^0 ,\delta\pmb{\beta}_\gamma^1)\| =O_P(\frac{\sqrt{\log T}}{\sqrt{T\delta}})$,

\item $\sup_{\gamma\in [\delta, 1-\delta]}\|\VEC(\widehat{\mathbf{B}}^0, \delta\widehat{\mathbf{B}}^1) -  \VEC(\tilde{\pmb{\beta}}_\gamma^0,\delta \tilde{\pmb{\beta}}_\gamma^1)\| =o_P(\frac{\sqrt{\log T}}{\sqrt{T\delta}})$.
\end{enumerate}

\item For $\forall \gamma\in (0,1)$,

\begin{enumerate}[leftmargin=*]
\item $\sqrt{T\delta}(\VEC(\tilde{\pmb{\beta}}_\gamma^0,\delta \tilde{\pmb{\beta}}_\gamma^1) -\VEC(\pmb{\beta}_\gamma^0 ,\delta\pmb{\beta}_\gamma^1)-v_2\pmb{\xi}_{\gamma,\delta} )\to_D N(\mathbf{0},\pmb{\Omega}_\gamma)$,
 
\item $\|\VEC(\widehat{\mathbf{B}}^0, \delta\widehat{\mathbf{B}}^1) -  \VEC(\tilde{\pmb{\beta}}_\gamma^0,\delta \tilde{\pmb{\beta}}_\gamma^1)\| =o_P(\frac{1}{\sqrt{T\delta}})$,
\end{enumerate}
where $\pmb\Omega_\gamma$ is defined in Theorem \ref{TM.2}.
\end{enumerate}
\end{lemma}

\subsection{Numerical Implementation}\label{Appen.A4}

As the objective function \eqref{def.obj1} is sufficiently smooth, the optimization involved in \eqref{def.est} can be implemented in a relatively easy manner by majority statistical software. In this study, we simply use ``fminunc'' of Matlab. 

There is only one minor issue left, as all of these statistical softwares require an initial value to invoke the optimization procedure. See \url{https://au.mathworks.com/help/optim/ug/fminunc.html} for instance. For practical implementation, we use the standard local-linear kernel regression to obtain the estimates of $(\pmb{\beta}_\gamma^0, \pmb{\beta}_\gamma^1)$ as the initial value to start the optimization. We refer readers to Chapter 6 of \cite{fan2003nonlinear} for a detailed introduction about the closed-from of the local-linear kernel regression. Apparently, these initial estimates do not vary with respect to $\tau$, and only vary with respect to $\gamma$.
\end{appendix}

{\footnotesize
\bibliography{Ref.bib}

@article{angrist2006quantile,
  author  = {Angrist, Joshua and Chernozhukov, Victor and Fern{\'a}ndez-Val, Iv{\'a}n},
  title   = {Quantile Regression under Misspecification, with an Application to the {U.S.} Wage Structure},
  journal = {Econometrica},
  year    = {2006},
  volume  = {74},
  number  = {2},
  pages   = {539-563},
  doi     = {10.1111/j.1468-0262.2006.00671.x}
}

@article{Paul2020,
    author = {Paul, Pascal},
    title = {The Time-Varying Effect of Monetary Policy on Asset Prices},
    journal = {The Review of Economics and Statistics},
    volume = {102},
    number = {4},
    pages = {690-704},
    year = {2020}
}

@article{Gao02012024,
author = {Jiti Gao and Bin Peng and Yayi Yan},
title = {Estimation, Inference, and Empirical Analysis for Time-Varying {VAR} Models},
journal = {Journal of Business \& Economic Statistics},
volume = {42},
number = {1},
pages = {310-321},
year = {2024}
}

@article{CM2024,
author = {Chavleishvili, Sulkhan and Manganelli, Simone},
title = {Forecasting and stress testing with quantile vector autoregression},
journal = {Journal of Applied Econometrics},
volume = {39},
number = {1},
pages = {66-85},
year = {2024}
}

@article{korobilis2025monitoring,
  title={Monitoring multi-country macroeconomic risk: A quantile factor-augmented vector autoregressive (QFAVAR) approach},
  author={Korobilis, Dimitris and Schr{\"o}der, Maximilian},
  journal={Journal of Econometrics},
  volume={249},
  pages={105730},
  year={2025}
}

@unpublished{ruzicka2024,
  author = {Ruzicka, Josef},
  title  = {Quantile Local Projections: Identification, Smooth Estimation, and Inference},
  year   = {2024},
  note   = {Working paper available at \url{https://sites.google.com/view/ruzicka/research}}
}

@article{JordaSchularickTaylor2017,
  author = {Jord{\`a}, {\`O}scar and Schularick, Moritz and Taylor, Alan M.},
  title = {Macrofinancial History and the New Business Cycle Facts},
  journal = {NBER Macroeconomics Annual},
  volume = {31},
  pages = {213-263},
  year = {2017},
  doi = {10.1086/690241}
}

@article{AdrianBoyarchenkoGiannone2019,
  author = {Adrian, Tobias and Boyarchenko, Nina and Giannone, Domenico},
  title = {Vulnerable Growth},
  journal = {American Economic Review},
  volume = {109},
  number = {4},
  pages = {1263-1289},
  year = {2019},
  doi = {10.1257/aer.20161923}
}

@incollection{BernankeGertlerGilchrist1999,
  author = {Bernanke, Ben S. and Gertler, Mark and Gilchrist, Simon},
  title = {The Financial Accelerator in a Quantitative Business Cycle Framework},
  booktitle = {Handbook of Macroeconomics},
  editor = {Taylor, John B. and Woodford, Michael},
  publisher = {Elsevier Science B.V.},
  volume = {1},
  pages = {1341-1393},
  year = {1999}
}

@article{Bernanke1983,
  author = {Bernanke, Ben S.},
  title = {Nonmonetary Effects of the Financial Crisis in the Propagation of the {Great Depression}},
  journal = {American Economic Review},
  volume = {73},
  number = {3},
  pages = {257-276},
  year = {1983}
}

@book{horn2012matrix,
  title={Matrix Analysis},
  author={Horn, Roger A and Johnson, Charles R},
  edition={2nd},
  year={2012},
  publisher={Cambridge University Press},
  address={Cambridge}
}

@article{YGP2025, 
title={ASYMPTOTICS FOR TIME-VARYING VECTOR {MA}($\infty$) PROCESSES}, 
volume={41}, 
number={3}, 
journal={Econometric Theory}, 
author={Yan, Yayi and Gao, Jiti and Peng, Bin}, 
year={2025}, 
pages={584-616}
}

@article{Horowitz1998,
 author = {Joel L. Horowitz},
 journal = {Econometrica},
 number = {6},
 pages = {1327-1351},
 title = {Bootstrap Methods for Median Regression Models},
 volume = {66},
 year = {1998}
}

@article{Primiceri2005,
    author = {Primiceri, Giorgio E.},
    title = {Time Varying Structural Vector Autoregressions and Monetary Policy},
    journal = {The Review of Economic Studies},
    volume = {72},
    number = {3},
    pages = {821-852},
    year = {2005}
}

@article{SW2001,
Author = {Stock, James H. and Watson, Mark W.},
Title = {Vector Autoregressions},
Journal = {Journal of Economic Perspectives},
Volume = {15},
Number = {4},
Year = {2001},
Pages = {101-115}
}

@article{Wu2005,
author = {Wei Biao Wu},
title = {Nonlinear system theory: Another look at dependence},
journal = {Proceedings of the National Academy of Sciences},
volume = {102},
number = {40},
pages = {14150-14154},
year = {2005},
doi = {10.1073/pnas.0506715102}
}

@article{ZhouWu2009,
author = {Zhou Zhou and Wei Biao Wu},
title = {Local linear quantile estimation for nonstationary time series},
journal = {The Annals of Statistics},
volume = {37},
number = {5B},
pages = {2696-2729},
year = {2009}
}

@book{Lut2005,
  title={New Introduction to Multiple Time Series Analysis},
  author={Helmut L{\"u}tkepohl},
  year={2005},
  publisher={Springer-Verlag Berlin Heidelberg}
}

@article{Koenker01092006,
  author = {Roger Koenker and Zhijie Xiao},
  title = {Quantile Autoregression},
  journal = {Journal of the American Statistical Association},
  volume = {101},
  number = {475},
  pages = {980-990},
  year = {2006},
}

@book{fan2003nonlinear,
  title={{Nonlinear Time Series: Nonparametric and Parametric Methods}},
  author={Fan, Jianqing and Yao, Qiwei},
  year={2003},
  publisher={Springer}
}

@article{HE2023367,
title = {Smoothed quantile regression with large-scale inference},
journal = {Journal of Econometrics},
volume = {232},
number = {2},
pages = {367-388},
year = {2023},
author = {Xuming He and Xiaoou Pan and Kean Ming Tan and Wen-Xin Zhou}
}

@article{GAO2025106035,
title = {Time-varying vector error-correction models: {E}stimation and inference},
journal = {Journal of Econometrics},
volume = {251},
pages = {106035},
year = {2025},
author = {Jiti Gao and Bin Peng and Yayi Yan}
}

@article{TWZ2022,
    author = {Tan, Kean Ming and Wang, Lan and Zhou, Wen-Xin},
    title = {High-Dimensional Quantile Regression: Convolution Smoothing and Concave Regularization},
    journal = {Journal of the Royal Statistical Society Series B: Statistical Methodology},
    volume = {84},
    number = {1},
    pages = {205-233},
    year = {2022}
}

@article{INOUE2024105726,
  title = {Local projections in unstable environments},
  journal = {Journal of Econometrics},
  volume = {244},
  number = {2},
  pages = {105726},
  year = {2024},
  author = {Atsushi Inoue and Barbara Rossi and Yiru Wang}
}

@unpublished{maung2025nonparametric,
  title={Nonparametric Time-Varying Local Projections},
  author={Maung, Kenwin and Yoshida, Nathaniel},
  note={Available at SSRN 5920403},
  year={2025}
}

@article{wojciechowski2025identifying,
  title={Identifying Unconditional Quantile Impulse Responses with an Application to Growth-at-Risk},
  author={Wojciechowski, Robert},
  note={\url{https://robertwojciechowski.github.io/research/}},
  year={2025}
}

@article{han2024estimation,
  title={Estimation and inference of quantile impulse response functions by local projections: With applications to {VaR} dynamics},
  author={Han, Heejoon and Jung, Whayoung and Lee, Ji Hyung},
  journal={Journal of Financial Econometrics},
  volume={22},
  number={1},
  pages={1-29},
  year={2024}
}

@article{Sims1980,
 author = {Christopher A. Sims},
 journal = {Econometrica},
 number = {1},
 pages = {1-48},
 title = {Macroeconomics and Reality},
 volume = {48},
 year = {1980}
}

@BOOK{kanwal1983,
  title = {{Generalized Fuctions: Theory and Technique}},
  publisher = {Academic Press},
  year = {1983},
  author = {Ram P Kanwal},
  address = {New York}
}

@article{Dahlhaus,
author = {R. Dahlhaus},
title = {{Fitting time series models to nonstationary processes}},
volume = {25},
journal = {The Annals of Statistics},
number = {1},
pages = {1-37},
year = {1997}
}

@ARTICLE {PS1992,
  author  = "Peter C. B. Phillips and Victor Solo",
  title   = "Asymptotics for linear processes",
  journal = "The Annals of Statistics",
  year    = "1992",
  volume  = "20",
  number  = "2",
  pages   = "971-1001"
}
}

\newpage 
\begin{appendix}

\small

\renewcommand{\thelemma}{B\arabic{lemma}}
\setcounter{lemma}{0} 
\renewcommand{\theproposition}{B\arabic{proposition}}
\setcounter{proposition}{0} 
\renewcommand{\theassumption}{B\arabic{assumption}}
\setcounter{assumption}{0} 
\renewcommand{\thesubsection}{B\arabic{subsection}}
\setcounter{subsection}{0} 
\renewcommand{\theequation}{B.\arabic{equation}}
\setcounter{equation}{0} 
\renewcommand{\thefigure}{B.\arabic{figure}}
\setcounter{figure}{0}
\renewcommand{\thetable}{B.\arabic{table}}
\setcounter{table}{0}

\setcounter{page}{1}

\linespread{1.1}\selectfont

\begin{center}

\begin{spacing}{1.5}
{\Large Online Supplementary Appendix B to \\``Time-Varying Quantile VAR Models: Estimation and Inference''}

\medskip
\medskip

NOT FOR PUBLICATION
\end{spacing}

\medskip
\medskip

{\sc Jiti Gao$^{\dag}$, Fei Liu$^\sharp$ and Bin Peng$^{\dag}$}

\medskip

$^\dag$Monash University and $^\sharp$University of Bath
\end{center}    

This supplementary document is organized as follows. Appendix \ref{Appen.B1} provides a jackknife method used for nonparametric bias correction. Appendix \ref{Appen.B2} introduces the notation and preliminary lemmas used throughout the mathematical proofs, which are contained in Appendix \ref{Appen.B3} and Appendix \ref{Appen.B4}.

\section*{Appendix B}

\subsection{Jackknife Bias Correction}\label{Appen.B1}

Write $\widehat\Psi^{(h)}_{ij,\delta}(\gamma)$ when it is necessary to display the bandwidth in the QIR estimator.  Theorem \ref{TM.4} shows that its leading bias is proportional to $\delta^2$.  Evaluate the same local-linear estimator at the two proportional bandwidths $\delta$ and $\delta/\sqrt2$, and define
\begin{eqnarray}\label{def.lp.jackknife}
 \widehat\Psi^{(h),{\rm bc}}_{ij}(\gamma)
 &\coloneqq&
 2\widehat\Psi^{(h)}_{ij,\delta/\sqrt2}(\gamma)
 -\widehat\Psi^{(h)}_{ij,\delta}(\gamma).
\end{eqnarray}
Using simple algebra, we can show that  the deterministic leading bias in \eqref{def.lp.jackknife} is
\begin{eqnarray*}
 2\left\{
 \frac{v_2}{2}\left(\frac{\delta}{\sqrt2}\right)^2
 \Psi^{(h,2)}_{ij}(\gamma)
 \right\}
 -\frac{v_2}{2}\delta^2
 \Psi^{(h,2)}_{ij}(\gamma)
 &=&0.
\end{eqnarray*}
Thus this correction avoids estimating $\Psi^{(h,2)}_{ij}(\gamma)$ directly.

Applying similar arguments in the proof of Theorem \ref{TM.4}, we can establish the debiased central limit theory. The details are omitted here due to page constraint.

\subsection{Notations and Preliminary Lemmas}\label{Appen.B2}

We define a few notations which will be repeatedly used throughout.

\textbf{Check Function} --- The first derivative of $\rho_{\tau,m} (u)$ is as follows:

\begin{eqnarray}\label{def.der} 
\rho_{\tau,m}^{(1)}(u)= -\int \rho_\tau(x)\phi_m^{(1)}(x-u)\mathrm{d}x=\int \psi_\tau(u+x)\phi_m(x)\mathrm{d}x.
\end{eqnarray}

\medskip

\textbf{Beveridge and Nelson (BN) Decomposition} --- Under the TV environment, the BN decomposition (\citealp{PS1992}) is as follows: 

\begin{eqnarray*} 
    \mathbf{c}_{t}(L) = \bar{\mathbf{c}}_t -(1-L)\tilde{\mathbf{c}}_t(L), 
\end{eqnarray*}
where $\bar{\mathbf{c}}_t\coloneqq \mathbf{c}_t(1) =\sum_{\ell=0}^\infty \mathbf{c}_{\ell, t}$, $\tilde{\mathbf{c}}_t(L) =\sum_{\ell=0}^\infty \tilde{\mathbf{c}}_{\ell, t} L^\ell$, and $\tilde{\mathbf{c}}_{\ell, t} =\sum_{k=\ell+1}^\infty \mathbf{c}_{k,t}$.

For $\forall r\ge 0$, we denote

\begin{eqnarray}\label{def.theta.r}
    \mathbf{c}_{t}^r (L) \coloneqq \sum_{\ell =0}^\infty \mathbf{c}_{\ell, t}^r L^\ell \coloneqq\sum_{\ell =0}^\infty (\mathbf{c}_{\ell+r, t} \otimes \mathbf{c}_{\ell, t})L^\ell.
\end{eqnarray}
Similarly, we also have 

\begin{eqnarray*}
    \mathbf{c}_{t}^r (L) =  \bar{\mathbf{c}}_t^r -(1-L)\tilde{\mathbf{c}}_t^r(L), 
\end{eqnarray*}
where $\bar{\mathbf{c}}_t^r\coloneqq \mathbf{c}_t^r(1) =\sum_{\ell=0}^\infty \mathbf{c}_{\ell, t}^r$, $\tilde{\mathbf{c}}_t^r(L) =\sum_{\ell=0}^\infty \tilde{\mathbf{c}}_{\ell, t}^r L^\ell$, and $\tilde{\mathbf{c}}_{\ell, t}^r =\sum_{k=\ell+1}^\infty \mathbf{c}_{k,t}^r$.

\medskip

Throughout the quantile VAR proofs, $E_{t-1}[\cdot]$ denotes $E[\cdot\mid\mathcal F_{t-1}]$.  In particular, $\mathbf x_{t-1}$ and each local parameter perturbation are $\mathcal F_{t-1}$-measurable, and $E_{t-1}[\psi_\tau(e_{\ell,t})]=0$ by Assumption \ref{Assumption1}. No conditional-mean restriction is imposed on $e_{\ell,t}$.

\begin{lemma} 
\label{lem:gelfand_bound}
Let $\tilde{\pmb{\phi}}$ be a fixed-dimension square matrix. Suppose the spectral radius of $\tilde{\pmb{\phi}}$ satisfies $\rho(\tilde{\pmb{\phi}}) < 1$. Then for any $\gamma \in (\rho(\tilde{\pmb{\phi}}), 1)$, there exists a finite constant $C_\phi \ge 1$ such that $\lVert \tilde{\pmb{\phi}}^k \rVert_2 \le C_\phi \gamma^k$ for all $k \ge 1$.
\end{lemma}

\begin{lemma}\label{Appendix.B1}
\item 
\begin{enumerate}[leftmargin=*]
\item Let $k_{\delta,t}^{(r)}(\gamma)= ((\gamma_t-\gamma)/\delta)^rK_\delta(\gamma_t-\gamma)$, for $r\in\{0,1,2,3,4\}$. Under Assumption \ref{Assumption3},
\begin{enumerate}
    \item $\max_t\sup_{\gamma\in[0,1]} |k_{\delta,t}^{(r)}(\gamma)| \le\frac{C}{\delta}$;
    \item $\sup_{\gamma\in[0,1]} \frac1T\sum_{t=1}^{T-1} |k_{\delta,t+1}^{(r)}(\gamma) -k_{\delta,t}^{(r)}(\gamma)| = O\!\left(\frac{1}{T\delta}\right)$. 
\end{enumerate}

\item Under Assumption \ref{Assumption1},
\begin{enumerate}[leftmargin=*]
\item $\max_{t\ge1}\sum_{\ell=0}^{\infty} \ell\|\mathbf c_{\ell,t}\|_2
 <\infty$;
\item $\max_{0\le s\le T-1}
 \sum_{t=0}^{s}\sum_{\ell=0}^{\infty} \ell\|\mathbf c_{\ell,t+1}-\mathbf c_{\ell,t}\|_2 <\infty$.
\end{enumerate}
For $r\ge0$, let
$\mathbf c_t^r(L)=\sum_{\ell=0}^{\infty}\mathbf c_{\ell,t}^rL^\ell$,
where $\mathbf c_{\ell,t}^r=\mathbf c_{\ell+r,t}\otimes
\mathbf c_{\ell,t}$.  Its Beveridge--Nelson decomposition is
$\mathbf c_t^r(L)=\bar{\mathbf c}_t^r-
(1-L)\tilde{\mathbf c}_t^r(L)$, where
$\bar{\mathbf c}_t^r=\sum_{\ell=0}^{\infty}\mathbf c_{\ell,t}^r$,
$\tilde{\mathbf c}_t^r(L)=\sum_{\ell=0}^{\infty}
\tilde{\mathbf c}_{\ell,t}^rL^\ell$, and
$\tilde{\mathbf c}_{\ell,t}^r=\sum_{k=\ell+1}^{\infty}
\mathbf c_{k,t}^r$.  Then
\begin{enumerate}[resume,leftmargin=*]
\item $\max_{t\ge1}\sum_{r=1}^{\infty}\sum_{\ell=0}^{\infty} \|\tilde{\mathbf c}_{\ell,t}^r\|_2 <\infty$;

\item $ \max_{t\ge1}\sum_{r=1}^{\infty} \|\bar{\mathbf c}_t^r\|_2 <\infty$;

\item $\max_{1\le s\le T-1} \sum_{t=1}^{s}\sum_{r=1}^{\infty}
 \|\tilde{\mathbf c}_{t+1}^r(1) -\tilde{\mathbf c}_{t}^r(1)\|_2 <\infty.$

\end{enumerate}

\item Uniformly in $\gamma\in[\delta,1-\delta]$,
\begin{enumerate}
    \item $\frac1T\sum_{t=1}^{T} \{\pmb\beta_\ell(\gamma_t)-\pmb\beta_\ell(\gamma) -(\gamma_t-\gamma)\pmb\beta_\ell^{(1)}(\gamma)\} K_\delta(\gamma_t-\gamma)= \frac{v_2\delta^2}{2}\pmb\beta_\ell^{(2)}(\gamma) +O(\delta^3+T^{-1})$.

 \item $\frac1T\sum_{t=1}^{T}
 \|\pmb\beta_\ell(\gamma_t)-\pmb\beta_\ell(\gamma)
 -(\gamma_t-\gamma)\pmb\beta_\ell^{(1)}(\gamma)\|_2^2
 K_\delta(\gamma_t-\gamma)= \frac{v_4\delta^4}{4}
 \|\pmb\beta_\ell^{(2)}(\gamma)\|_2^2 +o(\delta^4).$
\end{enumerate}

\end{enumerate}
\end{lemma}

\medskip

\begin{lemma}\label{Appendix.B2}
Under Assumptions \ref{Assumption1} and \ref{Assumption2} and for a sufficiently large $m$, the following results hold uniformly in $t\in[T]$ and $\ell\in[p]$,
\begin{enumerate}[leftmargin=*]
\item $\sup_u\left| \int\{\psi_\tau(x+u+w)-\psi_\tau(x+u)\} \phi_m(x)\mathrm dx \right|  \le C\min\{m|w|,1\};$

\item $E_{t-1}\!\left[ \{\rho_{\tau,m}^{(1)}(e_{\ell,t}+w) -\rho_{\tau,m}^{(1)}(e_{\ell,t})\}^2 \right]  \le C|w|;$

\item
as $|\xi|\to0$, $ E_{t-1}\!\left[
 \rho_{\tau,m}(e_{\ell,t}+\xi) -\rho_{\tau,m}(e_{\ell,t})
 -\rho_{\tau,m}^{(1)}(e_{\ell,t})\xi \right]= \frac12f_{\ell,t}(0\mid\mathcal F_{t-1})\xi^2 +O(|\xi|^3+\xi^2/m).$

\end{enumerate}
\end{lemma}

\medskip

\begin{lemma}\label{Appendix.B3}
Under Assumptions \ref{Assumption1}--\ref{Assumption3}, the following results hold for $r\in\{0,1\}$: 
\begin{enumerate}[leftmargin=*]
\item 
$\frac1T\sum_{t=1}^{T} \mathbf q_t^Qk_{\delta,t}^{(r)}(\gamma) = O_P((T\delta)^{-1/2})$, for $\gamma \in(0,1)$;  

$\sup_{\gamma \in[\delta,1-\delta]}\left\|\frac1T\sum_{t=1}^{T} \mathbf q_t^Qk_{\delta,t}^{(r)}(\gamma)\right\| = O_P(\sqrt{\log T/(T\delta)})$ with $\mathbf q_t^Q \coloneqq \pmb\psi_t\otimes\mathbf x_{t-1}$,

\item  if $\sqrt{T\delta}/(m\sqrt{\log T})\to0$, the rates established in Lemma \ref{Appendix.B3} also hold when replacing $\pmb\psi_t$  by $\pmb\rho_{m,t}^{(1)}$. Moreover, if $\sqrt{T\delta}/m\to0$, for  $\gamma\in(0,1)$, we have $\frac1T\sum_{t=1}^{T} \{ (\pmb\rho_{m,t}^{(1)}-\pmb\psi_t) \otimes\mathbf x_{t-1}  \}k_{\delta,t}^{(r)}(\gamma) =o_P ((T\delta)^{-1/2} )$.
\end{enumerate}
\end{lemma}

\begin{lemma}\label{Appendix.B4}
Under  Assumptions \ref{Assumption1}-\ref{Assumption5},  
\begin{enumerate}
    \item[(a)] $\max_{t\in[T]} \|\mathbf y_t^* -\tilde{\mathbf y}_t(\gamma_t)\|_\nu =O(T^{-1})$;
    \item[(b)] $ \max_{t\in[T]} \|\mathbf y_t -\tilde{\mathbf y}_t(\gamma_t)\|_\nu =O(T^{-1})$;
    \item[(c)] $ \max_{t\in[T]} \|\tilde{\mathbf y}_t(\gamma) -\tilde{\mathbf y}_t(\gamma')\|_\nu \le C|\gamma-\gamma'|$, for any $\gamma,\gamma'\in(0,1)$;
    \item[(d)] $\sup_{\gamma\in[0,1]} \|\tilde{\mathbf y}_t(\gamma) -\tilde{\mathbf y}_t^{*(k)}(\gamma)\|_\nu \le C\rho_x^k$, for some $\rho_x\in(0,1)$, where $\tilde{\mathbf y}_t^{*(k)}(\gamma)$  is the coupled stationary approximation formed by replacing $\pmb\eta_{t-k}$ with its independent copy.
\end{enumerate}
\end{lemma}

\medskip

\begin{lemma}\label{Appendix.B5}
For deterministic $\mathbf a\in\mathbb R^{j+pq}$, let $v^{(h)}_{i,t,T}(\mathbf a)\coloneqq y_{i,t+h}-\mathbf x_t^{[j]\top}\mathbf a $ and $\tilde v^{(h)}_{i,t}(\mathbf a;\gamma) \coloneqq \tilde y_{i,t+h}(\gamma) -\tilde{\mathbf x}_t^{[j]}(\gamma)^\top\mathbf a.$ Under the innovation-density condition in Assumption \ref{Assumption1} and Assumption \ref{Assumption5}, for every $b\in\mathbb R$ and $\epsilon>0$, there exists a positive number $\bar C_f$ such that
\begin{eqnarray*}
 P\!\left(
 |v^{(h)}_{i,t,T}(\mathbf a)-b|\le\epsilon
 \,\middle|\,\mathbf x_t^{[j]}
 \right)
 &\le&2\bar C_f\epsilon,\nonumber\\
 P\!\left(
 |\tilde v^{(h)}_{i,t}(\mathbf a;\gamma)-b|\le\epsilon
 \,\middle|\,\tilde{\mathbf x}_t^{[j]}(\gamma)
 \right)
 &\le&2\bar C_f\epsilon.
\end{eqnarray*}
\end{lemma}

\medskip

For the remaining QIR lemmas and proofs, fix $(i,j,h,\tau,\gamma)$ and write
\begin{eqnarray*}
 \pmb\alpha(\gamma)
 &\coloneqq&
 \pmb\alpha^{(h)}_{ij}(\gamma),\qquad
 \pmb\alpha_t\coloneqq\pmb\alpha(\gamma_t),
 \qquad
f_t^*\coloneqq
 g^{(h)}_{ij,t,T}(0\mid\mathbf x_t^{[j]}),
 \nonumber\\
u^{(h)_{i,t}}
 &\coloneqq&
 y_{i,t+h}-\mathbf x_t^{[j]\top}\pmb\alpha_t,\qquad
\mathbf s^{(h)}_{ij,t,T}
 \coloneqq
 \psi_\tau(u^{(h)}_{i,t})\mathbf x_t^{[j]}.
\end{eqnarray*}
For generic $\mathbf a^0$ and $\mathbf a^1$, define
\begin{eqnarray*}
\Delta\mathbf a_\delta
 &\coloneqq&
 \VEC\!\left(
 \mathbf a^0-\pmb\alpha(\gamma),
 \delta\{\mathbf a^1-\pmb\alpha^{(1)}(\gamma)\}
 \right),\quad
\Delta\bar y_t
 \coloneqq
 \mathbf x_t^{[j]\top}\pmb\alpha_t
 -\mathbf x_t^{[j]\top}\mathbf a^0
 -\mathbf z_t^{[j]\top}\mathbf a^1,\nonumber\\
 \pmb\xi^{(h)}_{ij,\gamma,\delta}
 &\coloneqq&
 \VEC\!\left(
 \frac{\delta^2}{2}\pmb\alpha^{(2)}(\gamma),\mathbf0
 \right),\quad
 r_{\delta,a}
 \coloneqq
 \|\Delta\mathbf a_\delta\|+\delta^2,\qquad
 \Lambda_{\delta,a}\coloneqq1+|\log r_{\delta,a}|,
 \nonumber\\
G_t
 &\coloneqq&
 \int_0^{\Delta\bar y_t}
 \{\rho^{(1)}_{\tau,m}(u^{(h)}_{i,t}+w)
 -\rho^{(1)}_{\tau,m}(u^{(h)}_{i,t})\}\mathrm dw,
 \qquad
\vartheta_t\coloneqq G_t-E[G_t].
\end{eqnarray*}

\medskip

\begin{lemma}\label{Appendix.B6}
Under the coefficient-smoothness, stability, and innovation-density conditions in Assumption \ref{Assumption1} and Assumptions \ref{Assumption4}--\ref{Assumption5},
\begin{eqnarray*}
 \max_{\substack{1\le t\le T-h:\\
 |\gamma_t-\gamma|\le\delta}}
 \left\|
 E\!\left[
 \psi_\tau(u^{(h)}_{i,t})\mathbf x_t^{[j]}
 \right]
 \right\|
 &=&
 O\!\left(T^{-(\nu-1)/\nu}\right).
\end{eqnarray*}
\end{lemma}

\medskip

\begin{lemma}\label{Appendix.B7}
Let $\mathbf s^{*(h,k)}_{ij,t,T}$ and $\tilde{\mathbf s}^{*(h,k)}_{ij,t}(\gamma)$ be the coupled versions of $\mathbf s^{(h)}_{ij,t,T}$ and $\tilde{\mathbf s}^{(h)}_{ij,t}(\gamma)$, respectively, obtained by replacing $\pmb\eta_{t+h-k}$ with an independent copy.  Define
\begin{eqnarray*}
\mathfrak d^{A}_{2,s,T}(k;\gamma)
 &\coloneqq&
 \max_{\substack{1\le t\le T-h:\\
 |\gamma_t-\gamma|\le\delta}}
 \left\|
 \mathbf s^{(h)}_{ij,t,T}
 -\mathbf s^{*(h,k)}_{ij,t,T}
 \right\|_2,\quad
\mathfrak d^{F}_{2,s}(k;\gamma)
 \coloneqq
 \sup_{t\in\mathbb Z}
 \left\|
 \tilde{\mathbf s}^{(h)}_{ij,t}(\gamma)
 -\tilde{\mathbf s}^{*(h,k)}_{ij,t}(\gamma)
 \right\|_2.
\end{eqnarray*}
Under the coefficient-smoothness, stability, and innovation-density conditions in Assumption \ref{Assumption1} and Assumptions \ref{Assumption4}--\ref{Assumption5}, there exist constants $C<\infty$ and $\varrho\in(0,1)$ such that, for all sufficiently large $T$,
\begin{eqnarray}\label{lp.score.fd}
 \mathfrak d^{A}_{2,s,T}(k;\gamma)
 +\mathfrak d^{F}_{2,s}(k;\gamma)
 &\le&
 C\mathbf 1\{k\le h\}
 +C\varrho^{\,k-h}\mathbf 1\{k>h\}.
\end{eqnarray}
Additionally,
\begin{eqnarray}\label{lp.cov.summable}
 \sum_{r=-\infty}^{\infty}
 \left\|
 \operatorname{Cov}\!\left(
 \tilde{\mathbf s}^{(h)}_{ij,t}(\gamma),
 \tilde{\mathbf s}^{(h)}_{ij,t+r}(\gamma)
 \right)
 \right\|
 &<&\infty.
\end{eqnarray}
\end{lemma}

\medskip

\begin{lemma}\label{Appendix.B8}
Under the coefficient-smoothness and stability conditions in Assumption \ref{Assumption1}, Assumption \ref{Assumption3}, and Assumption \ref{Assumption4}.2, for $n\in\{0,1,2,3,4\}$ and $\gamma\in(0,1)$, let
\begin{eqnarray*}
 \pmb\Sigma_\gamma^{[j]}
 &\coloneqq&
 E\!\left[
 \tilde{\mathbf x}_t^{[j]}(\gamma)
 \tilde{\mathbf x}_t^{[j]}(\gamma)^\top
 \right].
\end{eqnarray*}
Then
\begin{eqnarray*}
 \frac1T\sum_{t=1}^{T-h}
 \mathbf x_t^{[j]}\mathbf x_t^{[j]\top}
 k_{\delta,t}^{(n)}(\gamma)
 &=&
 v_n\pmb\Sigma_\gamma^{[j]}
 +O_P\!\left((T\delta)^{-1/2}+\delta+(T\delta)^{-1}\right),
\end{eqnarray*}
\end{lemma}

\medskip

\begin{lemma}\label{Appendix.B9}
Under Assumptions \ref{Assumption1}, \ref{Assumption2}, and \ref{Assumption5}, for a sufficiently large $m$, the following results hold almost surely in $\mathbf x_t^{[j]}$ and uniformly over $t$.

\begin{enumerate}[label=(\alph*),leftmargin=*]
\item
$E[\{\rho_{\tau,m}^{(1)}(u^{(h)}_{i,t}+w)-\rho_{\tau,m}^{(1)}(u^{(h)}_{i,t})\}^2 \mid\mathbf x_t^{[j]}] \le C|w|;$
\item
$ \left| E[\rho_{\tau,m}^{(1)}(u^{(h)}_{i,t}+w)-\rho_{\tau,m}^{(1)}(u^{(h)}_{i,t}) \mid\mathbf x_t^{[j]}] \right| \le C|w|;$

\item $E[\rho_{\tau,m}^{(1)}(u^{(h)}_{i,t}+w) -\rho_{\tau,m}^{(1)}(u^{(h)}_{i,t}) \mid\mathbf x_t^{[j]}] =f_t^*w+O(w^2+|w|/m)$;

\item $E[\rho_{\tau,m}(u^{(h)}_{i,t}+\xi)-\rho_{\tau,m}(u^{(h)}_{i,t}) -\rho_{\tau,m}^{(1)}(u^{(h)}_{i,t})\xi \mid\mathbf x_t^{[j]}] =\frac12f_t^*\xi^2+O(|\xi|^3+\xi^2/m)$;
\end{enumerate}
\end{lemma}

\medskip

\begin{lemma}\label{Appendix.B10}
Let $1<r\le2$ and consider the centred array $\pmb\Xi_{t,T}\coloneqq H_{t,T}(\mathcal E_{t+\kappa_T}) -E[H_{t,T}(\mathcal E_{t+\kappa_T})]$, where $\kappa_T\ge0$ is deterministic and may depend on $T$.  Let $\pmb\Xi_{t,T}^{*(k)}$ be obtained by replacing $\pmb\eta_{t+\kappa_T-k}$ with an independent copy and define
\begin{eqnarray*}
\mathfrak d_{r,\Xi}(k)
 &\coloneqq&
 \max_t\|\pmb\Xi_{t,T}-\pmb\Xi_{t,T}^{*(k)}\|_r.
\end{eqnarray*}
Then, for every deterministic scalar array $\{a_{t,T}\}$,
\begin{eqnarray}\label{lp.projection.Lr}
 \left\|\sum_ta_{t,T}\pmb\Xi_{t,T}\right\|_r
 &\le&
 C_r\left(\sum_t|a_{t,T}|^r\right)^{1/r}
 \sum_{k=0}^{\infty}\mathfrak d_{r,\Xi}(k).
\end{eqnarray}
In particular, when $r=2$, one may take $C_2=1$, and hence
\begin{eqnarray}\label{lp.projection.L2}
 E\left\|\sum_ta_{t,T}\pmb\Xi_{t,T}\right\|^2
 &\le&
 \left\{\sum_{k=0}^{\infty}\mathfrak d_{2,\Xi}(k)\right\}^2
 \sum_ta_{t,T}^2.
\end{eqnarray}
\end{lemma}

\medskip

\begin{lemma}\label{Appendix.B11}
Under the coefficient-smoothness and stability conditions in Assumption \ref{Assumption1} and Assumption \ref{Assumption4}.2, let $\mathbf y_{s,T}^{[r]}$ denote the coupled version of $\mathbf y_{s,T}$ obtained by replacing $\pmb\eta_r$ by an independent copy. There exist constants $C<\infty$ and $\rho_0\in(0,1)$ such that, uniformly in $T$, $s$, and $r\le s$,
\begin{eqnarray*}
 \|\mathbf y_{s,T}-\mathbf y_{s,T}^{[r]}\|_4
 &\le&C\rho_0^{s-r}.
\end{eqnarray*}
\end{lemma}

\medskip

\begin{lemma}\label{ap.tm5c.lemD}
For generic $\mathbf a^0$ and $\mathbf a^1$ and any deterministic scalar series $\{a_{t,T}\}$, the QIR perturbation defined above satisfies
\begin{eqnarray*}
 \operatorname{Var} \left(\sum_ta_t\vartheta_t\right)
 &\le&
 Cr_{\delta,a}^{3}
 \{1+|\log r_{\delta,a}|\}^2
 \sum_ta_t^2,
\end{eqnarray*}
\end{lemma}

\medskip

\begin{lemma}\label{ap.tm5c.lemL}
Let the coefficient-smoothness, stability, and innovation-density conditions in Assumption \ref{Assumption1} and Assumptions \ref{Assumption2}--\ref{Assumption5} hold.
\begin{enumerate}[label=(\roman*),leftmargin=*]
\item Suppose $H(\gamma,\mathbf x)$ satisfies
\begin{eqnarray*}
 \|H(\gamma,\mathbf x)\|
 &\le&C(1+\|\mathbf x\|^2),\nonumber\\
 \|H(\gamma,\mathbf x)-H(\gamma',\mathbf x')\|
 &\le&
 C(1+\|\mathbf x\|+\|\mathbf x'\|)
 \{|\gamma-\gamma'|+\|\mathbf x-\mathbf x'\|\}.
\end{eqnarray*}
Then, for $n=0,1,2,3,4$ and each $\gamma\in(0,1)$,
\begin{eqnarray*}
 \frac1T\sum_{t=1}^{T-h}
 H(\gamma_t,\mathbf x_t^{[j]})
 k_{\delta,t}^{(n)}(\gamma)
 &\to_P&
 v_nE [
 H(\gamma,\tilde{\mathbf x}_t^{[j]}(\gamma))
  ].
\end{eqnarray*}

\item If $\sqrt{T\delta}/m\to0$, then, for $n\in\{0,1\}$,
\begin{eqnarray*}
 \frac1T\sum_{t=1}^{T-h}
 \rho_{\tau,m}^{(1)}(u^{(h)}_{i,t})
 \mathbf x_t^{[j]}
 k_{\delta,t}^{(n)}(\gamma)
 &=&O_P((T\delta)^{-1/2}).
\end{eqnarray*}
\end{enumerate}
\end{lemma}






\subsection{Proof of Main Results}\label{Appen.B3}

\begin{proof}[Proof of Proposition \ref{prop.1}]
    \item 

We write
\begin{eqnarray*}
 \max_t\|\mathbf y_t-\mathbf y_t^*\|_\nu
 &\lesssim&
 \max_t\sum_{\ell=0}^{\infty}
 \|\mathbf c_{\ell,t}-\mathbf c_\ell(\gamma_t)\|_2
 \|\mathbf e_{t-\ell}\|_\nu\nonumber\\
 &\lesssim&
 \max_t\sum_{\ell=1}^{\infty}
 \left\|
 \prod_{j=0}^{\ell-1}\tilde{\mathbf A}_{t-j}
 -\tilde{\mathbf A}(\gamma_t)^\ell
 \right\|_2\nonumber\\
 &\lesssim&
 \max_t\sum_{\ell=1}^{\infty}\sum_{j=0}^{\ell-1}
 \left\|\prod_{k=0}^{j-1}\tilde{\mathbf A}_{t-k}\right\|_2
 \|\tilde{\mathbf A}_{t-j}
 -\tilde{\mathbf A}(\gamma_t)\|_2
 \left\|\prod_{k=j+1}^{\ell-1}
 \tilde{\mathbf A}(\gamma_t)\right\|_2\nonumber\\
 &\lesssim&
 \max_t\sum_{\ell=1}^{\infty}\sum_{j=0}^{\ell-1}
 \tilde{\rho}_y^{\ell-1}\frac{j}{T}
 =O(T^{-1}),
\end{eqnarray*}
where the first inequality follows from \eqref{def.yt.ma} and the
definition of $\mathbf y_t^*$, the second inequality follows from the
construction of $\mathbf c_{\ell,t}$ and $\mathbf c_\ell(\gamma_t)$,
the third inequality follows from \eqref{def.ineq.prod}, and
the fourth inequality follows from Assumption \ref{Assumption1}.2.(a) and Lemma \ref{lem:gelfand_bound} by taking $\tilde{\rho}_y\in (\rho_y, 1)$.

The proof is now completed.
\end{proof}

\begin{proof}[Proof of Lemma \ref{LM1}]
\item

(1). For any $\lambda\in [0,1]$ and $u,v\in \mathbb{R}$,
\begin{eqnarray*}
\rho_{\tau,m}(\lambda u+(1-\lambda)v) &=&\int \rho_\tau(\lambda u+(1-\lambda)v+m^{-1}x)\phi(x)\mathrm{d}x\notag \\
&\le&\lambda\int \rho_\tau( u+m^{-1}x)\phi(x)\mathrm{d}x + (1-\lambda)\int \rho_\tau( v+m^{-1}x)\phi(x)\mathrm{d}x \notag\\
&=&\lambda\rho_{\tau,m}(u)+(1-\lambda)\rho_{\tau,m}(v),
\end{eqnarray*}
where the inequality follows from $\rho_\tau(\cdot)$ being convex and $m^{-1}x = (\lambda + 1-\lambda)m^{-1}x$.

\medskip

(2). Note that $\phi_m(x)=m\phi(mx)$, and write

\begin{eqnarray*}
&&\rho_{\tau,m}(u)-\rho_\tau(u) \notag \\
&=& \int_{-1}^1 [\rho_\tau(x/m+u)-\rho_\tau(u)] \phi(x)\mathrm{d}x \notag \\
&=& \int_{-1}^1 [(x/m+u)(\tau - I(x/m+u<0)) -u(\tau - I(u<0)) ] \phi(x)\mathrm{d}x \notag \\
&=& \int_{-1}^1 [x/m\cdot (\tau - I(x<-um)) - u I(x<-um)+u I(u<0) ] \phi(x)\mathrm{d}x ,
\end{eqnarray*}
where the first equality follows from Assumption \ref{Assumption2}.

In what follows, we consider two cases: (i) $u\ge 0$ and (ii) $u< 0$. Start with case (i). By the definition of $\phi(x)$ in Assumption \ref{Assumption2}, if $-um\le -1$ (i.e., $u\ge m^{-1}$),

\begin{eqnarray*}
\rho_{\tau,m}(u)-\rho_\tau(u)= \int_{-1}^{1} x/m\cdot \tau  \phi(x)\mathrm{d}x =0.
\end{eqnarray*}

We then focus on the case with $-um\ge -1$, which yields that $0\le u\le m^{-1}$. By direct calculation, we write

\begin{eqnarray*}
\rho_{\tau,m}(u)-\rho_\tau(u) &=& \int_{-1}^{-um} [x/m\cdot (\tau - I(x<-um)) - u I(x<-um)] \phi(x)\mathrm{d}x \notag \\
&&+\int_{-um}^1 [x/m\cdot (\tau - I(x<-um)) - u I(x<-um)] \phi(x)\mathrm{d}x \notag \\
&=& \int_{-1}^{-um} [x/m\cdot (\tau - 1) - u ] \phi(x)\mathrm{d}x  +\int_{-um}^1 x/m\cdot \tau  \phi(x)\mathrm{d}x  \notag \\
&=& -\int_{-1}^{-um} (x/m + u) \phi(x)\mathrm{d}x  +\int_{-1}^1 x/m\cdot \tau  \phi(x)\mathrm{d}x \notag \\
&=& -\int_{-1}^{-um} \frac{x+um}{m} \phi(x)\mathrm{d}x ,
\end{eqnarray*}
where the fourth equality  follows from $\phi(\cdot)$ being symmetric of Assumption \ref{Assumption2}.

Note that

\begin{eqnarray*}
 -1\le x\le -um \ \Leftrightarrow \ \frac{-1+um}{m}\le \frac{x+um}{m}\le 0,
\end{eqnarray*}
which in connection with $0\le u\le m^{-1}$ further yields that

\begin{eqnarray*}
0\le -\frac{x+um}{m}\le \frac{1}{m}.
\end{eqnarray*}
Thus, we conclude that for case (i)

\begin{eqnarray*}
    0\le \rho_{\tau,m}(u)-\rho_\tau(u)\le \frac{1}{m} \int_{-1}^{0}\phi(x)\mathrm{d}x=\frac{1}{2m},
\end{eqnarray*}
where the last step follows from $\phi(\cdot)$ being symmetric of Assumption \ref{Assumption2}.

\medskip

Consider Case (ii). By the definition of $\phi(x)$ in Assumption \ref{Assumption2}, if $-um> 1$ (i.e., $u\le -m^{-1}$),

\begin{eqnarray*}
\rho_{\tau,m}(u)-\rho_\tau(u) =\int_{-1}^1 [x/m\cdot (\tau - 1) - u +u] \phi(x)\mathrm{d}x  =0.
\end{eqnarray*}

Thus we focus on the case with $-um \le 1$, which yields that $-m^{-1}\le u\le 0$. By direct calculation, we write

\begin{eqnarray*}
\rho_{\tau,m}(u)-\rho_\tau(u) &=& \int_{-1}^{-um} [x/m\cdot (\tau - I(x<-um)) - u I(x<-um)+u] \phi(x)\mathrm{d}x \notag \\
&&+\int_{-um}^1 [x/m\cdot (\tau - I(x<-um)) - u I(x<-um)+u] \phi(x)\mathrm{d}x \notag \\
&=& \int_{-1}^{-um} x/m\cdot (\tau - 1) \phi(x)\mathrm{d}x  +\int_{-um}^1 [x/m\cdot \tau +u] \phi(x)\mathrm{d}x \notag \\
&=&-\int_{-1}^{-um} x/m  \phi(x)\mathrm{d}x  +u\int_{-um}^1  \phi(x)\mathrm{d}x \notag \\
&=&-\int_{-1}^{um} x/m  \phi(x)\mathrm{d}x  +u\int_{-um}^1  \phi(x)\mathrm{d}x \notag \\
&=&-\int_{-1}^{um} x/m  \phi(x)\mathrm{d}x  -u\int_{um}^{-1}  \phi(x)\mathrm{d}x \notag \\
&=&-\int_{-1}^{um} x/m  \phi(x)\mathrm{d}x  +u\int_{-1}^{um}  \phi(x)\mathrm{d}x ,
\end{eqnarray*}
where third equality follows form $\phi(\cdot)$ being symmetric of Assumption \ref{Assumption2}, and the fourth equality follows from $\int_{um}^{-um}x/m  \phi(x)\mathrm{d}x  =0$ by Assumption \ref{Assumption2}.

Note that
 
\begin{eqnarray*}
-1\le x\le um \quad \Leftrightarrow  \quad -um\le -x\le 1 \quad \Leftrightarrow  \quad 0\le -x/m+u\le 1/m+u
\end{eqnarray*}
which in connection with $-m^{-1}\le u\le 0$ further yields that

\begin{eqnarray*}
0\le \rho_{\tau,m}(u)-\rho_\tau(u) \le   \frac{1}{m} \int_{-1}^{0}  \phi(x)\mathrm{d}x =\frac{1}{2m},
\end{eqnarray*}
where the last step follows from $\phi(\cdot)$ being symmetric of Assumption \ref{Assumption2}.

Putting the above results together, the proof is now completed.    
\end{proof}

\begin{proof}[Proof of Theorem \ref{TM.1}]
\item 

Before proceeding, we introduce some notations to facilitate the
development.  Let

\begin{eqnarray*}
    &&\Delta Q(\mathbf{B}^0, \mathbf{B}^1) \coloneqq Q(\mathbf{B}^0, \mathbf{B}^1) - Q_{\mathbf{e}} +Q_1(\mathbf{B}^0, \mathbf{B}^1),\quad
    \Delta\bar{y}_{\ell, t} \coloneqq \mathbf{x}_{t-1}^\top \pmb{\beta}_{\ell, t} -  \mathbf{x}_{t-1}^\top \mathbf{b}_{\ell}^0 -\mathbf{z}_{t-1}^\top \mathbf{b}_{\ell}^1 ,\notag \\
    &&\Delta \pmb{\beta}_{\ell}(\gamma) \coloneqq \pmb{\beta}_{\ell}(\gamma) -\mathbf{b}_{\ell}^0 ,\quad\Delta\pmb{\beta}_{\ell}^{(1)}(\gamma) \coloneqq \pmb{\beta}_{\ell}^{(1)}(\gamma)-\mathbf{b}_{\ell}^1.
\end{eqnarray*}
Here $\mathbf B_\delta$ and $\pmb\beta_{\gamma,\delta}$ are as defined in Theorem \ref{TM.1}.

We start our investigation by focusing on
$\Delta Q(\mathbf B^0,\mathbf B^1)$. Note that
\begin{eqnarray*}
 \Delta Q(\mathbf B^0,\mathbf B^1) 
 &=&\frac1T\sum_{t=1}^{T}\sum_{\ell=1}^{p}
 \left\{
 \rho_{\tau,m}(e_{\ell,t}+\Delta\bar y_{\ell,t})
 -\rho_{\tau,m}(e_{\ell,t})
 -\rho_{\tau,m}^{(1)}(e_{\ell,t})\Delta\bar y_{\ell,t}
 \right\}K_\delta(\gamma_t-\gamma)\nonumber\\
 &=&\frac1T\sum_{t=1}^{T}\sum_{\ell=1}^{p}
 \int_0^{\Delta\bar y_{\ell,t}}
 \left\{
 \rho_{\tau,m}^{(1)}(e_{\ell,t}+w)
 -\rho_{\tau,m}^{(1)}(e_{\ell,t})
 \right\}\mathrm dw\,K_\delta(\gamma_t-\gamma).
\end{eqnarray*}

Define $R_{\ell,t} \coloneqq \int_0^{\Delta\bar y_{\ell,t}} \left\{ \rho_{\tau,m}^{(1)}(e_{\ell,t}+w) -\rho_{\tau,m}^{(1)}(e_{\ell,t})  \right\}\mathrm dw.$ Since $\Delta\bar y_{\ell,t}$ is $\mathcal F_{t-1}$-measurable, $R_{\ell,t}-E_{t-1}[R_{\ell,t}]$ is a martingale difference. By Cauchy-Schwarz inequality and Lemma \ref{Appendix.B2}.2,

\begin{eqnarray}\label{rate.BH}
 E_{t-1}[R_{\ell,t}^2]
 &\le&
 |\Delta\bar y_{\ell,t}|
 \int_0^{|\Delta\bar y_{\ell,t}|}
 E_{t-1}\!\left[
 \left\{
 \rho_{\tau,m}^{(1)}(e_{\ell,t}+w)
 -\rho_{\tau,m}^{(1)}(e_{\ell,t})
 \right\}^2
 \right]\mathrm dw\le C|\Delta\bar y_{\ell,t}|^3.
\end{eqnarray}
For $|\gamma_t-\gamma|\le\delta$, Taylor expansion gives
\begin{eqnarray}\label{proof.tm1.dy}
 \Delta\bar y_{\ell,t}
 &=&
 \mathbf x_{t-1}^{\top}
 \left\{
 \Delta\pmb\beta_\ell(\gamma)
 +(\gamma_t-\gamma)\Delta\pmb\beta_\ell^{(1)}(\gamma)
 +\frac{(\gamma_t-\gamma)^2}{2}
 \pmb\beta_\ell^{(2)}(\gamma)
 +\mathbf r_{\ell,t}
 \right\},
\end{eqnarray}
where $\sup_t\|\mathbf r_{\ell,t}\|=O(\delta^3)$. Consequently,
\begin{eqnarray*}
 \frac1{T^2}\sum_{t=1}^{T}
 E|\Delta\bar y_{\ell,t}|^3
 K_\delta^2(\gamma_t-\gamma)
 &=&
 O\!\left(
 \frac{\|\Delta\mathbf B_\delta\|^3+\delta^6}
 {T\delta}
 \right).
\end{eqnarray*}

Thus, we can conclude that for $\|\Delta\mathbf B_\delta\|=o(1)$,
\begin{eqnarray}\label{def.rate.deltaQ}
  \left|
 \Delta Q(\mathbf B^0,\mathbf B^1)
 -\frac1T\sum_{t=1}^{T}\sum_{\ell=1}^{p}
 E_{t-1}[R_{\ell,t}]K_\delta(\gamma_t-\gamma)
 \right|
 &=&o_P\!\left(
 \|\Delta\mathbf B_\delta\|^2+\delta^4
 +\frac{1}{T\delta}
 \right).
\end{eqnarray}

 Using Lemma \ref{Appendix.B3} and the arguments that are similar to those in the proof of \eqref{def.rate.deltaQ}, we can establish the following uniform convergence result
\begin{eqnarray*}
 \sup_{\gamma\in[\delta,1-\delta]}
 \left|
 \Delta Q(\mathbf B^0,\mathbf B^1)
 -\frac1T\sum_{t=1}^{T}\sum_{\ell=1}^{p}
 E_{t-1}[R_{\ell,t}]K_\delta(\gamma_t-\gamma)
 \right|
 &=&o_P\!\left(
 \|\Delta\mathbf B_\delta\|^2+\delta^4
 +\frac{\log T}{T\delta}
 \right).
\end{eqnarray*}

We then focus on $\frac1T\sum_{t=1}^{T}\sum_{\ell=1}^{p} E_{t-1}[R_{\ell,t}]K_\delta(\gamma_t-\gamma)$. \eqref{proof.tm1.dy} and Lemma \ref{Appendix.B2} give
\begin{eqnarray}\label{proof.tm1.predictable}
 &&\frac1T\sum_{t=1}^{T}\sum_{\ell=1}^{p}
 E_{t-1}[R_{\ell,t}]K_\delta(\gamma_t-\gamma)\nonumber\\
 &=&
 \frac1{2T}\sum_{t=1}^{T}\sum_{\ell=1}^{p}
 f_{\ell,t}(0\mid\mathcal F_{t-1})
 (\Delta\bar y_{\ell,t})^2K_\delta(\gamma_t-\gamma)\nonumber\\
 &&+O_P\!\left[
 \frac1T\sum_{t=1}^{T}\sum_{\ell=1}^{p}
 \left\{|\Delta\bar y_{\ell,t}|^3
 +m^{-1}|\Delta\bar y_{\ell,t}|^2\right\}
 K_\delta(\gamma_t-\gamma)
 \right]\nonumber\\
 &=&
 \frac1{2T}\sum_{t=1}^{T}\sum_{\ell=1}^{p}
 f_{\ell,t}(0\mid\mathcal F_{t-1})
 (\Delta\bar y_{\ell,t})^2K_\delta(\gamma_t-\gamma)
 +o_P(\|\Delta\mathbf B_\delta\|^2+\delta^4).
\end{eqnarray}

To evaluate the leading quadratic term, let $w_t=(\gamma_t-\gamma)/\delta$ and write
\begin{eqnarray*}
 \mathbf h_{\ell,t}
 &\coloneqq&
 \mathbf b_\ell^0-\pmb\beta_\ell(\gamma)
 +\delta w_t\{\mathbf b_\ell^1
 -\pmb\beta_\ell^{(1)}(\gamma)\}
 -\frac{\delta^2w_t^2}{2}\pmb\beta_\ell^{(2)}(\gamma).
\end{eqnarray*}

By \eqref{proof.tm1.dy} and Assumption \ref{Assumption1},   we obtain 
\begin{eqnarray}\label{def.rate.Q1}
 &&\frac1{2T}\sum_{t=1}^{T}\sum_{\ell=1}^{p}
 f_{\ell,t}(0\mid\mathcal F_{t-1})
 (\Delta\bar y_{\ell,t})^2K_\delta(\gamma_t-\gamma)\nonumber\\
 &=&
 \frac12\int_{-1}^{1}
 (\Delta\mathbf B_\delta-w^2\pmb\xi_{\gamma,\delta})^{\top}
 \diag\{\pmb\Sigma_{\mathbf f,\gamma},
 w^2\pmb\Sigma_{\mathbf f,\gamma}\}
 (\Delta\mathbf B_\delta-w^2\pmb\xi_{\gamma,\delta})
 K(w)\mathrm dw +o_P \left(
 \|\Delta\mathbf B_\delta\|^2+\delta^4
 +\frac{\log T}{T\delta}
 \right)\nonumber\\
 &=&Q_2(\mathbf B^0,\mathbf B^1)
 +o_P\!\left(
 \|\Delta\mathbf B_\delta\|^2+\delta^4
 +\frac{\log T}{T\delta}
 \right).
\end{eqnarray}
Putting \eqref{def.rate.deltaQ}, \eqref{proof.tm1.predictable}, and \eqref{def.rate.Q1} together
completes the proof.

The proof is now completed.
\end{proof}

\begin{proof}[Proof of Lemma \ref{Appendix.A1}]
\item    

(1). To proceed, recall the notations defined in the proof of Theorem \ref{TM.1}. Firstly, we rewrite the following term in Theorem \ref{TM.1} 

\begin{eqnarray*}
    &&Q(\mathbf{B}^0, \mathbf{B}^1) -Q_{\mathbf{e}}+Q_1(\mathbf{B}^0, \mathbf{B}^1) - Q_2(\mathbf{B}^0, \mathbf{B}^1)\notag \\
    &=& Q(\mathbf{B}^0, \mathbf{B}^1) -Q(\pmb{\beta}_\gamma^0, \pmb{\beta}_\gamma^1) - Q_2(\mathbf{B}^0, \mathbf{B}^1) +Q_1(\mathbf{B}^0, \mathbf{B}^1)- \Delta Q_{\mathbf{e}},
\end{eqnarray*}
where 

\begin{eqnarray*}
    \Delta Q_{\mathbf{e}} &\coloneqq & Q_{\mathbf{e}} - Q(\pmb{\beta}_{\gamma}^0, \pmb{\beta}_{\gamma}^1) =\frac{1}{T}\sum_{t=1}^{T}\sum_{\ell=1}^p \{\rho_{\tau, m}\left(e_{\ell,t} \right) - \rho_{\tau, m} (e_{\ell,t}+\Delta\bar{y}_{\ell, t\mid \pmb{\beta}}  ) \} \cdot K_\delta(\gamma_t -\gamma),\notag \\
    \Delta\bar{y}_{\ell, t\mid \pmb{\beta}} &\coloneqq & \mathbf{x}_{t-1}^\top \pmb{\beta}_{\ell, t} - \mathbf{x}_{t-1}^\top \pmb{\beta}_{\ell}(\gamma) -\mathbf{z}_{t-1}^\top \pmb{\beta}_{\ell}^{(1)}(\gamma).
\end{eqnarray*}

We now study $Q_1(\mathbf{B}^0, \mathbf{B}^1)-\Delta Q_{\mathbf{e}}$.

\begin{eqnarray*}
    Q_1(\mathbf{B}^0, \mathbf{B}^1)-\Delta Q_{\mathbf{e}} &=& -\frac{1}{T}\sum_{t=1}^{T}\sum_{\ell=1}^p \rho_{\tau, m}^{(1)} (e_{\ell,t}  ) \cdot  (\mathbf{x}_{t-1}^\top\pmb{\beta}_{\ell, t} -  \mathbf{x}_{t-1}^\top \mathbf{b}_{\ell}^0 -\mathbf{z}_{t-1}^\top \mathbf{b}_{\ell}^1) \cdot K_\delta(\gamma_t -\gamma)\notag\\
    &&+\frac{1}{T}\sum_{t=1}^{T}\sum_{\ell=1}^p \{ \rho_{\tau, m} (e_{\ell,t}+\Delta\bar{y}_{\ell, t\mid \pmb{\beta}}) - \rho_{\tau, m}\left(e_{\ell,t} \right) \} \cdot K_\delta(\gamma_t -\gamma) \notag \\
    &=& \frac{1}{T}\sum_{t=1}^{T}\sum_{\ell=1}^p\int_{\Delta\bar{y}_{\ell, t} }^0 \rho_{\tau, m}^{(1)} (e_{\ell,t}  ) \mathrm{d}w \cdot K_\delta(\gamma_t -\gamma)\notag \\
    &&+\frac{1}{T}\sum_{t=1}^{T}\sum_{\ell=1}^p \int_0^{\Delta\bar{y}_{\ell, t\mid \pmb{\beta}}}\rho_{\tau, m}^{(1)} (e_{\ell,t} +w ) \mathrm{d}w \cdot K_\delta(\gamma_t -\gamma)\notag \\
    &=& \frac{1}{T}\sum_{t=1}^{T}\sum_{\ell=1}^p\int_{\Delta\bar{y}_{\ell, t} }^{\Delta\bar{y}_{\ell, t\mid \pmb{\beta}}} \rho_{\tau, m}^{(1)} (e_{\ell,t}  ) \mathrm{d}w\cdot K_\delta(\gamma_t -\gamma) \notag \\
    &&+\frac{1}{T}\sum_{t=1}^{T}\sum_{\ell=1}^p \int_0^{\Delta\bar{y}_{\ell, t\mid \pmb{\beta}}}\{\rho_{\tau, m}^{(1)} (e_{\ell,t} +w ) -\rho_{\tau, m}^{(1)} (e_{\ell,t} )\}\mathrm{d}w \cdot K_\delta(\gamma_t -\gamma).
\end{eqnarray*}
By the conditional version of the development in \eqref{rate.BH}, Lemma \ref{Appendix.B2}.3, we have uniformly in $\gamma\in[\delta,1-\delta]$
\begin{eqnarray*}
 \frac{1}{T}\sum_{t=1}^{T}\sum_{\ell=1}^p
 \int_0^{\Delta\bar{y}_{\ell,t\mid\pmb\beta}}
 \{\rho_{\tau,m}^{(1)}(e_{\ell,t}+w)
 -\rho_{\tau,m}^{(1)}(e_{\ell,t})\}\mathrm dw
 K_\delta(\gamma_t-\gamma)
 &=&O_P\!\left(\delta^4
 +\delta^3\sqrt{\frac{\log T}{T\delta}}\right).
\end{eqnarray*}

We then focus on $\frac{1}{T}\sum_{t=1}^{T}\sum_{\ell=1}^p\int_{\Delta\bar{y}_{\ell, t} }^{\Delta\bar{y}_{\ell, t\mid \pmb{\beta}}} \rho_{\tau, m}^{(1)} (e_{\ell,t}  )\mathrm{d}w$, and write

\begin{eqnarray}\label{rate.e.beta}
    &&\left|\frac{1}{T}\sum_{t=1}^{T}\sum_{\ell=1}^p\int_{\Delta\bar{y}_{\ell, t} }^{\Delta\bar{y}_{\ell, t\mid \pmb{\beta}}} \rho_{\tau, m}^{(1)} (e_{\ell,t}  )\mathrm{d}w \cdot K_\delta(\gamma_t -\gamma)\notag \right| \notag \\
    &=&\left| \frac{1}{T}\sum_{t=1}^{T}\sum_{\ell=1}^p \rho_{\tau, m}^{(1)} (e_{\ell,t}  )(\Delta \pmb{\beta}_{\ell}(\gamma) +(\gamma_t-\gamma)\Delta \pmb{\beta}_{\ell}^{(1)}(\gamma) ) \cdot K_\delta(\gamma_t -\gamma)\right| \notag \\
    &\le &\frac{\sqrt{\log T}}{\sqrt{T\delta}}\cdot \|\mathbf{B}_{\delta}-\pmb{\beta}_{\gamma,\delta}\|_2,
\end{eqnarray}
where last step holds uniformly in $\gamma\in[\delta, 1-\delta]$ by Lemma \ref{Appendix.B3}.

Up to this point, we can conclude that uniformly in $\gamma\in[\delta, 1-\delta]$

\begin{eqnarray*}
 |Q_1(\mathbf{B}^0,\mathbf{B}^1)-\Delta Q_{\mathbf e}|
 &\le&
 O_P(1)\frac{\sqrt{\log T}}{\sqrt{T\delta}}
 \|\mathbf B_\delta-\pmb\beta_{\gamma,\delta}\|
 +O_P\!\left(\delta^4
 +\delta^3\sqrt{\frac{\log T}{T\delta}}\right).
\end{eqnarray*}
 
To complete the proof, we denote

\begin{eqnarray*}
    \bar{\mathbb{B}}  =\left\{(\mathbf{B}^0, \mathbf{B}^1)\mid \frac{\sqrt{T\delta}}{\sqrt{\log T}}\|\mathbf{B}_{\delta}-\pmb{\beta}_{\gamma,\delta}\| \ge c\right\},
\end{eqnarray*}
wherein $c$ is a sufficiently large constant. Recall that
$\delta\lesssim T^{-1/5}$, which implies
$\delta^2\lesssim\sqrt{\log T/(T\delta)}$. Thus, by the quadratic form
derived in Theorem \ref{TM.1}, we know that uniformly in
$\gamma\in[\delta,1-\delta]$,

\begin{eqnarray*}
    \Pr\left(\inf_{\bar{\mathbb{B}}}Q(\mathbf{B}^0, \mathbf{B}^1) \le Q(\pmb{\beta}_{\gamma}^0, \pmb{\beta}_{\gamma}^1)\right)\to 0.
\end{eqnarray*}

The proof of (1) is now completed.

\medskip

(2). For $\forall\gamma \in (0,1)$, the investigation of \eqref{rate.e.beta} reduces to 

\begin{eqnarray*}
    &&\left|\frac{1}{T}\sum_{t=1}^{T}\sum_{\ell=1}^p\int_{\Delta\bar{y}_{\ell, t} }^{\Delta\bar{y}_{\ell, t\mid \pmb{\beta}}} \rho_{\tau, m}^{(1)} (e_{\ell,t}  )\mathrm{d}w \cdot K_\delta(\gamma_t -\gamma)\notag \right| \notag \\
    &=&\left| \frac{1}{T}\sum_{t=1}^{T}\sum_{\ell=1}^p \rho_{\tau, m}^{(1)} (e_{\ell,t}  )(\Delta \pmb{\beta}_{\ell}(\gamma) +(\gamma_t-\gamma)\Delta \pmb{\beta}_{\ell}^{(1)}(\gamma) ) \cdot K_\delta(\gamma_t -\gamma)\right| \notag \\
    &\le &\frac{1}{\sqrt{T\delta}}\cdot \|\mathbf{B}_{\delta}-\pmb{\beta}_{\gamma,\delta}\|_2,
\end{eqnarray*}
where the last step holds for $\forall\gamma \in (0,1)$ by Lemma \ref{Appendix.B3}. The rest is identical to the proof in (1). Thus, the result follows. 
\end{proof}

\begin{proof}[Proof of Lemma \ref{Appendix.A2}]
\item    

Firstly, recall those notations defined in Theorem \ref{TM.1} and its proof. Also, let $\widehat{\mathbf B}_\delta\coloneqq \VEC(\widehat{\mathbf B}^0,\delta\widehat{\mathbf B}^1)$ and $\tilde{\pmb\beta}_{\gamma,\delta}\coloneqq \VEC(\tilde{\pmb\beta}_\gamma^0,
\delta\tilde{\pmb\beta}_\gamma^1)$ for notational simplicity. 

For both results, it is sufficient to focus on $\mathbf J_\gamma$.

In what follows, we prove (1.a) and (2.a) first, and then move on to
(1.b) and (2.b).

\medskip

(1.a). The result follows from Lemma \ref{Appendix.B3} immediately.

\medskip

(2.a). Write
\begin{eqnarray*}
 \sqrt{T\delta}\mathbf J_\gamma
 &=&\frac1{\sqrt{T\delta}}\sum_{t=1}^{T}
 [\pmb\psi_t\otimes
 \VEC(\mathbf x_{t-1},\delta^{-1}\mathbf z_{t-1})]
 K((\gamma_t-\gamma)/\delta)\nonumber\\
 &&+\frac1{\sqrt{T\delta}}\sum_{t=1}^{T}
 [(\pmb\rho_{m,t}^{(1)}-\pmb\psi_t)\otimes
 \VEC(\mathbf x_{t-1},\delta^{-1}\mathbf z_{t-1})]
 K((\gamma_t-\gamma)/\delta)\nonumber\\
 &=&\frac1{\sqrt{T\delta}}\sum_{t=1}^{T}\mathbf w_t
 +O_P(\sqrt{T\delta}/m),
\end{eqnarray*}
where the last equality follows from Lemma \ref{Appendix.B3} and
$\sqrt{T\delta}/m\to0$, and the definition of $\mathbf w_t$ is evident.

The variable $\mathbf w_t$ is $\mathcal F_t$-measurable, and  Assumption \ref{Assumption1} gives
\begin{eqnarray*}
 E[\mathbf w_t\mid\mathcal F_{t-1}]
 &=&
 \left[
 E[\pmb\psi_t\mid\mathcal F_{t-1}]
 \otimes\VEC(\mathbf x_{t-1},\delta^{-1}\mathbf z_{t-1})
 \right]K((\gamma_t-\gamma)/\delta)
 =\mathbf0.
\end{eqnarray*}
Thus $\{\mathbf w_t,\mathcal F_t\}$ is a martingale difference array. By Assumption \ref{Assumption1},
\begin{eqnarray}\label{proof.qvar.pqv}
 \frac1{T\delta}\sum_{t=1}^{T}
 E[\mathbf w_t\mathbf w_t^{\top}
 \mid\mathcal F_{t-1}]
 &\to_P&
 \diag\{a_1\mathbf M_\gamma^Q,a_2\mathbf M_\gamma^Q\},
\end{eqnarray}
where $a_1$ and $a_2$ have been defined in the body of this assumption.

For notational simplicity, let $w_t=\mathbf d^{\top}\mathbf w_t$, where $\|\mathbf d\|=1$ is an arbitrary deterministic vector used in the Cram\'er--Wold device. Since $|\pmb\psi_t|$ and $K(\cdot)$ are bounded, conditional Markov inequality gives, for every $\epsilon>0$,
\begin{eqnarray*}
 &&\frac1{T\delta}\sum_{t=1}^{T}
 E\!\left[
 |w_t|^2
 I\{|w_t|>\epsilon\sqrt{T\delta}\}
 \mid\mathcal F_{t-1}
 \right]\nonumber\\
 &\le&
 \frac{C}{(T\delta)^{\nu/2}}
 \sum_{t=1}^{T}
 E[\|\mathbf x_{t-1}\|^\nu\mid\mathcal F_{t-1}]
 |K(w_{t,\gamma})|^\nu
 =o_P(1).
\end{eqnarray*}
The last equality follows because only $O(T\delta)$ terms are nonzero and the moving-average representation and Assumption \ref{Assumption1} imply $\sup_{T,t}\|\mathbf x_{t-1}\|_\nu<\infty$.  The martingale central limit theorem, \eqref{proof.qvar.pqv}, and the Cram\'er--Wold device therefore yield
\begin{eqnarray}\label{proof.qvar.exact.clt}
 \frac1{\sqrt{T\delta}}\sum_{t=1}^{T}\mathbf w_t
 \to_D
 N\!\left(
 \mathbf0,
 \diag\{a_1\mathbf M_\gamma^Q,a_2\mathbf M_\gamma^Q\}
 \right).
\end{eqnarray}
By \eqref{def.beta.tilde},
\begin{eqnarray}\label{proof.clt.bridge}
 \tilde{\pmb\beta}_{\gamma,\delta}
 -\pmb\beta_{\gamma,\delta}
 &=&
 v_2\pmb\xi_{\gamma,\delta}
 +\diag\{\pmb\Sigma_{\mathbf f,\gamma},
 v_2\pmb\Sigma_{\mathbf f,\gamma}\}^{-1}
 \mathbf J_\gamma.
\end{eqnarray}
\eqref{proof.qvar.exact.clt} and \eqref{proof.clt.bridge} prove (2.a) with the covariance matrix
defined in Theorem \ref{TM.2}.

\medskip

(1.b) \& (2.b). We write

\begin{eqnarray}\label{def.diff}
    &&\big| Q(\mathbf{B}^0, \mathbf{B}^1) -Q_{\mathbf{e}}+Q_1(\mathbf{B}^0, \mathbf{B}^1) - Q_2(\mathbf{B}^0, \mathbf{B}^1) \notag\\
    &&-Q(\tilde{\pmb{\beta}}_\gamma^0 , \tilde{\pmb{\beta}}_\gamma^1) +Q_{\mathbf{e}}-Q_1(\tilde{\pmb{\beta}}_\gamma^0 ,  \tilde{\pmb{\beta}}_\gamma^1) + Q_2(\tilde{\pmb{\beta}}_\gamma^0 ,\tilde{\pmb{\beta}}_\gamma^1) \big| \notag\\
    &=&\big|Q(\mathbf{B}^0, \mathbf{B}^1)-Q(\tilde{\pmb{\beta}}_\gamma^0 , \tilde{\pmb{\beta}}_\gamma^1) +\mathbf{J}_\gamma^\top (\mathbf{B}_\delta - \tilde{\pmb{\beta}}_{\gamma,\delta})\notag \\
    &&-\frac{1}{2}\int_{-1}^{1} (\mathbf{B}_\delta -\pmb{\beta}_{\gamma,\delta}-w^2\pmb{\xi}_{\gamma,\delta})^\top \diag\{\pmb{\Sigma}_{\mathbf{f},\gamma}, w^2\pmb{\Sigma}_{\mathbf{f},\gamma}\} (\mathbf{B}_\delta -\pmb{\beta}_{\gamma,\delta}-w^2\pmb{\xi}_{\gamma,\delta}) K(w)\mathrm{d}w \notag \\
    &&+\frac{1}{2} \int_{-1}^{1} (\tilde{\pmb{\beta}}_{\gamma,\delta} -\pmb{\beta}_{\gamma,\delta}-w^2\pmb{\xi}_{\gamma,\delta})^\top \diag\{\pmb{\Sigma}_{\mathbf{f},\gamma}, w^2\pmb{\Sigma}_{\mathbf{f},\gamma}\} (\tilde{\pmb{\beta}}_{\gamma,\delta} -\pmb{\beta}_{\gamma,\delta}-w^2\pmb{\xi}_{\gamma,\delta}) K(w)\mathrm{d}w \big|\notag \\
    &=&\big|Q(\mathbf{B}^0, \mathbf{B}^1)-Q(\tilde{\pmb{\beta}}_\gamma^0 , \tilde{\pmb{\beta}}_\gamma^1) +\mathbf{J}_\gamma^\top (\mathbf{B}_\delta - \tilde{\pmb{\beta}}_{\gamma,\delta})\notag \\
    &&-\frac{1}{2}\int_{-1}^{1} (\mathbf{B}_\delta-\tilde{\pmb{\beta}}_{\gamma,\delta}+\tilde{\pmb{\beta}}_{\gamma,\delta} -\pmb{\beta}_{\gamma,\delta}-w^2\pmb{\xi}_{\gamma,\delta})^\top \diag\{\pmb{\Sigma}_{\mathbf{f},\gamma}, w^2\pmb{\Sigma}_{\mathbf{f},\gamma}\} \notag \\
    &&\quad \cdot (\mathbf{B}_\delta-\tilde{\pmb{\beta}}_{\gamma,\delta}+\tilde{\pmb{\beta}}_{\gamma,\delta}-\pmb{\beta}_{\gamma,\delta}-w^2\pmb{\xi}_{\gamma,\delta}) K(w)\mathrm{d}w \notag \\
    &&+\frac{1}{2} \int_{-1}^{1} (\tilde{\pmb{\beta}}_{\gamma,\delta} -\pmb{\beta}_{\gamma,\delta}-w^2\pmb{\xi}_{\gamma,\delta})^\top \diag\{\pmb{\Sigma}_{\mathbf{f},\gamma}, w^2\pmb{\Sigma}_{\mathbf{f},\gamma}\} (\tilde{\pmb{\beta}}_{\gamma,\delta} -\pmb{\beta}_{\gamma,\delta}-w^2\pmb{\xi}_{\gamma,\delta}) K(w)\mathrm{d}w \big|\notag \\
    &=&\big|Q(\mathbf{B}^0, \mathbf{B}^1)-Q(\tilde{\pmb{\beta}}_\gamma^0 , \tilde{\pmb{\beta}}_\gamma^1) +\mathbf{J}_\gamma^\top (\mathbf{B}_\delta - \tilde{\pmb{\beta}}_{\gamma,\delta})\notag \\
    &&-\frac{1}{2}\int_{-1}^{1} (\mathbf{B}_\delta -\tilde{\pmb{\beta}}_{\gamma,\delta} )^\top \diag\{\pmb{\Sigma}_{\mathbf{f},\gamma}, w^2\pmb{\Sigma}_{\mathbf{f},\gamma}\} (\mathbf{B}_\delta -\tilde{\pmb{\beta}}_{\gamma,\delta}) K(w)\mathrm{d}w \notag \\
    &&-\int_{-1}^{1} (\tilde{\pmb{\beta}}_{\gamma,\delta} -\pmb{\beta}_{\gamma,\delta}-w^2\pmb{\xi}_{\gamma,\delta})^\top \diag\{\pmb{\Sigma}_{\mathbf{f},\gamma}, w^2\pmb{\Sigma}_{\mathbf{f},\gamma}\} (\mathbf{B}_\delta -\tilde{\pmb{\beta}}_{\gamma,\delta}) K(w)\mathrm{d}w  \big|.
\end{eqnarray}

We now show that 

\begin{eqnarray*}
     \mathbf{J}_\gamma^\top (\mathbf{B}_\delta - \tilde{\pmb{\beta}}_{\gamma,\delta}) &=&\int_{-1}^{1} (\tilde{\pmb{\beta}}_{\gamma,\delta} -\pmb{\beta}_{\gamma,\delta}-w^2\pmb{\xi}_{\gamma,\delta})^\top \diag\{\pmb{\Sigma}_{\mathbf{f},\gamma}, w^2\pmb{\Sigma}_{\mathbf{f},\gamma}\} (\mathbf{B}_\delta -\tilde{\pmb{\beta}}_{\gamma,\delta}) K(w)\mathrm{d}w 
\end{eqnarray*}
in order to further simplify \eqref{def.diff}. It is sufficient to show that

\begin{eqnarray*}
    \mathbf{J}_\gamma =\int_{-1}^{1}   \diag\{\pmb{\Sigma}_{\mathbf{f},\gamma}, w^2\pmb{\Sigma}_{\mathbf{f},\gamma}\} (\tilde{\pmb{\beta}}_{\gamma,\delta} -\pmb{\beta}_{\gamma,\delta}-w^2\pmb{\xi}_{\gamma,\delta}) K(w)\mathrm{d}w ,
\end{eqnarray*}
which is precisely how $\mathbf{J}_\gamma$, $\tilde{\pmb{\beta}}_{\gamma,\delta}$, $\pmb{\xi}_{\gamma,\delta}$ are constructed in \eqref{def.beta.tilde}. Therefore,

\begin{eqnarray*}
    &&\big| Q(\mathbf{B}^0, \mathbf{B}^1) -Q_{\mathbf{e}}+Q_1(\mathbf{B}^0, \mathbf{B}^1) - Q_2(\mathbf{B}^0, \mathbf{B}^1) \notag\\
    &&-Q(\tilde{\pmb{\beta}}_\gamma^0 , \tilde{\pmb{\beta}}_\gamma^1) +Q_{\mathbf{e}}-Q_1(\tilde{\pmb{\beta}}_\gamma^0 ,  \tilde{\pmb{\beta}}_\gamma^1) + Q_2(\tilde{\pmb{\beta}}_\gamma^0 ,\tilde{\pmb{\beta}}_\gamma^1) \big| \notag\\
    &=&\big|Q(\mathbf{B}^0, \mathbf{B}^1)-Q(\tilde{\pmb{\beta}}_\gamma^0 , \tilde{\pmb{\beta}}_\gamma^1) -\frac{1}{2}\int_{-1}^{1} (\mathbf{B}_\delta -\tilde{\pmb{\beta}}_{\gamma,\delta} )^\top \diag\{\pmb{\Sigma}_{\mathbf{f},\gamma}, w^2\pmb{\Sigma}_{\mathbf{f},\gamma}\} (\mathbf{B}_\delta -\tilde{\pmb{\beta}}_{\gamma,\delta}) K(w)\mathrm{d}w  \big|.
\end{eqnarray*}

(1.b). In connection with Lemma \ref{Appendix.A1}, we obtain uniformly in $\gamma\in [\delta, 1-\delta]$

\begin{eqnarray*}
    o_P \left(\frac{\log T}{T\delta}\right)&=&\big| Q(\widehat{\mathbf{B}}^0, \widehat{\mathbf{B}}^1) -Q_{\mathbf{e}}+Q_1(\widehat{\mathbf{B}}^0,\widehat{\mathbf{B}}^1) - Q_2(\widehat{\mathbf{B}}^0, \widehat{\mathbf{B}}^1) \notag\\
    &&-Q(\tilde{\pmb{\beta}}_\gamma^0 , \tilde{\pmb{\beta}}_\gamma^1) +Q_{\mathbf{e}}-Q_1(\tilde{\pmb{\beta}}_\gamma^0 ,  \tilde{\pmb{\beta}}_\gamma^1) + Q_2(\tilde{\pmb{\beta}}_\gamma^0 ,\tilde{\pmb{\beta}}_\gamma^1) \big| \notag\\
    &=&\big|Q(\widehat{\mathbf{B}}^0, \widehat{\mathbf{B}}^1)-Q(\tilde{\pmb{\beta}}_\gamma^0 , \tilde{\pmb{\beta}}_\gamma^1) -\frac{1}{2}\int_{-1}^{1} (\mathbf{B}_\delta -\tilde{\pmb{\beta}}_{\gamma,\delta} )^\top \diag\{\pmb{\Sigma}_{\mathbf{f},\gamma}, w^2\pmb{\Sigma}_{\mathbf{f},\gamma}\} (\mathbf{B}_\delta -\tilde{\pmb{\beta}}_{\gamma,\delta}) K(w)\mathrm{d}w\big| .
\end{eqnarray*}
Thus, uniformly in $\gamma\in [\delta, 1-\delta]$,

\begin{eqnarray*}
    |Q(\widehat{\mathbf{B}}^0, \widehat{\mathbf{B}}^1)-Q(\tilde{\pmb{\beta}}_\gamma^0, \tilde{\pmb{\beta}}_\gamma^1)  | > \frac{1}{4} \big| \int_{-1}^{1} (\widehat{\mathbf{B}}_\delta -\tilde{\pmb{\beta}}_{\gamma,\delta} )^\top \diag\{\pmb{\Sigma}_{\mathbf{f},\gamma}, w^2\pmb{\Sigma}_{\mathbf{f},\gamma}\} (\widehat{\mathbf{B}}_\delta -\tilde{\pmb{\beta}}_{\gamma,\delta}) K(w)\mathrm{d}w \big|
\end{eqnarray*}
which infers that $\sqrt{T\delta/\log T}\|\widehat{\mathbf{B}}_\delta -\tilde{\pmb{\beta}}_{\gamma, \delta}  \|\gtrsim  1$ cannot hold uniformly in $\gamma\in [\delta, 1-\delta]$.

The proof of (1.b) is now completed.

\medskip

(2.b). The result follows from a argument similar to (1.b).

\end{proof}

\begin{proof}[Prooof of Theorem \ref{TM.3}]
\item     

In what follows, we specifically consider two cases: $\tilde{q}<q$ and $\tilde{q}>q$, and show that both cannot hold with probability approaching 1.

\medskip

Case 1 --- Without loss of generality, suppose that $\tilde{q}=q-1$. We compare 

\begin{eqnarray*}
    \frac{1}{L}\sum_{\ell=1}^L Q_{\tilde{\gamma}_\ell\mid \tilde{q}}(\mathbf{B}_{ \mid \tilde{q}}^0, \mathbf{B}_{ \mid \tilde{q}}^1)\quad \text{and}\quad \frac{1}{L}\sum_{\ell=1}^L Q_{\tilde{\gamma}_\ell\mid q}(\pmb{\beta}_{\tilde{\gamma}_\ell}^0, \pmb{\beta}_{\tilde{\gamma}_\ell}^1),
\end{eqnarray*}
in which $\pmb{\beta}_{\tilde{\gamma}_\ell}^0$ and $\pmb{\beta}_{\tilde{\gamma}_\ell}^1$ are defined in \eqref{def.beta0} and \eqref{def.beta12} already.

Note that for $\forall \ell \in [L]$,

\begin{eqnarray*}
    Q_{\tilde{\gamma}_\ell\mid \tilde{q}}(\mathbf{B}_{\mid \tilde{q}}^0, \mathbf{B}_{\mid \tilde{q}}^1) \equiv Q_{\tilde{\gamma}_\ell\mid q}(\mathbf{B}_{\mid q}^0, \mathbf{B}_{\mid q}^1),
\end{eqnarray*}
in which $\mathbf{B}_{\mid q}^0$ and $\mathbf{B}_{\mid q}^1$ take the following forms:

\begin{eqnarray}\label{def.B.con}
    \mathbf{B}_{\mid q}^0 \coloneqq (\mathbf{b}_{1}^0,\ldots, \mathbf{b}_{p-1}^0, \mathbf{0}_{pq\times 1})\quad\text{and}\quad \mathbf{B}_{\mid q}^1 \coloneqq (\mathbf{b}_{1}^0,\ldots, \mathbf{b}_{p-1}^1, \mathbf{0}_{pq\times 1}).
\end{eqnarray}
Thus, in what follows, we consider $Q_{\tilde{\gamma}_\ell\mid q}(\mathbf{B}_{\mid q}^0, \mathbf{B}_{\mid q}^1)$ under the condition \eqref{def.B.con}. Moreover, it is easy to show that we only need to restrict our investigation on the set $\{(\mathbf{B}_{\mid q}^0, \mathbf{B}_{\mid q}^1)\mid \|(\mathbf{B}_{\mid q}^0, \mathbf{B}_{\mid q}^1) \|\le \bar{c}<\infty\}$, in which $\bar{c}$ is a sufficiently large constant.

As in the proof of Theorem \ref{TM.1}, we define for $\ell \in [L]$

\begin{eqnarray*}
    \Delta Q_{\tilde{\gamma}_\ell\mid q}(\mathbf{B}_{\mid q}^0, \mathbf{B}_{\mid q}^1) \coloneqq Q_{\tilde{\gamma}_\ell\mid q}(\mathbf{B}_{\mid q}^0, \mathbf{B}_{\mid q}^1) - Q_{\mathbf{e}} +Q_{1,\tilde{\gamma}_\ell\mid q}(\mathbf{B}_{\mid q}^0, \mathbf{B}_{\mid q}^1),
\end{eqnarray*}
where $Q_{1, \tilde{\gamma}_\ell\mid q}(\mathbf{B}_{\mid q}^0, \mathbf{B}_{\mid q}^1)$ is the counterpart of $Q_{1}(\mathbf{B}^0, \mathbf{B}^1)$ defined in Theorem \ref{TM.1}. By the development for \eqref{def.rate.deltaQ}, it is easy to know that for all $\ell\in [L]$

\begin{eqnarray*}
    &&\Delta Q_{\tilde{\gamma}_\ell\mid q}(\mathbf{B}_{\mid q}^0, \mathbf{B}_{\mid q}^1) -E_{\mathbf{e}}[\Delta Q_{\tilde{\gamma}_\ell\mid q}(\mathbf{B}_{\mid q}^0, \mathbf{B}_{\mid q}^1)] =O_P\left(\frac{1}{\sqrt{T\delta}}\right),\notag \\
    &&Q_{1, \tilde{\gamma}_\ell\mid q}(\mathbf{B}_{\mid q}^0, \mathbf{B}_{\mid q}^1) = O_P\left(\frac{1}{\sqrt{T\delta}}\right)
\end{eqnarray*}
provided $\|(\mathbf{B}_{\mid q}^0, \mathbf{B}_{\mid q}^1) \|\le \bar{c}$. 

We now pay attention to $E_{\mathbf{e}}[\Delta Q_{\tilde{\gamma}_\ell\mid q}(\mathbf{B}_{\mid q}^0, \mathbf{B}_{\mid q}^1)]$. In view of the proofs of Theorem \ref{TM.1} and Lemma \ref{Appendix.B3} and the condition in the body of this lemma, we know that under \eqref{def.B.con} there is at least one $\ell \in [L]$ such that

\begin{eqnarray*}
    &&E_{\mathbf{e}}[\Delta Q_{\tilde{\gamma}_\ell\mid q}(\mathbf{B}_{\mid q}^0, \mathbf{B}_{\mid q}^1)] \gtrsim \|\mathbf{A}_{q}(\gamma_\ell) \|^2>0.
\end{eqnarray*}

Putting everything together, it immediately gives that with probability approaching 1

\begin{eqnarray*}
    \frac{1}{L}\sum_{\ell=1}^L Q_{\tilde{\gamma}_\ell\mid \tilde{q}}(\mathbf{B}_{ \mid \tilde{q}}^0, \mathbf{B}_{ \mid \tilde{q}}^1)> \frac{1}{L}\sum_{\ell=1}^L Q_{\tilde{\gamma}_\ell\mid q}(\pmb{\beta}_{\tilde{\gamma}_\ell}^0, \pmb{\beta}_{\tilde{\gamma}_\ell}^1).
\end{eqnarray*}
Thus, it rules out Case 1 with probability approaching 1.

\medskip
 
Case 2 --- Suppose that we let $\tilde{q}=q+1$ without loss of generality. In this case, by the proof of Theorem \ref{TM.1} and the statement of Lemma \ref{Appendix.A1}, it is easy to know that for all $\ell \in [L]$

\begin{eqnarray*}
    Q_{\tilde{\gamma}_\ell\mid \tilde{q}}(\widehat{\mathbf{B}}_{\mid \tilde{q}}^0, \widehat{\mathbf{B}}_{\mid \tilde{q}}^1)  -Q_{\mathbf{e}} =O_P\left(\frac{1}{T\delta}\right)\quad\text{and}\quad Q_{\tilde{\gamma}_\ell\mid q}(\pmb{\beta}_{\tilde{\gamma}_\ell}^0, \pmb{\beta}_{\tilde{\gamma}_\ell}^1) -Q_{\mathbf{e}} =O_P\left(\frac{1}{T\delta}\right).
\end{eqnarray*}
However, by the condition in the body of this theorem,

\begin{eqnarray*}
    \omega_T\tilde{q} - \omega_Tq = \omega_T\gg \frac{1}{T\delta}.
\end{eqnarray*}
It again gives that with probability approaching 1

\begin{eqnarray*}
    \frac{1}{L}\sum_{\ell=1}^L Q_{\tilde{\gamma}_\ell\mid \tilde{q}}(\mathbf{B}_{ \mid \tilde{q}}^0, \mathbf{B}_{ \mid \tilde{q}}^1)> \frac{1}{L}\sum_{\ell=1}^L Q_{\tilde{\gamma}_\ell\mid q}(\pmb{\beta}_{\tilde{\gamma}_\ell}^0, \pmb{\beta}_{\tilde{\gamma}_\ell}^1).
\end{eqnarray*}
Thus, it rules out Case 2 with probability approaching 1.

Based on the development of Cases 1 and 2, the proof is now completed.
\end{proof}

\begin{proof}[Proof of Theorem \ref{TM.4}]
\item 

For brevity, we use $\mathbf{D}\coloneqq\mathbf{D}^{(h)}_{ij}(\gamma)$, and $K_t(\gamma)\coloneqq K\bigl(\tfrac{\gamma_t-\gamma}{\delta}\bigr)$ when no misunderstanding arises. We also suppress $(i,j,h,\tau,\gamma)$ on the estimated coefficients whenever no confusion can arise.  All other QIR proof notation is the same as those defined before Lemma \ref{Appendix.B6}. Let $C$ and $C'$ denote generic finite constants and $\varrho\in(0,1)$ be a generic geometric rate. They may take different values at each occurrence.

Recall that $\pmb\alpha(\gamma)$ minimizes the  population objective \eqref{def.alpha.proj} and   $\tilde u^{(h)}_{i,t}(\gamma)= \tilde y_{i,t+h}(\gamma) -\tilde{\mathbf x}_t^{[j]}(\gamma)^\top \pmb\alpha(\gamma).$  With $\psi_\tau(u)=\tau-\mathbf 1\{u\le0\}$, the subgradient first-order condition gives
\begin{eqnarray}\label{ap.tm5c.foc}
 E\!\left[
 \psi_\tau\!\left(
 \tilde u^{(h)}_{i,t}(\gamma)
 \right)
 \tilde{\mathbf x}_t^{[j]}(\gamma)
 \right]
 &=&\mathbf0.
\end{eqnarray}
Additionally, by Lemma \ref{Appendix.B6}, we obtain
\begin{eqnarray}\label{ap.tm5c.foc_2}
 \max_{\substack{1\le t\le T-h:\\
 |\gamma_t-\gamma|\le\delta}}
 \left\|
 E\!\left[
 \psi_\tau(u^{(h)}_{i,t})
 \mathbf x_t^{[j]}
 \right]
 \right\|
 &=&O\!\left(T^{-(\nu-1)/\nu}\right),
\end{eqnarray}
In fact, by writing $g^{(h)}_{i,t} =E[\psi_\tau(u^{(h)}_{i,t})\mid\mathbf x_t^{[j]}]$ and $\zeta_t=\psi_\tau(u^{(h)}_{i,t})-g^{(h)}_{i,t}$, \eqref{ap.tm5c.foc_2} yields $E[\zeta_t\mathbf x_t^{[j]}]=\mathbf0$ and $E[g^{(h)}_{i,t}\mathbf x_t^{[j]}] =O\!\left(T^{-(\nu-1)/\nu}\right)$.

Define the QIR expansion terms

\begin{eqnarray*}
&&Q^{(h)}_{u,ij}\coloneqq\frac{1}{T-h}\!\sum_{t=1}^{T-h}\rho_{\tau,m}(u^{(h)}_{i,t}) K_\delta(\gamma_t-\gamma),\quad
Q^{(h)}_{1,ij}(\mathbf{a}^0, \mathbf{a}^1)\coloneqq-\frac{1}{T-h}\!\sum_{t=1}^{T-h}\rho^{(1)}_{\tau,m}(u^{(h)}_{i,t})\Delta\bar y_t  K_\delta(\gamma_t-\gamma),
\nonumber\\
&&Q^{(h)}_{2,ij}(\mathbf{a}^0, \mathbf{a}^1)\coloneqq\frac12\!\int_{-1}^{1}(\Delta\mathbf{a}_\delta-w^2\pmb\xi^{(h)}_{ij,\gamma,\delta})^\top\diag\{\mathbf{D},w^2\mathbf{D}\}(\Delta\mathbf{a}_\delta-w^2\pmb\xi^{(h)}_{ij,\gamma,\delta})K(w)\mathrm{d}w.
\end{eqnarray*}

We consider the expansion of $ Q^{(h)}_{ij}(\mathbf{a}^0, \mathbf{a}^1)$, where $\mathbf{a}^0$ and $\mathbf{a}^1$ satisfy $\|\Delta\mathbf{a}_\delta\|=o(1)$. Since the fundamental theorem of calculus gives $\rho_{\tau,m}(u+\Delta\bar y_t)-\rho_{\tau,m}(u)-\rho^{(1)}_{\tau,m}(u)\Delta\bar y_t=\int_0^{\Delta\bar y_t}\{\rho^{(1)}_{\tau,m}(u+w)-\rho^{(1)}_{\tau,m}(u)\}\mathrm{d}w$,

\begin{eqnarray*}
    Q^{(h)}_{ij}(\mathbf{a}^0, \mathbf{a}^1)
    &=& \frac{1}{T-h}\sum_{t=1}^{T-h}
      \rho_{\tau,m}\!\left(
        u^{(h)}_{i,t} + \Delta\bar y_t
      \right)
      K_\delta(\gamma_t-\gamma)
      \nonumber\\
      &=&Q^{(h)}_{u,ij}-Q^{(h)}_{1,ij}(\mathbf{a}^0, \mathbf{a}^1)
      +\frac{1}{T-h}\sum_{t=1}^{T-h}\int_0^{\Delta\bar y_t}\{\rho^{(1)}_{\tau,m}( u^{(h)}_{i,t} +w)-\rho^{(1)}_{\tau,m}( u^{(h)}_{i,t} )\}\mathrm{d}w  K_\delta(\gamma_t-\gamma).
\end{eqnarray*}

Let $\Delta Q^{(h)}_{ij}(\mathbf{a}^0, \mathbf{a}^1) \coloneqq Q^{(h)}_{ij}(\mathbf{a}^0, \mathbf{a}^1)-Q^{(h)}_{u,ij} +Q^{(h)}_{1,ij}(\mathbf{a}^0, \mathbf{a}^1)$. The argument of Lemma~\ref{ap.tm5c.lemD}, together with the fact $\sum_{t=1}^{T-h} K^2_\delta(\gamma_t-\gamma)=O(T/\delta)$, gives

\begin{eqnarray*}
    &&{\rm Var}(\Delta Q^{(h)}_{ij}(\mathbf{a}^0, \mathbf{a}^1)-E[\Delta Q^{(h)}_{ij}(\mathbf{a}^0, \mathbf{a}^1)])={\rm Var}\left(\frac{1}{T-h}\sum_{t=1}^{T-h}K_\delta(\gamma_t-\gamma)\vartheta_t\right)
    \nonumber\\
    &\leq&
    Cr_{\delta,a}^3\Lambda_{\delta,a}^2
    \frac{1}{(T-h)^2}\sum_{t=1}^{T-h}
    K^2_\delta(\gamma_t-\gamma)
    = 
    O\left(
    \frac{r_{\delta,a}^3\Lambda_{\delta,a}^2}
    {T\delta}
    \right),
\end{eqnarray*}
By Chebyshev inequality and Young's inequality,  it implies 
\begin{eqnarray}\label{ap.tm5c.fluct}
 \Delta Q^{(h)}_{ij}(\mathbf{a}^0, \mathbf{a}^1)
 -E[\Delta Q^{(h)}_{ij}(\mathbf{a}^0, \mathbf{a}^1)]
 &=&
 O_P\!\left(
 \frac{r_{\delta,a}^{3/2}\Lambda_{\delta,a}}
 {\sqrt{T\delta}}
 \right)
 =o_P\!\left(
 \|\Delta\mathbf{a}_\delta\|^2+\delta^4+\frac1{T\delta}
 \right),
\end{eqnarray}
for $\mathbf{a}^0$ and $\mathbf{a}^1$ that satisfy $\|\Delta\mathbf{a}_\delta\|=o(1)$.

For $E[\Delta Q^{(h)}_{ij}(\mathbf{a}^0, \mathbf{a}^1)]$, iterated expectations and Lemma~\ref{Appendix.B9} give
\begin{eqnarray*}
    E[\Delta Q^{(h)}_{ij}(\mathbf{a}^0, \mathbf{a}^1)]
    &=&
    \frac{1}{T-h}\sum_{t=1}^{T-h}E[G_t]
    K_\delta(\gamma_t-\gamma)
    \nonumber\\
    &=&
    \frac{1}{2(T-h)}\sum_{t=1}^{T-h}
    E[f_t^*\Delta\bar y_t^2]
    K_\delta(\gamma_t-\gamma)\nonumber\\
    &&+
    O\!\left[
    \frac{1}{T-h}\sum_{t=1}^{T-h}
    E\!\left(
    |\Delta\bar y_t|^3+\Delta\bar y_t^2/m
    \right)
    K_\delta(\gamma_t-\gamma)
    \right]
    \nonumber\\
    &=&
    \frac{1}{2(T-h)}\sum_{t=1}^{T-h}
    E[f_t^*\Delta\bar y_t^2]
    K_\delta(\gamma_t-\gamma)
    +o(\|\Delta\mathbf{a}_\delta\|^2+\delta^4).
\end{eqnarray*}

For its leading-order term, define $\mathbf D_{t,T}^*\coloneqq E [ f_t^*\mathbf x_t^{[j]}\mathbf x_t^{[j]\top} ].$ Assumption \ref{Assumption5}, H\"older's inequality, Lemma \ref{Appendix.B4}.(b), and the boundedness and Lipschitz continuity of $\tilde g^{(h)}_{ij}(0\mid\mathbf x;\gamma)$ give
\begin{eqnarray*}
 &&\max_{\substack{1\le t\le T-h:\\
 |\gamma_t-\gamma|\le\delta}}
 \left\|
 \mathbf D_{t,T}^*
 -E\!\left[
 \tilde g^{(h)}_{ij}\!\left(
 0\mid\tilde{\mathbf x}_t^{[j]}(\gamma_t);\gamma_t
 \right)
 \tilde{\mathbf x}_t^{[j]}(\gamma_t)
 \tilde{\mathbf x}_t^{[j]}(\gamma_t)^\top
 \right]
 \right\|
 =o(1).
\end{eqnarray*}
Inserting the Taylor expansion of $\Delta\bar y_t$ gives
\begin{eqnarray}\label{tm5_ex2}
   &&\frac{1}{2(T-h)}\sum_{t=1}^{T-h}
   E[f^*_t\Delta\bar y_t^2] K_\delta(\gamma_t-\gamma)
   \nonumber\\
   &=&\frac{1}{2(T-h)}\sum_{t=1}^{T-h} K_\delta(\gamma_t-\gamma)(\pmb{\alpha}(\gamma)- \mathbf{a}^0) ^\top\mathbf D_{t,T}^*(\pmb{\alpha}(\gamma)- \mathbf{a}^0)
   \nonumber\\
   &&+\frac{1}{(T-h)}\sum_{t=1}^{T-h} K_\delta(\gamma_t-\gamma)(\pmb{\alpha}(\gamma)- \mathbf{a}^0) ^\top\mathbf D_{t,T}^*(\pmb{\alpha}^{(1)}(\gamma)- \mathbf{a}^1)(\gamma_t-\gamma)
   \nonumber\\
   &&+\frac{1}{2(T-h)}\sum_{t=1}^{T-h} K_\delta(\gamma_t-\gamma)(\pmb{\alpha}(\gamma)- \mathbf{a}^0) ^\top\mathbf D_{t,T}^*\pmb{\alpha}^{(2)}(\tilde{\gamma}_t)(\gamma_t-\gamma)^2
   \nonumber\\
 &&+\frac{1}{2(T-h)}\sum_{t=1}^{T-h}K_\delta(\gamma_t-\gamma)(\pmb{\alpha}^{(1)}(\gamma)- \mathbf{a}^1) ^\top\mathbf D_{t,T}^*(\pmb{\alpha}^{(1)}(\gamma)- \mathbf{a}^1)(\gamma_t-\gamma)^2
 \nonumber\\
 &&+\frac{1}{2(T-h)}\sum_{t=1}^{T-h} K_\delta(\gamma_t-\gamma)(\pmb{\alpha}^{(1)}(\gamma)- \mathbf{a}^1) ^\top\mathbf D_{t,T}^*\pmb{\alpha}^{(2)}(\tilde{\gamma}_t)(\gamma_t-\gamma)^3
 \nonumber\\
 &&+\frac{1}{8(T-h)}\sum_{t=1}^{T-h} K_\delta(\gamma_t-\gamma)\pmb{\alpha}^{(2)}(\tilde{\gamma}_t)^\top\mathbf D_{t,T}^*\pmb{\alpha}^{(2)}(\tilde{\gamma}_t)(\gamma_t-\gamma)^4.
\end{eqnarray}

Because $\pmb\alpha^{(h)}_{ij}(\cdot)$ is twice continuously differentiable, $\max_{\substack{1\le t\le T-h:\\ |\gamma_t-\gamma|\le\delta}} \left\| \pmb\alpha^{(2)}(\tilde\gamma_t) -\pmb\alpha^{(2)}(\gamma)\right\|=o(1)$. The Lipschitz property of $\tilde g^{(h)}_{ij}(0\mid\mathbf x;\gamma)$, the rescaled-time Lipschitz bound in Lemma \ref{Appendix.B4}.(c), and H\"older's inequality further imply
\begin{eqnarray*}
 \max_{\substack{1\le t\le T-h:\\
 |\gamma_t-\gamma|\le\delta}}
 \|\mathbf D_{t,T}^*-\mathbf D\|
 &=&o(1)+O(\delta)=o(1).
\end{eqnarray*}
Consequently, the kernel Riemann-sum calculation gives, for $n\in\{0,1,2,3,4\}$,
\begin{eqnarray}\label{ap.tm5c.curvature.kernel}
 &&\frac1{T-h}\sum_{t=1}^{T-h}
 \mathbf D_{t,T}^*
 \left(\frac{\gamma_t-\gamma}{\delta}\right)^n
 K_\delta(\gamma_t-\gamma)
 =v_n\mathbf D+o(1).
\end{eqnarray}
Applying \eqref{ap.tm5c.curvature.kernel} to each term on the right-hand side of \eqref{tm5_ex2}, we obtain
\begin{eqnarray}\label{tm5_ex3}
    \frac{1}{2(T-h)}\sum_{t=1}^{T-h}
    E[f^*_t\Delta\bar y_t^2]K_\delta(\gamma_t-\gamma)
    &=&Q^{(h)}_{2,ij}(\mathbf{a}^0, \mathbf{a}^1)
    +o(\|\Delta\mathbf{a}_\delta\|^2+\delta^4).
\end{eqnarray}

Combining \eqref{ap.tm5c.fluct}-\eqref{tm5_ex3} proves
\begin{eqnarray*}
    Q^{(h)}_{ij}(\mathbf{a}^0, \mathbf{a}^1)-Q^{(h)}_{u,ij}
    =-Q^{(h)}_{1,ij}(\mathbf{a}^0, \mathbf{a}^1)+Q^{(h)}_{2,ij}(\mathbf{a}^0, \mathbf{a}^1)
    +o_P\bigl(\|\Delta\mathbf{a}_\delta\|^2+\delta^4+\tfrac{1}{T\delta}\bigr).
\end{eqnarray*}

For notational simplicity, let $s_T\coloneqq\delta^2+(T\delta)^{-1/2}$. For a fixed vector $\mathbf b$, let $(\mathbf a_T^0(\mathbf b),\mathbf a_T^1(\mathbf b))$ satisfy $\Delta\mathbf a_\delta=s_T\mathbf b$, and  define
\begin{eqnarray*}
 \mathcal G_T(\mathbf b)
 &\coloneqq&
 s_T^{-2}\left\{
 Q^{(h)}_{ij}(\mathbf a_T^0(\mathbf b),\mathbf a_T^1(\mathbf b))-Q^{(h)}_{u,ij}
 +Q^{(h)}_{1,ij}(\mathbf a_T^0(\mathbf b),\mathbf a_T^1(\mathbf b))
 \right\}.
\end{eqnarray*}
Because $ Q^{(h)}_{ij}$ is convex and $Q^{(h)}_{1,ij}$ is affine, $\mathcal G_T(\mathbf b)$ is convex in $\mathbf b$.  Equations \eqref{ap.tm5c.fluct}--\eqref{tm5_ex3} give, for each fixed $\mathbf b$,
\begin{eqnarray*}
 \mathcal G_T(\mathbf b)
 -s_T^{-2}Q^{(h)}_{2,ij}(\mathbf a_T^0(\mathbf b),
 \mathbf a_T^1(\mathbf b))
 &=&o_P(1).
\end{eqnarray*}
The second term is a continuous convex quadratic function of $\mathbf b$, uniformly on every compact set.  Therefore, we are ready to establish
\begin{eqnarray}\label{ap.tm5c.bah.uniform}
 &&\sup_{\|\Delta\mathbf a_\delta\|\le Ms_T}
 \left| Q^{(h)}_{ij}(\mathbf a^0,\mathbf a^1)-Q^{(h)}_{u,ij}
 +Q^{(h)}_{1,ij}(\mathbf a^0,\mathbf a^1)
 -Q^{(h)}_{2,ij}(\mathbf a^0,\mathbf a^1)
 \right|
 =o_P(s_T^2),
\end{eqnarray}
for every fixed $M<\infty$.  

Let $\mathbf{S}^{(h)}_{ij,\gamma}\coloneqq\tfrac{1}{T-h}\sum_t\rho^{(1)}_{\tau,m}(u^{(h)}_{i,t})\VEC(\mathbf x_t^{[j]},\delta^{-1}\mathbf z_t^{[j]})K_\delta(\gamma_t-\gamma)$. By Lemma~\ref{ap.tm5c.lemL}, we have $\|\mathbf{S}^{(h)}_{ij,\gamma}\|=O_P((T\delta)^{-1/2})$. On the sphere $\|\Delta\mathbf{a}_\delta\|=M(T\delta)^{-1/2}$, \eqref{ap.tm5c.bah.uniform} gives, 
\begin{eqnarray*}
    Q^{(h)}_{ij}(\mathbf{a}^0, \mathbf{a}^1)-Q^{(h)}_{u,ij}&\ge&-\|\mathbf{S}^{(h)}_{ij,\gamma}\|\|\Delta\mathbf{a}_\delta\|+\tfrac12\lambda_{\min}(\mathbf{D})\|\Delta\mathbf{a}_\delta\|^2-C\delta^2\|\Delta\mathbf{a}_\delta\|
    \nonumber\\
    &&+o_P\bigl(\|\Delta\mathbf{a}_\delta\|^2+\delta^4+\tfrac{1}{T\delta}\bigr).
\end{eqnarray*}
Hence, under the condition $\delta^2\lesssim(T\delta)^{-1/2}$  required by Assumption~\ref{Assumption3}, for a sufficiently large $M$,
\[
\inf_{\|\Delta\mathbf{a}_{\delta}\|=M(T\delta)^{-1/2}}
\left\{
 Q^{(h)}_{ij}(\mathbf{a}^0, \mathbf{a}^1)-Q^{(h)}_{u,ij}
\right\}
>0
\]
with probability tending to one. Since $ Q^{(h)}_{ij}$ is convex as discussed in Lemma~\ref{LM1}, its minimizer must lie inside the ball $\{\|\Delta\mathbf{a}_{\delta}\|\le M(T\delta)^{-1/2}\}$. Therefore,
\begin{eqnarray}\label{ap.tm5c.rate}
\left\|
\VEC\!\left(
\widehat{\mathbf{a}}^0-\pmb{\alpha}(\gamma),
\delta\bigl(
\widehat{\mathbf{a}}^{1}
-\pmb{\alpha}^{(1)}(\gamma)
\bigr)
\right)
\right\|
=
O_P\!\left((T\delta)^{-1/2}\right).
\end{eqnarray}

We now introduce $\tilde{\pmb{\alpha}}=(\tilde{\pmb{\alpha}}^0,\tilde{\pmb{\alpha}}^1)$ that minimizes $-Q^{(h)}_{1,ij}(\mathbf{a}^0, \mathbf{a}^1)+Q^{(h)}_{2,ij}(\mathbf{a}^0, \mathbf{a}^1)$. In what follows, we show that $\widehat{\mathbf{a}}^0$ and $\tilde{\pmb{\alpha}}^0$ differ by $o_P((T\delta)^{-1/2})$ so that the two share the same limit distribution.

It is straightforward to see that $Q^{(h)}_{1,ij}(\mathbf{a}^0, \mathbf{a}^1)$ is affine in $\Delta\mathbf{a}_\delta$ with gradient $\mathbf{S}^{(h)}_{ij,\gamma}$ and $Q^{(h)}_{2,ij}(\mathbf{a}^0, \mathbf{a}^1)$ is quadratic. Hence

\begin{eqnarray*}
    \frac{\partial}{\partial\Delta\mathbf{a}_\delta}\left\{-Q^{(h)}_{1,ij}(\mathbf{a}^0, \mathbf{a}^1)+Q^{(h)}_{2,ij}(\mathbf{a}^0, \mathbf{a}^1)\right\}
    =-\mathbf{S}^{(h)}_{ij,\gamma}+\int_{-1}^{1}\diag\{\mathbf{D},w^2\mathbf{D}\}\left(\Delta\mathbf{a}_\delta-w^2\pmb\xi^{(h)}_{ij,\gamma,\delta}\right)K(w)\mathrm{d}w ,
\end{eqnarray*}
so that the minimizer $\tilde{\pmb{\alpha}}$, written through $\Delta\tilde{\pmb{\alpha}}_\delta\coloneqq\VEC(\tilde{\pmb{\alpha}}^0-\pmb{\alpha}(\gamma),\delta(\tilde{\pmb{\alpha}}^1-\pmb{\alpha}^{(1)}(\gamma)))$, is characterized by the first-order condition

\begin{eqnarray}\label{ap.tm5c.bridgefoc}
    \mathbf{S}^{(h)}_{ij,\gamma}&=&\int_{-1}^1\diag\{\mathbf{D},w^2\mathbf{D}\}\left(\Delta\tilde{\pmb{\alpha}}_\delta-w^2\pmb\xi^{(h)}_{ij,\gamma,\delta}\right)K(w)\mathrm{d}w 
    \nonumber\\
    &=&\diag\{\mathbf{D},v_2\mathbf{D}\}\left(\Delta\tilde{\pmb{\alpha}}_\delta -v_2\pmb\xi^{(h)}_{ij,\gamma,\delta} \right),
\end{eqnarray}
where the second equality holds because the terms associated with the fourth kernel moment vanishes after they are multiplied by the zero elements in $\pmb\xi^{(h)}_{ij,\gamma,\delta}$. 

Solving for $\VEC(\tilde{\pmb{\alpha}}^0,\delta\tilde{\pmb{\alpha}}^1)$ from \eqref{ap.tm5c.bridgefoc}, we obtain  
\begin{eqnarray}\label{ap.tm5c.bridge}
    \VEC(\tilde{\pmb{\alpha}}^0,\delta\tilde{\pmb{\alpha}}^1)=\VEC(\pmb{\alpha}(\gamma),\delta\pmb{\alpha}^{(1)}(\gamma))+v_2\pmb\xi^{(h)}_{ij,\gamma,\delta}+\diag\{\mathbf{D},v_2\mathbf{D}\}^{-1}\mathbf{S}^{(h)}_{ij,\gamma}.
\end{eqnarray}
For the second term on the right-hand side of \eqref{ap.tm5c.bridge}, $\|v_2\pmb\xi^{(h)}_{ij,\gamma,\delta}\|\le C\delta^2$. For the last term,
\begin{eqnarray*}
    \|\diag\{\mathbf{D},v_2\mathbf{D}\}^{-1}\mathbf{S}^{(h)}_{ij,\gamma}\|&\le&\|\diag\{\mathbf{D},v_2\mathbf{D}\}^{-1}\|_2\,\|\mathbf{S}^{(h)}_{ij,\gamma}\|
    \nonumber\\
    &\leq&\bigl(\min\{1,v_2\}\,\lambda_{\min}(\mathbf{D})\bigr)^{-1}\|\mathbf{S}^{(h)}_{ij,\gamma}\|
    \nonumber\\
    &=&O_P((T\delta)^{-1/2}).
\end{eqnarray*}
In summary, we obtain 
\begin{eqnarray}\label{ap.tm5c.bridgerate}
    \|\VEC(\tilde{\pmb{\alpha}}^0,\delta\tilde{\pmb{\alpha}}^1)-\VEC(\pmb{\alpha}(\gamma),\delta\pmb{\alpha}^{(1)}(\gamma))\|=O_P\left((T\delta)^{-1/2}\right).
\end{eqnarray}

To proceed, we further simplify the notations. Let $\Delta\widehat{\mathbf{a}}_\delta=\VEC(\widehat{\mathbf{a}}^0-\pmb{\alpha}(\gamma),\delta(\widehat{\mathbf{a}}^1-\pmb{\alpha}^{(1)}(\gamma)))$, $\mathcal R^{(h)}_{ij}(\mathbf{a}^0, \mathbf{a}^1)\coloneqq Q^{(h)}_{ij}(\mathbf{a}^0, \mathbf{a}^1)-Q^{(h)}_{u,ij} +Q^{(h)}_{1,ij}(\mathbf{a}^0, \mathbf{a}^1)-Q^{(h)}_{2,ij}(\mathbf{a}^0, \mathbf{a}^1)$, and $\mathbf{d}\coloneqq\Delta\widehat{\mathbf{a}}_\delta-\Delta\tilde{\pmb{\alpha}}_\delta$. We now evaluate $\mathcal R^{(h)}_{ij}(\widehat{\mathbf{a}}^0, \widehat{\mathbf{a}}^{1})-\mathcal R^{(h)}_{ij}(\tilde{\pmb{\alpha}}^0, \tilde{\pmb{\alpha}}^1)$. Specifically, it is clear to see that
\begin{eqnarray*}
    \mathcal R^{(h)}_{ij}(\widehat{\mathbf{a}}^0, \widehat{\mathbf{a}}^{1})-\mathcal R^{(h)}_{ij}(\tilde{\pmb{\alpha}}^0, \tilde{\pmb{\alpha}}^1)&=& Q^{(h)}_{ij}(\widehat{\mathbf{a}}^0, \widehat{\mathbf{a}}^{1})- Q^{(h)}_{ij}(\tilde{\pmb{\alpha}}^0, \tilde{\pmb{\alpha}}^1)+Q^{(h)}_{1,ij}(\widehat{\mathbf{a}}^0, \widehat{\mathbf{a}}^{1})-Q^{(h)}_{1,ij}(\tilde{\pmb{\alpha}}^0, \tilde{\pmb{\alpha}}^1)
    \nonumber\\
    &&-(Q^{(h)}_{2,ij}(\widehat{\mathbf{a}}^0, \widehat{\mathbf{a}}^{1})-Q^{(h)}_{2,ij}(\tilde{\pmb{\alpha}}^0, \tilde{\pmb{\alpha}}^1))
\end{eqnarray*}
For $Q^{(h)}_{1,ij}(\widehat{\mathbf{a}}^0, \widehat{\mathbf{a}}^{1})-Q^{(h)}_{1,ij}(\tilde{\pmb{\alpha}}^0, \tilde{\pmb{\alpha}}^1)$, we can easily obtain
\begin{eqnarray}\label{ap.tm5c.q1diff}
    Q^{(h)}_{1,ij}(\widehat{\mathbf{a}}^0, \widehat{\mathbf{a}}^{1})-Q^{(h)}_{1,ij}(\tilde{\pmb{\alpha}}^0, \tilde{\pmb{\alpha}}^1)
    &=&\frac{1}{T-h}\sum_{t=1}^{T-h}\rho^{(1)}_{\tau,m}(u^{(h)}_{i,t})\VEC\left(\mathbf x_t^{[j]},\delta^{-1}\mathbf z_t^{[j]}\right)^\top\mathbf{d}\;K_t(\gamma)
    \nonumber\\
    &=&\left(\mathbf{S}^{(h)}_{ij,\gamma}\right)^\top\mathbf{d}.
\end{eqnarray}
For $Q^{(h)}_{2,ij}(\widehat{\mathbf{a}}^0, \widehat{\mathbf{a}}^{1})$, it follows \eqref{ap.tm5c.bridgefoc} that 
\begin{eqnarray}\label{ap.tm5c.q2diff}
    Q^{(h)}_{2,ij}(\widehat{\mathbf{a}}^0, \widehat{\mathbf{a}}^1)&=&\tfrac12\!\int_{-1}^{1}(\mathbf{d}+\Delta\tilde{\pmb{\alpha}}_\delta-w^2\pmb\xi^{(h)}_{ij,\gamma,\delta})^\top\diag\{\mathbf{D},w^2\mathbf{D}\}(\mathbf{d}+\Delta\tilde{\pmb{\alpha}}_\delta-w^2\pmb\xi^{(h)}_{ij,\gamma,\delta})K(w)\mathrm{d}w
    \nonumber\\
    &=&Q^{(h)}_{2,ij}(\tilde{\pmb{\alpha}}^0, \tilde{\pmb{\alpha}}^1)+\left(\mathbf{S}^{(h)}_{ij,\gamma}\right)^\top\mathbf{d}+\tfrac12\!\int_{-1}^{1}\mathbf{d}^\top\diag\{\mathbf{D},w^2\mathbf{D}\}\mathbf{d}K(w)\mathrm{d}w
\end{eqnarray}
By \eqref{ap.tm5c.q1diff} and \eqref{ap.tm5c.q2diff}, we have

\begin{eqnarray}\label{ap.tm5c.eq1}
    &&\mathcal R^{(h)}_{ij}(\widehat{\mathbf{a}}^0, \widehat{\mathbf{a}}^{1})-\mathcal R^{(h)}_{ij}(\tilde{\pmb{\alpha}}^0, \tilde{\pmb{\alpha}}^1) \notag \\
    &=& Q^{(h)}_{ij}(\widehat{\mathbf{a}}^0, \widehat{\mathbf{a}}^{1})- Q^{(h)}_{ij}(\tilde{\pmb{\alpha}}^0, \tilde{\pmb{\alpha}}^1)-\tfrac12\!\int_{-1}^{1}\mathbf{d}^\top\diag\{\mathbf{D},w^2\mathbf{D}\}\mathbf{d}K(w)\mathrm{d}w.
\end{eqnarray}

We further investigate the quadratic term. Additionally, by \eqref{ap.tm5c.rate} and \eqref{ap.tm5c.bridgerate},  both $\widehat{\mathbf{a}}$  and $\tilde{\pmb{\alpha}}$ lie in the ball $\|\Delta\mathbf{a}_\delta\|=O_P((T\delta)^{-1/2})$, which together with \eqref{ap.tm5c.bah.uniform} implies that 
\begin{eqnarray}\label{ap.tm5c.eq2}
     \left|\mathcal R^{(h)}_{ij}(\widehat{\mathbf{a}})-\mathcal R^{(h)}_{ij}(\tilde{\pmb{\alpha}})\right|\le\left|\mathcal R^{(h)}_{ij}(\widehat{\mathbf{a}})\right|+\left|\mathcal R^{(h)}_{ij}(\tilde{\pmb{\alpha}})\right|=o_P\left(\tfrac{1}{T\delta}\right).
\end{eqnarray}
Since $ Q^{(h)}_{ij}(\widehat{\mathbf{a}}^0, \widehat{\mathbf{a}}^{1})- Q^{(h)}_{ij}(\tilde{\pmb{\alpha}}^0, \tilde{\pmb{\alpha}}^1)\le0$, in light of \eqref{ap.tm5c.eq1} and \eqref{ap.tm5c.eq2}, we can formulate the following upper bound for the quadratic term
\begin{eqnarray}\label{ap.tm5c.eq3}
    \tfrac12\!\int_{-1}^{1}\mathbf{d}^\top\diag\{\mathbf{D},w^2\mathbf{D}\}\mathbf{d}K(w)\mathrm{d}w\leq -(\mathcal R^{(h)}_{ij}(\widehat{\mathbf{a}}^0, \widehat{\mathbf{a}}^{1})-\mathcal R^{(h)}_{ij}(\tilde{\pmb{\alpha}}^0, \tilde{\pmb{\alpha}}^1))=o_P\left(\tfrac{1}{T\delta}\right).
\end{eqnarray}
For the lower bound, we have
\begin{eqnarray}\label{ap.tm5c.eq4}
    \int_{-1}^{1}\mathbf{d}^\top\diag\{\mathbf{D},w^2\mathbf{D}\}\mathbf{d}\,K(w)\mathrm{d}w
    &=&\mathbf{d}^{0,\top}\mathbf{D}\mathbf{d}^0+v_2\mathbf{d}^{1,\top}\mathbf{D}\mathbf{d}^1
    \nonumber\\
    &\ge&\lambda_{\min}(\mathbf{D})\min\{1,v_2\}\,\|\mathbf{d}\|^2.
\end{eqnarray}
In summary of the results that have been established in \eqref{ap.tm5c.eq3} and \eqref{ap.tm5c.eq4}, we are ready to obtain 
\begin{eqnarray}\label{ap.tm5c.close}
    \|\VEC(\widehat{\mathbf{a}}^{0},\delta\widehat{\mathbf{a}}^{1})-\VEC(\tilde{\pmb{\alpha}}^0,\delta\tilde{\pmb{\alpha}}^1)\|=o_P((T\delta)^{-1/2}).
\end{eqnarray}

We then replace the smoothed score in the  Bahadur expansion by the exact quantile score before establishing the CLT, and  show that the cost of replacement is asymptotically negligible. By Lemma~\ref{LM1}(2) the smoothed check function $\rho_{\tau,m}(u)$ coincides with $\rho_\tau(u)$ outside the band $\{|u|\le1/m\}$, so their derivatives automatically agree there.  Inside this band, by the construction of $\rho_{\tau,m}(u)$, we can easily show that $|\rho^{(1)}_{\tau,m}(u)-\psi_\tau(u)|\le 1$. In summary, we obtain
\begin{eqnarray}\label{ap.tm5c.desmooth.pt}
    |\rho^{(1)}_{\tau,m}(u)-\psi_\tau(u)|\le\mathbf{1}\{|u|\le1/m\} .
\end{eqnarray}

By Lemma \ref{Appendix.B5}, it is clear to see that 
\begin{eqnarray*}
 P\!\left(
 |u^{(h)}_{i,t}|\le m^{-1}
 \,\middle|\,\mathbf x_t^{[j]}
 \right)
 &\le&\frac{2\bar C_f}{m}.
\end{eqnarray*}
Multiplying by $\|\mathbf x_t^{[j]}\|$ and taking expectations gives
\begin{eqnarray}\label{ap.tm5c.desmooth.exp}
 E\!\left[
 \mathbf1\{|u^{(h)}_{i,t}|\le m^{-1}\}
 \|\mathbf x_t^{[j]}\|
 \right]
 &\le&
 \frac{2\bar C_f}{m}E\|\mathbf x_t^{[j]}\|.
\end{eqnarray}
Combining \eqref{ap.tm5c.desmooth.pt} and \eqref{ap.tm5c.desmooth.exp}, we obtain
\begin{eqnarray*}
 &&E\left\|
 \frac{1}{\sqrt{T\delta}}
 \sum_{t=1}^{T-h}
 \{\rho_{\tau,m}^{(1)}(u^{(h)}_{i,t})
 -\psi_\tau(u^{(h)}_{i,t})\}
 \mathbf x_t^{[j]}K_t(\gamma)
 \right\|\nonumber\\
 &\le&
 \frac{1}{\sqrt{T\delta}}
 \sum_{t=1}^{T-h}
 K_t(\gamma)
 E\!\left[
 \mathbf1\{|u^{(h)}_{i,t}|\le m^{-1}\}
 \|\mathbf x_t^{[j]}\|
 \right]\nonumber\\
 &\le&
 \frac{C}{m\sqrt{T\delta}}
 \sum_{t=1}^{T-h}K_t(\gamma)
 =O\!\left(\frac{\sqrt{T\delta}}{m}\right)=o(1).
\end{eqnarray*}
It follows that the normalized score representation is
\begin{eqnarray}\label{ap.tm5c.desmooth}
 \sqrt{T\delta}\,\mathbf{S}^{(h)}_{ij,\gamma}
 &=&
 \frac{1}{\sqrt{T\delta}}
 \sum_{t=1}^{T-h}
 \psi_\tau(u^{(h)}_{i,t})
 \VEC\!\left(
 \mathbf x_t^{[j]},
 \delta^{-1}\mathbf z_t^{[j]}
 \right)
 K_t(\gamma)+o_P(1).
\end{eqnarray}

We next establish the central limit theorem for the leading-order term.  Write
\begin{eqnarray*}
 \mathbf s_{t,T}
 &\coloneqq&
 \psi_\tau(u^{(h)}_{i,t})\mathbf x_t^{[j]},\qquad
 \overline{\mathbf s}_{t,T}
 \coloneqq
 \mathbf s_{t,T}-E[\mathbf s_{t,T}],\qquad
 \tilde{\mathbf s}_t(\gamma)
 \coloneqq
 \tilde{\mathbf s}^{(h)}_{ij,t}(\gamma).
\end{eqnarray*}

By Lemma \ref{Appendix.B6}, we obtain 
\begin{eqnarray*}
 \left\|
 \frac{1}{\sqrt{T\delta}}
 \sum_{t=1}^{T-h}
 E[\mathbf s_{t,T}]K_t(\gamma)
 \right\|
 &\le&\frac{1}{\sqrt{T\delta}}
 \sum_{t=1}^{T-h}K_t(\gamma)\cdot\max_{\substack{1\le t\le T-h:\\
 |\gamma_t-\gamma|\le\delta}}
 \left\| 
 E\!\left[
 \psi_\tau(u^{(h)}_{i,t})
 \mathbf x_t^{[j]}
 \right]
 \right\|
\nonumber\\
 &\le&
 \frac{CT^{-(\nu-1)/\nu}}{\sqrt{T\delta}}
 \sum_{t=1}^{T-h}K_t(\gamma)
=O\!\left(
 \sqrt{T\delta}\,T^{-(\nu-1)/\nu}
 \right)
 =o(1),
\end{eqnarray*}
where the last equality holds by the fact $\sqrt{T\delta}\,T^{-(\nu-1)/\nu}= T^{-1/2+1/\nu-\vartheta/2} \longrightarrow0$, with $\delta\asymp T^{-\vartheta}$ and $\vartheta<1-\nu^{-1}$ .

Let $a_T\coloneqq\delta+T^{-1}$.  For $|\gamma_t-\gamma|\le\delta$ and $|\gamma_{t+h}-\gamma|\le\delta+h/T$,  Lemma \ref{Appendix.B4} gives
\begin{eqnarray}\label{ap.tm5c.freeze.state}
 &&\|\mathbf x_t^{[j]}
 -\tilde{\mathbf x}_t^{[j]}(\gamma)\|_4
 \le Ca_T,\quad\|y_{i,t+h}
 -\tilde y_{i,t+h}(\gamma)\|_4
 \le Ca_T,\quad|u_{i,t}^{(h)}
 -\tilde u_{i,t}^{(h)}(\gamma)\|_4
 \le Ca_T.
\end{eqnarray}

For any $\eta>0$, Lemma \ref{Appendix.B5}, Markov's inequality, and \eqref{ap.tm5c.freeze.state} jointly imply
\begin{eqnarray*}
 P\!\left(
 \psi_\tau(u_{i,t}^{(h)})
 \ne
 \psi_\tau(\tilde u_{i,t}^{(h)}(\gamma))
 \right)
 &\le&
 P\!\left(
 |\tilde u_{i,t}^{(h)}(\gamma)|\le\eta
 \right)
 +P\!\left(
 |u_{i,t}^{(h)}-\tilde u_{i,t}^{(h)}(\gamma)|>\eta
 \right)\le
 C\eta+C\eta^{-4}a_T^4.
\end{eqnarray*}
Taking $\eta=a_T^{4/5}$, we obtain  
\begin{eqnarray}\label{ap.tm5c.freeze.indicator}
 \left\|
 \psi_\tau(u_{i,t}^{(h)})
 -\psi_\tau(\tilde u_{i,t}^{(h)}(\gamma))
 \right\|_4=P^{1/4}\!\left(
 \psi_\tau(u_{i,t}^{(h)})
 \ne
 \psi_\tau(\tilde u_{i,t}^{(h)}(\gamma))
 \right)
 &\le&Ca_T^{1/5}.
\end{eqnarray}
It then follows  \eqref{ap.tm5c.freeze.state}, \eqref{ap.tm5c.freeze.indicator} and  H\"older's inequality that 
\begin{eqnarray*}
 \|\mathbf s_{t,T}-\tilde{\mathbf s}_t(\gamma)\|_2
 &\le&\|\psi_\tau(u^{(h)}_{i,t})(\mathbf x_t^{[j]}-\tilde{\mathbf x}_t^{[j]}(\gamma))\|_2
 + \|(\psi_\tau(u_{i,t}^{(h)})-\psi_\tau(\tilde u_{i,t}^{(h)}(\gamma)))\tilde{\mathbf x}_t^{[j]}(\gamma)\|_2
 \nonumber\\
 &\le&
 \|\mathbf x_t^{[j]}
 -\tilde{\mathbf x}_t^{[j]}(\gamma)\|_2
 +\left\|
 \psi_\tau(u_{i,t}^{(h)})
 -\psi_\tau(\tilde u_{i,t}^{(h)}(\gamma))
 \right\|_4
 \|\tilde{\mathbf x}_t^{[j]}(\gamma)\|_4\le Ca_T^{1/5}.
\end{eqnarray*}

Since $E[\tilde{\mathbf s}_t(\gamma)]=\mathbf0$ by \eqref{ap.tm5c.foc},
\begin{eqnarray}\label{ap.tm5c.freeze.pointwise}
 \|\overline{\mathbf s}_{t,T}
 -\tilde{\mathbf s}_t(\gamma)\|_2
 &\le&
 2\|\mathbf s_{t,T}
 -\tilde{\mathbf s}_t(\gamma)\|_2
 \le Ca_T^{1/5}.
\end{eqnarray}

For notational simplicity, define $\pmb\Xi_{t,T}=\overline{\mathbf s}_{t,T} -\tilde{\mathbf s}_t(\gamma)$ and $\pmb\Xi_{t,T}^{*(k)}$ as the coupled version by replacing $\pmb\eta_{t+h-k}$ with its independent copy. Additionally, define
\begin{eqnarray*}
\mathfrak d_{2,\Xi,T}(k)
 &\coloneqq&
 \max_{\substack{1\le t\le T-h:\\K_t(\gamma)\ne0}}
 \|\pmb\Xi_{t,T}-\pmb\Xi_{t,T}^{*(k)}\|_2.
\end{eqnarray*}
Applying \eqref{ap.tm5c.freeze.pointwise} to the original and coupled variables gives $\mathfrak d_{2,\Xi,T}(k)\le Ca_T^{1/5}$.  Moreover, the original and coupled scores have the same marginal expectations.  Hence centering does not alter their coupling difference, and Lemma \ref{Appendix.B7} gives
\begin{eqnarray*}
 \|\pmb\Xi_{t,T}-\pmb\Xi_{t,T}^{*(k)}\|_2
 &\le&
 \|\mathbf s^{(h)}_{ij,t,T}
 -\mathbf s^{*(h,k)}_{ij,t,T}\|_2
 +\|\tilde{\mathbf s}^{(h)}_{ij,t}(\gamma)
 -\tilde{\mathbf s}^{*(h,k)}_{ij,t}(\gamma)\|_2\le
 C\varrho^{(k-h)_+}.
\end{eqnarray*}
Consequently, $\mathfrak d_{2,\Xi,T}(k)\le C\varrho^{(k-h)_+}$. In summary, we are ready to establish the following bound:

\begin{eqnarray}\label{ap.tm5c.freeze.fd.sum}
 \sum_{k=0}^{\infty}\mathfrak d_{2,\Xi,T}(k)
 &\le&
 C\sum_{k=0}^{\infty}
 \min\!\left\{
 a_T^{1/5},\varrho^{(k-h)_+}
 \right\}\le
 Ca_T^{1/5}\{1+|\log a_T|\}.
\end{eqnarray}

Lemma \ref{ap.tm5c.lemD}, \eqref{ap.tm5c.freeze.fd.sum}, and the fact $(T\delta)^{-1}\sum_tK_t^2(\gamma)=O(1)$ therefore yield
\begin{eqnarray*}
 &&\left\|
 \frac{1}{\sqrt{T\delta}}
 \sum_{t=1}^{T-h}
 K_t(\gamma)
 \{\overline{\mathbf s}_{t,T}
 -\tilde{\mathbf s}_t(\gamma)\}
 \right\|_2
 \le
 Ca_T^{1/5}\{1+|\log a_T|\}
 =o(1).
\end{eqnarray*}

We now consider a further decomposition of  $\tilde{\mathbf s}_t(\gamma)$.  Let $M_T\coloneqq\lceil c_0\log(T\delta)\rceil$, $q_T\coloneqq M_T+h$ and

\begin{eqnarray*}
 \tilde{\mathbf s}_t^{[M_T]}(\gamma)
 &\coloneqq&
 E\!\left[
 \tilde{\mathbf s}_t(\gamma)
 \,\middle|\,
 \pmb\eta_{t-M_T},\ldots,\pmb\eta_{t+h}
 \right].
\end{eqnarray*}
Additionally, define the expanding finite-window filtration $\mathcal G_{t,k} \coloneqq \sigma(\pmb\eta_{t+h-k},\ldots,\pmb\eta_{t+h})$, for $k\ge0$, and $\mathcal Q_{t,k}Z \coloneqq E[Z\mid\mathcal G_{t,k}] -E[Z\mid\mathcal G_{t,k-1}]$,  for $k\ge1$. Since $\tilde{\mathbf s}_t(\gamma)$ is $\mathcal H_{t+h}$-measurable, we can write
\begin{eqnarray}\label{ap.tm5c.window.projection}
 &&\tilde{\mathbf s}_t(\gamma)
 -\tilde{\mathbf s}_t^{[M_T]}(\gamma)
 =\sum_{k>q_T} \mathcal Q_{t,k}\tilde{\mathbf s}_t(\gamma).
\end{eqnarray}
Let $\tilde{\mathbf s}^{*(h,k)}_{ij,t}(\gamma)$ be the coupled version of $\tilde{\mathbf s}_t(\gamma)$ obtained by replacing $\pmb\eta_{t+h-k}$ with its independent copy.  Its $L^2$ coupling coefficient is $\mathfrak d^F_{2,s}(k;\gamma)$ from Lemma \ref{Appendix.B7}. By the construction of $\tilde{\mathbf s}^{*(h,k)}_{ij,t}(\gamma)$, it is clear to see that
\begin{eqnarray*}
 E[\tilde{\mathbf s}^{*(h,k)}_{ij,t}(\gamma)
 \mid\mathcal G_{t,k}]
 &=&
 E[\tilde{\mathbf s}_t(\gamma)
 \mid\mathcal G_{t,k-1}],
\end{eqnarray*}
and hence conditional Jensen's inequality gives
\begin{eqnarray*}
 \|\mathcal Q_{t,k}
 \tilde{\mathbf s}_t(\gamma)\|_2
 &=&
 \left\|
 E[\tilde{\mathbf s}_t(\gamma)
 -\tilde{\mathbf s}^{*(h,k)}_{ij,t}(\gamma)
 \mid\mathcal G_{t,k}]
 \right\|_2\le\mathfrak d^F_{2,s}(k;\gamma).
\end{eqnarray*}
Combining this result with \eqref{ap.tm5c.window.projection} and Lemma \ref{Appendix.B7} yields the following bound:
\begin{eqnarray}\label{ap.tm5c.mapprox.pointwise}
 \left\|
 \tilde{\mathbf s}_t(\gamma)
 -\tilde{\mathbf s}_t^{[M_T]}(\gamma)
 \right\|_2
 &\le&
 \sum_{k>q_T}\mathfrak d^F_{2,s}(k;\gamma)
 \le C\varrho^{M_T}.
\end{eqnarray}
Choosing $c_0$ in  $M_T$ such that $c_0>1/(2|\log\varrho|)$, \eqref{ap.tm5c.mapprox.pointwise} gives
\begin{eqnarray}\label{ap.tm5c.mapprox}
 &&\left\|
 \frac{1}{\sqrt{T\delta}}
 \sum_{t=1}^{T-h}
 K_t(\gamma)
 \{\tilde{\mathbf s}_t(\gamma)
 -\tilde{\mathbf s}_t^{[M_T]}(\gamma)\}
 \right\|_2
 \le
 C\sqrt{T\delta}\,\varrho^{M_T}
 =o(1).
\end{eqnarray}

It remains to establish a CLT for the term with $\tilde{\mathbf s}_t^{[M_T]}(\gamma)$.  For fixed interior $\gamma$, let
\begin{eqnarray*}
 \mathcal T_T(\gamma)
 &\coloneqq&
 \{1\le t\le T-h:K_t(\gamma)\ne0\},\qquad
 N_T\coloneqq\#\mathcal T_T(\gamma).
\end{eqnarray*}
Additionally, define
\begin{eqnarray*}
\mathfrak b_T&\coloneqq&\lfloor(T\delta)^{1/2}\rfloor,\qquad
 \ell_T\coloneqq2q_T+1,\quad
 L_T\coloneqq
 \left\lfloor\frac{N_T}{\mathfrak b_T+\ell_T}\right\rfloor,\qquad
 \underline t_T\coloneqq\min\mathcal T_T(\gamma).
\end{eqnarray*}
For $1\le\ell\le L_T$, define the big block and its following gap by
\begin{eqnarray*}
 \mathcal B_{\ell,T}
 &\coloneqq&
 \{\underline t_T+(\ell-1)(\mathfrak b_T+\ell_T),\ldots,
 \underline t_T+(\ell-1)(\mathfrak b_T+\ell_T)+\mathfrak b_T-1\},\nonumber\\
 \mathcal J_{\ell,T}
 &\coloneqq&
 \{\underline t_T+(\ell-1)(\mathfrak b_T+\ell_T)+\mathfrak b_T,\ldots,
 \underline t_T+\ell(\mathfrak b_T+\ell_T)-1\}.
\end{eqnarray*}
Let $\mathcal J_T=\bigcup_{\ell=1}^{L_T}\mathcal J_{\ell,T}$ and let $\mathcal R_T$ contain the unassigned terminal indices.  Then
\begin{eqnarray*}
 L_T&\le&C\frac{T\delta}{\mathfrak b_T},\qquad
 \#\mathcal J_T\le C\frac{T\delta\,\ell_T}{\mathfrak b_T},\qquad
 \#\mathcal R_T\le \mathfrak b_T+\ell_T.
\end{eqnarray*}

Since $ \tilde{\mathbf s}_t^{[M_T]}(\gamma)$ is $q_T$-dependent and has a uniformly bounded fourth moment, for any fixed conformable vector $\pmb\lambda$, we have
\begin{eqnarray*}
 &&\operatorname{Var}\!\left(
 \frac{1}{\sqrt{T\delta}}
 \sum_{t\in\mathcal J_T}K_t(\gamma)
 \pmb\lambda^\top\tilde{\mathbf s}_t^{[M_T]}(\gamma)
 \right)
 \le C\frac{q_T\ell_T}{\mathfrak b_T}=o(1),\nonumber\\
 &&\operatorname{Var}\!\left(
 \frac{1}{\sqrt{T\delta}}
 \sum_{t\in\mathcal R_T}K_t(\gamma)
 \pmb\lambda^\top\tilde{\mathbf s}_t^{[M_T]}(\gamma)
 \right)
 \le C\frac{q_T(\mathfrak b_T+\ell_T)}{T\delta}=o(1).
\end{eqnarray*}
Since $\ell_T>q_T$, the innovation windows belonging to different big blocks are disjoint.  Hence the big-block sums $B_{\ell,T} \coloneqq \frac{1}{\sqrt{T\delta}} \sum_{t\in\mathcal B_{\ell,T}} K_t(\gamma) \pmb\lambda^\top \tilde{\mathbf s}_t^{[M_T]}(\gamma)$, are independent and centred, for  $1\le\ell\le L_T$.

We now verify the Lindeberg condition.  Within each $\mathcal B_{\ell,T}$, split the indices into the $q_T+1$ classes such that  the variables within each class are independent. Using Minkowski's inequality, we can obtain 
\begin{eqnarray*}
 \left\|
 \sum_{t\in\mathcal B_{\ell,T}}K_t(\gamma)\pmb\lambda^\top
 \tilde{\mathbf s}_t^{[M_T]}(\gamma)
 \right\|_4
 &\le&
 C\{(q_T+1)\mathfrak b_T\}^{1/2},
\end{eqnarray*}
and 
\begin{eqnarray*}
 \sum_{\ell=1}^{L_T}E|B_{\ell,T}|^4
 &\le&
 C\frac{T\delta}{\mathfrak b_T}
 \frac{\mathfrak b_T^2(q_T+1)^2}{(T\delta)^2}\le
 C\frac{\mathfrak b_TM_T^2}{T\delta}
 =o(1),
\end{eqnarray*}
where the second inequality uses $q_T=M_T+h$.

 Consequently, for every $\epsilon>0$,
\begin{eqnarray}\label{ap.tm5c.block.lindeberg}
 \sum_{\ell=1}^{L_T}
 E\!\left[
 B_{\ell,T}^2\mathbf1\{|B_{\ell,T}|>\epsilon\}
 \right]
 &\le&
 \epsilon^{-2}\sum_{\ell=1}^{L_T}E|B_{\ell,T}|^4
 =o(1).
\end{eqnarray}
Thus Lindeberg  condition is proved.

We then calculate the limiting variance.  By Lemma \ref{Appendix.B7},
\begin{eqnarray}\label{ap.tm5c.lrv.limit}
 \frac{1}{T\delta}
 \operatorname{Var}\!\left(
 \sum_{t=1}^{T-h}
 K_t(\gamma)\tilde{\mathbf s}_t(\gamma)
 \right)
 &=&
 \sum_{r=-\infty}^{\infty}
 \left\{
 \frac{1}{T\delta}
 \sum_{t=1}^{T-h}
 K_t(\gamma)K_{t+r}(\gamma)
 \right\}
 \operatorname{Cov}\!\left(
 \tilde{\mathbf s}_0(\gamma),
 \tilde{\mathbf s}_r(\gamma)
 \right)\nonumber\\
 &\longrightarrow&
 a_1
\mathbf V^{(h)}_{ij}(\gamma).
\end{eqnarray}
The $L^2$ approximation in \eqref{ap.tm5c.mapprox} and \eqref{ap.tm5c.lrv.limit} imply
\begin{eqnarray*}
 \sum_{\ell=1}^{L_T}\operatorname{Var}(B_{\ell,T})
 &\longrightarrow&
 a_1\pmb\lambda^\top
 \mathbf V^{(h)}_{ij}(\gamma)\pmb\lambda.
\end{eqnarray*}
Using \eqref{ap.tm5c.block.lindeberg} and the Lindeberg--Feller theorem for independent  arrays, we obtain
\begin{eqnarray*}
 \sum_{\ell=1}^{L_T}B_{\ell,T}
 &\to_D&
 N\!\left(
 0,
 a_1\pmb\lambda^\top
 \mathbf V^{(h)}_{ij}(\gamma)\pmb\lambda
 \right).
\end{eqnarray*}

Applying the same argument jointly to the two components of $\mathbf{S}^{(h)}_{ij,\gamma}$, we obtain
\begin{eqnarray*}
 \sqrt{T\delta}\,\mathbf{S}^{(h)}_{ij,\gamma}
 &\to_D&
 N\!\left(
 \mathbf0,
 \diag\{
 a_1\mathbf V^{(h)}_{ij}(\gamma),
 a_2\mathbf V^{(h)}_{ij}(\gamma)
 \}
 \right).
\end{eqnarray*}

Combining this  with \eqref{ap.tm5c.bridge}, \eqref{ap.tm5c.close}, and \eqref{ap.tm5c.desmooth}, and then selecting the coefficient on $y_{j,t}$, gives
\begin{eqnarray*}
 &&\sqrt{T\delta}
 \left\{
 \widehat\Psi^{(h)}_{ij}(\gamma)
 -\Psi^{(h)}_{ij}(\gamma)
 -\frac{v_2\delta^2}{2}
 \Psi^{(h),2}_{ij}(\gamma)
 \right\}
 \to_D
 N\!\left(
 0,
 a_1\pmb\iota_j^\top
 \mathbf D^{-1}
 \mathbf V^{(h)}_{ij}(\gamma)
 \mathbf D^{-1}
 \pmb\iota_j
 \right),
\end{eqnarray*}
which is the stated result in Theorem \ref{TM.4}.
\end{proof}

\begin{proof}[Proof of Thereom \ref{TM.5}]

\item

For brevity, suppress $(i,j,h,\tau,\gamma)$ whenever no confusion can arise, and define $r_T\coloneqq \delta^2+(T\delta)^{-1/2}$ and $u_t\coloneqq y_{i,t+h}-\mathbf x_t^{[j]\top} \pmb\alpha^{(h)}_{ij}(\gamma_t)$.

By Taylor's expansion and \eqref{ap.tm5c.rate}, we obtain  
\begin{eqnarray}\label{proof.lp.inference.residual.rate}
 \max_{|\gamma_t-\gamma|\le\delta}
 \frac{\left|
 \widehat u^{(h)}_{ij,t}(\gamma)-u_t
 \right|}{1+\|\mathbf x_t^{[j]}\|}
 &=&O_P(r_T).
\end{eqnarray}
Define the infeasible density-kernel average
\begin{eqnarray*}
 \widehat{\mathbf D}^{\,\circ}
 &\coloneqq&
 \frac1{T-h}\sum_{t=1}^{T-h}
 \mathcal L_{b_T}(u_t)
 \mathbf x_t^{[j]}\mathbf x_t^{[j]\top}
 K_\delta(\gamma_t-\gamma).
\end{eqnarray*}
Since $\mathcal L$ is Lipschitz continuous,
\begin{eqnarray*}
 |\mathcal L_{b_T}(v)-\mathcal L_{b_T}(u)|
 &\le&
 Cb_T^{-2}|v-u|.
\end{eqnarray*}
The uniform fourth-moment bound, the compact support kernel function, and Markov's inequality jointly give
\begin{eqnarray*}
 \frac1{T-h}\sum_{t=1}^{T-h}
 (1+\|\mathbf x_t^{[j]}\|^3)
 K_\delta(\gamma_t-\gamma)
 &=&O_P(1).
\end{eqnarray*}
Then, it follows from \eqref{proof.lp.inference.residual.rate} that
\begin{eqnarray}\label{proof.lp.inference.D.residual}
 \left\|
 \widehat{\mathbf D}^{(h)}_{ij}(\gamma)
 -\widehat{\mathbf D}^{\,\circ}
 \right\|
 &\le&
 O_P\!\left(\frac{r_T}{b_T^2}\right)
 =o_P(1).
\end{eqnarray}

We next control the centred part of $\widehat{\mathbf D}^{\,\circ}$.  Define $\mathbf H_{t,T}(b_T) \coloneqq \mathcal L_{b_T}(u_t) \mathbf x_t^{[j]}\mathbf x_t^{[j]\top}$ and $\pmb\Xi_{t,T}(b_T) \coloneqq \mathbf H_{t,T}(b_T)-E[\mathbf H_{t,T}(b_T)].$ Additionally,  let $\mathbf H_{t,T}^{*(k)}(b_T)$ and $\pmb\Xi_{t,T}^{*(k)}(b_T)$ be the coupled versions of $\mathbf H_{t,T}(b_T)$ and $\pmb\Xi_{t,T}(b_T)$, respectively, by replacing $\pmb\eta_{t+h-k}$ with an independent copy.  Conditional on $\mathbf x_t^{[j]}$,  Lemma \ref{Appendix.B5}  yields

 \begin{eqnarray*}
 \|\mathbf H_{t,T}(b_T)\|_{4/3}^{4/3}
 &=& 
E\left[
\left\|
\mathcal L_{b_T}(u_t)
\mathbf x_t^{[j]}\mathbf x_t^{[j]\top}
\right\|^{4/3}
\right]
 \nonumber\\
 &\le&
 \frac{C}{b_T^{4/3}}
 E\!\left[
 \|\mathbf x_t^{[j]}\|^{8/3}
 \mathbf1\{|u_t|\le b_T\}
 \right]\nonumber\\
&=&\frac{C}{b_T^{4/3}}
E\left[
\|\mathbf x_t^{[j]}\|^{8/3}
P\left(
|u_t|\le b_T
\mid\mathbf x_t^{[j]}
\right)
\right].
 \nonumber\\
 &\le&
 Cb_T^{-1/3}
 E\|\mathbf x_t^{[j]}\|^{8/3}
 \le Cb_T^{-1/3}.
\end{eqnarray*}
The same bound holds for the coupled variable.  Hence,
\begin{eqnarray}\label{proof.lp.inference.D.envelope}
 \|\pmb\Xi_{t,T}(b_T)
 -\pmb\Xi_{t,T}^{*(k)}(b_T)\|_{4/3}
 &\le&Cb_T^{-1/4}.
\end{eqnarray}
On the other hand, we simply add and subtract $\mathcal L_{b_T}(u_t) \mathbf x_t^{[j],*(k)}\mathbf x_t^{[j],*(k)\top}$.  The boundedness and Lipschitz property of kernel function, H\"older's inequality, and Lemma \ref{Appendix.B11} jointly give, for $k>h$,
\begin{eqnarray*}
\|\mathbf H_{t,T}(b_T)
 -\mathbf H_{t,T}^{*(k)}(b_T)\|_{4/3} &\le&
 \frac{C}{b_T}
 \left\|
 \mathbf x_t^{[j]}\mathbf x_t^{[j]\top}
 -\mathbf x_t^{[j],*(k)}
 \mathbf x_t^{[j],*(k)\top}
 \right\|_{4/3}\nonumber\\
 &&+
 \frac{C}{b_T^2}
 \left\|
 (u_t-u_t^{*(k)})
 \mathbf x_t^{[j],*(k)}
 \mathbf x_t^{[j],*(k)\top}
 \right\|_{4/3}\nonumber\\
 &\le&
 Cb_T^{-2}\rho_D^{\,k-h}
\end{eqnarray*}
for some $\rho_D\in(0,1)$. Combining it with \eqref{proof.lp.inference.D.envelope} gives
\begin{eqnarray}\label{proof.lp.inference.D.coupling}
 \mathfrak d_{4/3,D}(k)
 &\coloneqq&
 \max_t
 \|\pmb\Xi_{t,T}(b_T)
 -\pmb\Xi_{t,T}^{*(k)}(b_T)\|_{4/3}\le
 C\min\left\{
 b_T^{-1/4},
 b_T^{-2}\rho_D^{(k-h)_+}
 \right\}.
\end{eqnarray}
Let $k_{b,T} \coloneqq h+\left\lceil \frac{7|\log b_T|}{4|\log\rho_D|} \right\rceil.$ Using the first bound in \eqref{proof.lp.inference.D.coupling} up to $k_{b,T}$ and the second bound thereafter,
\begin{eqnarray}\label{proof.lp.inference.D.theta}
 \sum_{k=0}^{\infty}\mathfrak d_{4/3,D}(k)
 \le
 C(k_{b,T}+1)b_T^{-1/4}
 +Cb_T^{-2}
 \sum_{k>k_{b,T}}\rho_D^{k-h}\le
 Cb_T^{-1/4}\{1+|\log b_T|\}.
\end{eqnarray}
Applying Lemma \ref{Appendix.B10} with $r=4/3$ and $a_{t,T}=(T-h)^{-1}K_\delta(\gamma_t-\gamma)$,  since $ \sum_{t=1}^{T-h}|a_{t,T}|^{4/3} \le C(T\delta)^{-1/3},$ \eqref{proof.lp.inference.D.theta} gives
\begin{eqnarray}\label{proof.lp.inference.D.centered}
 &&\left\|
 \frac1{T-h}\sum_{t=1}^{T-h}
 \{\mathbf H_{t,T}(b_T)-E\mathbf H_{t,T}(b_T)\}
 K_\delta(\gamma_t-\gamma)
 \right\|_{4/3}
 \le
 C\frac{1+|\log b_T|}{(T\delta b_T)^{1/4}}
 =o(1).
\end{eqnarray}

It remains to study the mean part.  Conditional on $\mathbf x_t^{[j]}$,
\begin{eqnarray*}
 E[\mathcal L_{b_T}(u_t)\mid\mathbf x_t^{[j]}]
 &=&
 \int \mathcal L(v)
 g^{(h)}_{ij,t,T}(b_Tv\mid\mathbf x_t^{[j]})\,\mathrm dv=
 g^{(h)}_{ij,t,T}(0\mid\mathbf x_t^{[j]})
 +O(b_T),
\end{eqnarray*}
where the second equality follows from Assumptions \ref{Assumption5} and \ref{Assumption6}.  Consequently,
\begin{eqnarray}\label{proof.lp.inference.D.mean}
 &&\frac1{T-h}\sum_{t=1}^{T-h}
 E[\mathbf H_{t,T}(b_T)]
 K_\delta(\gamma_t-\gamma)\nonumber\\
 &=&
 \frac1{T-h}\sum_{t=1}^{T-h}
 E\!\left[
 g^{(h)}_{ij,t,T}(0\mid\mathbf x_t^{[j]})
 \mathbf x_t^{[j]}\mathbf x_t^{[j]\top}
 \right]K_\delta(\gamma_t-\gamma)
 +O(b_T)\nonumber\\
 &=&
 \mathbf D^{(h)}_{ij}(\gamma)+o(1),
\end{eqnarray}
where the last equality follows from \eqref{ap.tm5c.curvature.kernel}. Equations \eqref{proof.lp.inference.D.residual}, \eqref{proof.lp.inference.D.centered}, and \eqref{proof.lp.inference.D.mean} imply
\begin{eqnarray}\label{proof.lp.inference.D.consistency}
 \widehat{\mathbf D}^{(h)}_{ij}(\gamma)
 &\to_P&
 \mathbf D^{(h)}_{ij}(\gamma),\qquad
 \widehat{\mathbf D}^{(h)}_{ij}(\gamma)^{-1}
 \to_P
 \mathbf D^{(h)}_{ij}(\gamma)^{-1}.
\end{eqnarray}

We next establish consistency of the covariance part.  Define $r_{s,T}\coloneqq\delta+(T\delta)^{-1/2}$ and $ \tilde{\mathbf s}_t \coloneqq\tilde{\mathbf s}^{(h)}_{ij,t}(\gamma)$ and let $\tilde{\mathbf\Gamma}_{k,T}(\gamma)$ be the sample autocovariance obtained from $\widehat{\mathbf\Gamma}^{(h)}_{ij,k}(\gamma)$ by replacing the estimated centred scores with $\tilde{\mathbf s}_t$.

Equation \eqref{ap.tm5c.rate}, \eqref{lp.indicator.coupling}, Lemma \ref{Appendix.B4}, and Lemma \ref{Appendix.B5} imply the score-replacement rate $O_P(r_{s,T}^{1/5})$.  Moreover, Lemmas \ref{Appendix.B7} and \ref{Appendix.B10} imply that the kernel-weighted mean of the stationary-approximation score is $O_P((T\delta)^{-1/2})$.  The Cauchy--Schwarz inequality therefore gives
\begin{eqnarray}\label{proof.lp.inference.HAC.replace.short}
 \left\|
 \widehat{\mathbf\Gamma}^{(h)}_{ij,k}(\gamma)
 -\tilde{\mathbf\Gamma}_{k,T}(\gamma)
 \right\|
 &=&O_P\{r_{s,T}^{1/5}+(T\delta)^{-1/2}\},
\end{eqnarray}
uniformly in $0\le k\le\ell_{H,T}$.

For each $r\ge0$, use the coupled score $\tilde{\mathbf s}^{*(h,r)}_{ij,v}(\gamma)$ from Lemma \ref{Appendix.B7} and define
\begin{eqnarray*}
\mathfrak d_{8/3,s}(r)
 &\coloneqq&
 \sup_{v\in\mathbb Z}
 \left\|
 \tilde{\mathbf s}^{(h)}_{ij,v}(\gamma)
 -\tilde{\mathbf s}^{*(h,r)}_{ij,v}(\gamma)
 \right\|_{8/3}.
\end{eqnarray*}
Using H\"older's inequality and similar arguments used in the proof of Lemma \ref{Appendix.B7}, we obtain $\mathfrak d_{8/3,s}(r) \le C\varrho^{(r-h)_+}.$

For each fixed $k\ge0$ and $r\ge0$, construct both coupled score factors using the same independent copy of $\pmb\eta_{t+k+h-r}$.  Define
\begin{eqnarray*}
 \pmb\Xi_{t,k}
 &\coloneqq&
 \tilde{\mathbf s}_t\tilde{\mathbf s}_{t+k}^{\top}
 -E[\tilde{\mathbf s}_t\tilde{\mathbf s}_{t+k}^{\top}],
 \nonumber\\
 \pmb\Xi_{t,k}^{*(r)}
 &\coloneqq&
 \left\{
 \begin{array}{ll}
 \tilde{\mathbf s}^{*(h,r-k)}_{ij,t}(\gamma)
 \tilde{\mathbf s}^{*(h,r)}_{ij,t+k}(\gamma)^\top
 -E\!\left[
 \tilde{\mathbf s}^{*(h,r-k)}_{ij,t}(\gamma)
 \tilde{\mathbf s}^{*(h,r)}_{ij,t+k}(\gamma)^\top
 \right],&r\ge k,\\
 \tilde{\mathbf s}_t
 \tilde{\mathbf s}^{*(h,r)}_{ij,t+k}(\gamma)^\top
 -E\!\left[
 \tilde{\mathbf s}_t
 \tilde{\mathbf s}^{*(h,r)}_{ij,t+k}(\gamma)^\top
 \right],&r<k.
 \end{array}
 \right.
\end{eqnarray*}
It is clear to see that the original and coupled products have the same marginal expectation. Moreover, by construction, we have
\begin{eqnarray*}
 \|\tilde{\mathbf s}_{t+k}
 -\tilde{\mathbf s}^{*(h,r)}_{ij,t+k}(\gamma)\|_{8/3}
 \le\mathfrak d_{8/3,s}(r),\quad
 \|\tilde{\mathbf s}_{t}
 -\tilde{\mathbf s}^{*(h,r-k)}_{ij,t}(\gamma)\|_{8/3}
 &\le&\mathfrak d_{8/3,s}(r-k),\qquad r\ge k.
\end{eqnarray*}
For $r<k$, the first score factor is unchanged. Expanding the product difference and applying H\"older's inequality therefore give
\begin{eqnarray*}
 \|\pmb\Xi_{t,k}-\pmb\Xi_{t,k}^{*(r)}\|_{4/3}
 &\le&
 C\left\{
 \mathfrak d_{8/3,s}(r)
 +\mathbf1\{r\ge k\}\mathfrak d_{8/3,s}(r-k)
 \right\}.
\end{eqnarray*}
Consequently,
\begin{eqnarray*}
 \sup_{k\ge0}\sum_{r=0}^{\infty}
 \|\pmb\Xi_{t,k}-\pmb\Xi_{t,k}^{*(r)}\|_{4/3}
 &\le&
 2C\sum_{r=0}^{\infty}\mathfrak d_{8/3,s}(r)
 <\infty.
\end{eqnarray*}
Applying Lemma \ref{Appendix.B10} with $r=4/3$ yields 
\begin{eqnarray*}
 \left\|
 \tilde{\mathbf\Gamma}_{k,T}(\gamma)
 -E\tilde{\mathbf\Gamma}_{k,T}(\gamma)
 \right\|_{4/3}
 &\le&C(T\delta)^{-1/4},
\end{eqnarray*}
uniformly in $0\le k\le\ell_{H,T}$.

Let $\tilde{\mathbf\Omega}_{T}^{B}(\gamma)$ denote the Bartlett sum obtained from $\widehat{\mathbf\Omega}^{(h)}_{ij}(\gamma)$ by replacing $\widehat{\mathbf\Gamma}^{(h)}_{ij,k}(\gamma)$ with $\tilde{\mathbf\Gamma}_{k,T}(\gamma)$.  Since the Bartlett weights are bounded by one, using \eqref{proof.lp.inference.HAC.replace.short}, Minkowski's inequality, and Assumption \ref{Assumption6} gives
\begin{eqnarray}\label{proof.lp.inference.HAC.errors}
 \left\|
 \widehat{\mathbf\Omega}^{(h)}_{ij}(\gamma)
 -\tilde{\mathbf\Omega}_{T}^{B}(\gamma)
 \right\|
 =o_P(1),\quad
 \left\|
 \tilde{\mathbf\Omega}_{T}^{B}(\gamma)
 -E\tilde{\mathbf\Omega}_{T}^{B}(\gamma)
 \right\|
 &=&o_P(1).
\end{eqnarray}

Let $\mathbf\Gamma_k(\gamma)\coloneqq E[\tilde{\mathbf s}_0\tilde{\mathbf s}_k^\top]$.
Stationarity gives $$E\tilde{\mathbf\Gamma}_{k,T}(\gamma)
 = \lambda_{k,T}(\gamma)\mathbf\Gamma_k(\gamma),$$
 where $\lambda_{k,T}(\gamma) \coloneqq \frac1{(T-h)\delta} \sum_{t=1}^{T-h-k}K_t(\gamma)K_{t+k}(\gamma)$. The compact support and Lipschitz continuity of kernel function imply
\begin{eqnarray*}
 \sup_{0\le k\le\ell_{H,T}}
 |\lambda_{k,T}(\gamma)-a_1|
 &\le&
 C\frac{\ell_{H,T}+1}{T\delta}
 =o(1).
\end{eqnarray*}

By Lemma \ref{Appendix.B7}, Assumptions \ref{Assumption5}  and \ref{Assumption6},  we obtain 
\begin{eqnarray}\label{proof.lp.inference.HAC.population}
 E\tilde{\mathbf\Omega}_{T}^{B}(\gamma)
 &\longrightarrow&
 a_1\mathbf V^{(h)}_{ij}(\gamma).
\end{eqnarray}
Combining \eqref{proof.lp.inference.HAC.errors} and \eqref{proof.lp.inference.HAC.population} gives
\begin{eqnarray}\label{proof.lp.inference.HAC.consistency}
 \widehat{\mathbf\Omega}^{(h)}_{ij}(\gamma)
 &\to_P&
 a_1\mathbf V^{(h)}_{ij}(\gamma).
\end{eqnarray}
Finally, \eqref{proof.lp.inference.D.consistency} and \eqref{proof.lp.inference.HAC.consistency} jointly prove Theorem \ref{TM.5}.
\end{proof}

\subsection{Proofs of the Preliminary Lemmas}\label{Appen.B4}

\begin{proof}[Proof of Lemma \ref{lem:gelfand_bound}]
    \item 

By Gelfand's formula \citep[Corollary 5.6.14]{horn2012matrix}, the spectral radius of $\tilde{\pmb{\phi}}$ satisfies
\begin{equation*}
    \rho(\tilde{\pmb{\phi}}) = \lim_{k \to \infty} \lVert \tilde{\pmb{\phi}}^k \rVert_2^{1/k}.
\end{equation*}
Since $\rho(\tilde{\pmb{\phi}}) < 1$, fix any $\gamma \in (\rho(\tilde{\pmb{\phi}}), 1)$ and let $\epsilon \coloneqq \gamma - \rho(\tilde{\pmb{\phi}}) > 0$. By the definition of the limit, there exists a finite integer $K(\epsilon) \ge 1$ such that for all $k > K(\epsilon)$,
\begin{equation*}
    \Bigl\lvert \lVert \tilde{\pmb{\phi}}^k \rVert_2^{1/k} - \rho(\tilde{\pmb{\phi}}) \Bigr\rvert < \epsilon 
    \implies \lVert \tilde{\pmb{\phi}}^k \rVert_2^{1/k} < \rho(\tilde{\pmb{\phi}}) + \epsilon = \gamma.
\end{equation*}
Raising both sides to the $k$-th power yields $\lVert \tilde{\pmb{\phi}}^k \rVert_2 < \gamma^k$ for all $k > K(\epsilon)$. 

For the finitely many initial powers $1 \le k \le K(\epsilon)$, define the constant
\begin{equation*}
    C_\phi := \max\left\{ 1, \, \max_{1 \le k \le K(\epsilon)} \frac{\lVert \tilde{\pmb{\phi}}^k \rVert_2}{\gamma^k} \right\} < \infty.
\end{equation*}
Combining both regimes ($1 \le k \le K(\epsilon)$ and $k > K(\epsilon)$), it follows that
\begin{equation*}
    \lVert \tilde{\pmb{\phi}}^k \rVert_2 \le C_\phi \gamma^k \quad \text{for all } k \ge 1,
\end{equation*}
which completes the proof.
\end{proof}

\begin{proof}[Proof of Lemma \ref{Appendix.B1}]
\item

(1). (a). When $r=0$, the result is obvious. Next, consider
\begin{eqnarray*}
 k_{\delta,t}^{(r)}(\gamma)
 &=&
 \left(\frac{\gamma_t-\gamma}{\delta}\right)^r
 K_\delta(\gamma_t-\gamma)
\end{eqnarray*}
with $r\in\{1,2,3,4\}$.

By design, we only need to consider two cases:
$|(\gamma_t-\gamma)/\delta|>1$ and
$|(\gamma_t-\gamma)/\delta|\le1$. When
$|(\gamma_t-\gamma)/\delta|>1$, $k_{\delta,t}^{(r)}(\gamma)=0$ by
Assumption \ref{Assumption3}. When
$|(\gamma_t-\gamma)/\delta|\le1$,
\begin{eqnarray*}
 &&\max_{t\ge1}\sup_{\gamma\in[0,1]}
 |k_{\delta,t}^{(r)}(\gamma)| \le
 \max_{t\ge1}\left|\frac{\gamma_t-\gamma}{\delta}\right|^r
 \sup_{\gamma\in[0,1]}\frac1\delta
 \left|K\left(\frac{\gamma_t-\gamma}{\delta}\right)\right|
 \le\frac C\delta.
\end{eqnarray*}

(b). When $r=0$, $k_{\delta,t}^{(0)}(\gamma)
=K_\delta(\gamma_t-\gamma)$.  The Lipschitz continuity and compact
support of $K$ give
\begin{eqnarray*}
 |k_{\delta,t+1}^{(0)}(\gamma)
 -k_{\delta,t}^{(0)}(\gamma)|
 &\le&
 \frac{C}{T\delta^2}
 \mathbf1\{|\gamma_t-\gamma|\le\delta+T^{-1}\}.
\end{eqnarray*}

Next, consider $r\in\{1,2,3,4\}$.
By adding and subtracting $w_{t+1,\gamma}^rK(w_{t,\gamma})$, write
\begin{eqnarray*}
 |k_{\delta,t+1}^{(r)}(\gamma)
 -k_{\delta,t}^{(r)}(\gamma)|
 \le
 \frac1\delta |K(w_{t,\gamma})|
 |w_{t+1,\gamma}^{r}-w_{t,\gamma}^{r}|
 +\frac1\delta |w_{t+1,\gamma}|^{r}
 |K(w_{t+1,\gamma})-K(w_{t,\gamma})|.
\end{eqnarray*}
Whenever either kernel weight is nonzero,
$|w_{t,\gamma}|\le1+(T\delta)^{-1}$ and
$|w_{t+1,\gamma}|\le1+(T\delta)^{-1}$.  Since
$|w_{t+1,\gamma}-w_{t,\gamma}|=(T\delta)^{-1}$, boundedness and
Lipschitz continuity of $K$ imply
\begin{eqnarray*}
 |k_{\delta,t+1}^{(r)}(\gamma)
 -k_{\delta,t}^{(r)}(\gamma)|
 &\le&
 \frac{C}{T\delta^2}
 \mathbf1\{|\gamma_t-\gamma|\le\delta+T^{-1}\},
\end{eqnarray*}
for $r\in\{1,2,3,4\}$.

Uniformly in $\gamma\in[0,1]$, at most $C(T\delta+1)$ terms in
these bounds are nonzero.  Hence, for every
$r\in\{0,1,2,3,4\}$,
\begin{eqnarray*}
 &&\sup_{\gamma\in[0,1]}\frac1T\sum_{t=1}^{T-1}
 |k_{\delta,t+1}^{(r)}(\gamma)
 -k_{\delta,t}^{(r)}(\gamma)| \le
 \frac{C(T\delta+1)}{T^2\delta^2}
 =O\left(\frac1{T\delta}\right),
\end{eqnarray*}
where the last equality follows from $T\delta\to\infty$.

\medskip

(2). (a). Write
\begin{eqnarray*}
 &&\max_{t\ge1}\sum_{\ell=0}^{\infty}
 \ell\|\mathbf c_{\ell,t}\|_2
 =\max_{t\ge1}\sum_{\ell=0}^{\infty}
 \ell\|\mathbf s\mathbf C_{\ell,t}\mathbf s^\top\|_2\le
 \max_{t\ge1}\sum_{\ell=0}^{\infty}
 \ell\|\mathbf C_{\ell,t}\|_2
 \le C\sum_{\ell=0}^{\infty}\ell\tilde{\rho}_y^\ell<\infty,
\end{eqnarray*}
where $\mathbf s$ and $\mathbf C_{\ell,t}$ are defined in \eqref{def.yt.ma}, and the last step follows from Assumption \ref{Assumption1}, and the last step follows from Lemma \ref{lem:gelfand_bound} by letting $\tilde{\rho}_y\in (\rho_y, 1)$.

\smallskip

(b). Note that for any conformable matrices
\begin{eqnarray}\label{def.ineq.prod}
 \prod_{i=1}^{r}\mathbf z_i-
 \prod_{i=1}^{r}\mathbf w_i
 &=&
 \sum_{j=1}^{r}
 \left(\prod_{k=1}^{j-1}\mathbf z_k\right)
 (\mathbf z_j-\mathbf w_j)
 \left(\prod_{k=j+1}^{r}\mathbf w_k\right).
\end{eqnarray}
Thus, we write
\begin{eqnarray*}
 &&\max_{0\le s\le T-1}\sum_{t=0}^{s}
 \sum_{\ell=0}^{\infty}
 \ell\|\mathbf c_{\ell,t+1}-\mathbf c_{\ell,t}\|_2\nonumber\\
 &\le&
 \max_{0\le s\le T-1}\sum_{t=0}^{s}
 \sum_{\ell=0}^{\infty}
 \ell\|\mathbf C_{\ell,t+1}-\mathbf C_{\ell,t}\|_2\nonumber\\
 &=&
 \max_{0\le s\le T-1}\sum_{t=0}^{s}
 \sum_{\ell=1}^{\infty}\ell
 \left\|
 \prod_{j=0}^{\ell-1}\tilde{\mathbf A}_{t+1-j}
 -\prod_{j=0}^{\ell-1}\tilde{\mathbf A}_{t-j}
 \right\|_2\nonumber\\
 &\le&
 \max_{0\le s\le T-1}
 C
 \sum_{\ell=1}^{\infty}\ell^2\tilde{\rho}_y^{\ell-1}<\infty.
\end{eqnarray*}
where the first inequality follows from the definition of $\mathbf{c}_{\ell,t}$ in \eqref{def.yt.ma}, and the second inequality follows from Assumption \ref{Assumption1} and \eqref{def.ineq.prod}.

\smallskip

In (c)--(e), the decomposition
$\mathbf c_t^r(L)=\bar{\mathbf c}_t^r-
(1-L)\tilde{\mathbf c}_t^r(L)$ is straightforward in view of
Lemma BN of \cite{PS1992}.

(c). We write
\begin{eqnarray*}
 &&\max_{t\ge1}\sum_{r=1}^{\infty}
 \sum_{\ell=0}^{\infty}
 \|\tilde{\mathbf c}_{\ell,t}^r\|_2\le
 \max_{t\ge1}\sum_{r=1}^{\infty}
 \sum_{\ell=0}^{\infty}\sum_{k=\ell+1}^{\infty}
 \|\mathbf c_{k+r,t}\otimes\mathbf c_{k,t}\|_2\nonumber\\
 &=&
 \max_{t\ge1}\sum_{r=1}^{\infty}
 \sum_{\ell=1}^{\infty}
 \ell\|\mathbf c_{\ell+r,t}\|_2
 \|\mathbf c_{\ell,t}\|_2\lesssim
 \max_{t\ge1}
 \left(\sum_{\ell=1}^{\infty}
 \ell\|\mathbf c_{\ell,t}\|_2\right)
 \left(\sum_{\ell=1}^{\infty}
 \|\mathbf c_{\ell,t}\|_2\right)<\infty,
\end{eqnarray*}
where the last step follows from Lemma \ref{Appendix.B1}.2(a).

\smallskip

(d). Note that
\begin{eqnarray*}
 &&\max_{t\ge1}\sum_{r=1}^{\infty}
 \|\bar{\mathbf c}_t^r\|_2
 \le\max_{t\ge1}\sum_{r=1}^{\infty}
 \sum_{\ell=0}^{\infty}
 \|\mathbf c_{\ell+r,t}\otimes\mathbf c_{\ell,t}\|_2\le
 \left(\max_{t\ge1}\sum_{\ell=0}^{\infty}
 \|\mathbf c_{\ell,t}\|_2\right)^2<\infty,
\end{eqnarray*}
where the last step follows from Lemma \ref{Appendix.B1}.2(a).

\smallskip

(e). Note that
\begin{eqnarray*}
 \sum_{t=1}^{s}
 \|\tilde{\mathbf c}_{t+1}^r(1)
 -\tilde{\mathbf c}_{t}^r(1)\|_2&\le&
 \sum_{t=1}^{s}\sum_{\ell=0}^{\infty}
 \sum_{k=\ell+1}^{\infty}
 \|\mathbf c_{k,t+1}^r-\mathbf c_{k,t}^r\|_2\nonumber\\
 &=&
 \sum_{t=1}^{s}\sum_{\ell=1}^{\infty}\ell
 \|\mathbf c_{\ell+r,t+1}\otimes\mathbf c_{\ell,t+1}
 -\mathbf c_{\ell+r,t}\otimes\mathbf c_{\ell,t}\|_2\nonumber\\
 &\le&
 \sum_{t=1}^{s}\sum_{\ell=1}^{\infty}\ell
 \|\mathbf c_{\ell+r,t+1}\|_2
 \|\mathbf c_{\ell,t+1}-\mathbf c_{\ell,t}\|_2\nonumber\\
 &&+
 \sum_{t=1}^{s}\sum_{\ell=1}^{\infty}\ell
 \|\mathbf c_{\ell+r,t+1}-\mathbf c_{\ell+r,t}\|_2
 \|\mathbf c_{\ell,t}\|_2.
\end{eqnarray*}
Thus, we can write
\begin{eqnarray*}
 &&\max_{1\le s\le T-1}\sum_{t=1}^{s}
 \sum_{r=1}^{\infty}
 \|\tilde{\mathbf c}_{t+1}^r(1)
 -\tilde{\mathbf c}_{t}^r(1)\|_2\nonumber\\
 &\lesssim&
 \left(\max_{1\le s\le T-1}\sum_{t=1}^{s}
 \sum_{\ell=1}^{\infty}\ell
 \|\mathbf c_{\ell,t+1}-\mathbf c_{\ell,t}\|_2\right)
 \left(\max_{t\ge1}\sum_{\ell=1}^{\infty}
 \|\mathbf c_{\ell,t}\|_2\right)\nonumber\\
 &&+
 \left(\max_{1\le s\le T-1}\sum_{t=1}^{s}
 \sum_{r=1}^{\infty}
 \|\mathbf c_{r,t+1}-\mathbf c_{r,t}\|_2\right)
 \left(\max_{t\ge1}\sum_{\ell=1}^{\infty}\ell
 \|\mathbf c_{\ell,t}\|_2\right)<\infty,
\end{eqnarray*}
where the last step follows from Lemma \ref{Appendix.B1}.2(a)--(b).

\medskip

(3). (a). Let $\beta_{\ell,j}(\cdot)$ stand for the $j^{th}$ element
of $\pmb\beta_\ell(\cdot)$. We then just focus on the $j^{th}$ element
without loss of generality. Write
\begin{eqnarray*}
 &&\frac1T\sum_{t=1}^{T}
 \{\beta_{\ell,j}(\gamma_t)-\beta_{\ell,j}(\gamma)
 -(\gamma_t-\gamma)\beta_{\ell,j}^{(1)}(\gamma)\}
 K_\delta(\gamma_t-\gamma)\nonumber\\
 &=&
 \frac1{2T}\sum_{t=1}^{T}(\gamma_t-\gamma)^2
 \beta_{\ell,j}^{(2)}(\gamma_t^*)K_\delta(\gamma_t-\gamma)
 \nonumber\\
 &=&
 \frac1{2T}\sum_{t=1}^{T}(\gamma_t-\gamma)^2
 \beta_{\ell,j}^{(2)}(\gamma)K_\delta(\gamma_t-\gamma)
 \nonumber\\
 &&+
 \frac1{2T}\sum_{t=1}^{T}(\gamma_t-\gamma)^2
 \{\beta_{\ell,j}^{(2)}(\gamma_t^*)
 -\beta_{\ell,j}^{(2)}(\gamma)\}K_\delta(\gamma_t-\gamma),
\end{eqnarray*}
where $\gamma_t^*$ is between $\gamma$ and $\gamma_t$. As
$\beta_{\ell,j}$ is twice continuously differentiable by Assumption
\ref{Assumption1}, we know that
\begin{eqnarray*}
 &&\frac1T\sum_{t=1}^{T}(\gamma_t-\gamma)^2
 |\beta_{\ell,j}^{(2)}(\gamma_t^*)
 -\beta_{\ell,j}^{(2)}(\gamma)|K_\delta(\gamma_t-\gamma)
 \nonumber\\
 &\lesssim&
 \frac1T\sum_{t=1}^{T}|\gamma_t-\gamma|^3
 K_\delta(\gamma_t-\gamma)=O(\delta^3).
\end{eqnarray*}
Also, by the fact that $\gamma\in[\delta,1-\delta]$,
\begin{eqnarray*}
 &&\frac1T\sum_{t=1}^{T}(\gamma_t-\gamma)^2
 \beta_{\ell,j}^{(2)}(\gamma)K_\delta(\gamma_t-\gamma)=
 \delta^2v_2\beta_{\ell,j}^{(2)}(\gamma)+O(T^{-1})
\end{eqnarray*}
uniformly in $\gamma\in[\delta,1-\delta]$. Then (a) follows.

(b). 
Similar to the development of (a), we can write

\begin{eqnarray*}
    &&\frac{1}{T}\sum_{t=1}^{T} |\beta_{\ell,j}(\gamma_t)-\beta_{\ell,j}(\gamma)-(\gamma_t-\gamma)\beta_{\ell,j}^{(1)}(\gamma) |^2 K_\delta(\gamma_t -\gamma) \notag\\
    &=&\frac{1}{4T}\sum_{t=1}^{T} | (\gamma_t-\gamma)^2 \beta_{\ell,j}^{(2)}(\gamma_t^*) |^2 K_\delta(\gamma_t -\gamma)\notag \\
    &\asymp &\frac{1}{4T}\sum_{t=1}^{T} |(\gamma_t-\gamma)^2 \beta_{\ell,j}^{(2)}(\gamma) |^2 K_\delta(\gamma_t -\gamma),
\end{eqnarray*}
which implies 
\begin{eqnarray*}
 \frac1T\sum_{t=1}^{T}
 \|\pmb\beta_\ell(\gamma_t)-\pmb\beta_\ell(\gamma)
 -(\gamma_t-\gamma)\pmb\beta_\ell^{(1)}(\gamma)\|_2^2
 K_\delta(\gamma_t-\gamma)
 &=&
 \frac{v_4\delta^4}{4}
 \|\pmb\beta_\ell^{(2)}(\gamma)\|_2^2+o(\delta^4).
\end{eqnarray*}
The proof is now completed.
\end{proof}

\medskip

\begin{proof}[Proof of Lemma \ref{Appendix.B2}]
\item

(1). For $\forall w\in \mathbb{R}$, write
\begin{eqnarray*}
 &&\sup_u\left|
 \int\{\psi_\tau(x+u+w)-\psi_\tau(x+u)\}
 \phi_m(x)\mathrm dx\right|\nonumber\\
 &=&
 \sup_u\left|
 \int_{-\infty}^{-u-w}(\tau-1)\phi_m(x)\mathrm dx
 +\int_{-u-w}^{\infty}\tau\phi_m(x)\mathrm dx
 -\int_{-\infty}^{-u}(\tau-1)\phi_m(x)\mathrm dx
 -\int_{-u}^{\infty}\tau\phi_m(x)\mathrm dx
 \right|\nonumber\\
 &=&
 \sup_u\left|\int_{-u-w}^{-u}\phi_m(x)\mathrm dx\right|.
\end{eqnarray*}
The same calculation applies with the integration limits reversed when $w<0$.  Since $\|\phi_m\|_\infty=m\|\phi\|_\infty$ and $\int\phi_m(x)\mathrm dx=1$, we obtain

\begin{eqnarray*}
 \sup_u\left|\int_{-u-w}^{-u}\phi_m(x)\mathrm dx\right|
 &\le&
 \min\{m\|\phi\|_\infty|w|,1\}
 \le C\min\{m|w|,1\},
\end{eqnarray*}
which proves part 1.

\medskip

(2). In this proof, we drop the indices of $e_{\ell,t}$, and write it as $e$ for simplicity. For $\forall w\in \mathbb{R}$, write
\begin{eqnarray*}
 \rho_{\tau,m}^{(1)}(e+w)-\rho_{\tau,m}^{(1)}(e)&=&
 \int\rho_\tau(x)
 \{\phi_m^{(1)}(x-e-w)-\phi_m^{(1)}(x-e)\}\mathrm dx\nonumber\\
 &=&
 \int\{\psi_\tau(x+e+w)-\psi_\tau(x+e)\}
 \phi_m(x)\mathrm dx.
\end{eqnarray*}
Conditional Jensen's inequality yields

\begin{eqnarray*}
 &&E_{t-1}\!\left[
 \{\rho_{\tau,m}^{(1)}(e+w)-\rho_{\tau,m}^{(1)}(e)\}^2
 \right]\nonumber\\
 &\le&
 \int E_{t-1}\!\left[
 \{\psi_\tau(e+w+x)-\psi_\tau(e+x)\}^2
 \right]\phi_m(x)\mathrm dx.
\end{eqnarray*}
For $w>0$, the squared score difference is the indicator of a zero crossing.  The uniform conditional-density bound in Assumption \ref{Assumption1} therefore gives
\begin{eqnarray*}
 &&\int E_{t-1}\!\left[
 \{\psi_\tau(e+w+x)-\psi_\tau(e+x)\}^2
 \right]\phi_m(x)\mathrm dx\nonumber\\
 &=&
 \int
 P(-x-w<e\le-x\mid\mathcal F_{t-1})
 \phi_m(x)\mathrm dx\le
 C|w|\int\phi_m(x)\mathrm dx
 =C|w|.
\end{eqnarray*}
The same result follows for $w<0$ after reversing the crossing interval.
This proves part 2.

\medskip

(3). Again write $e=e_{\ell,t}$.  By the convolution representation and
Fubini's theorem, subtracting the linear term before taking the
conditional expectation gives

\begin{eqnarray*}
 &&E_{t-1}\!\left[
 \rho_{\tau,m}(e+\xi)-\rho_{\tau,m}(e)
 -\rho_{\tau,m}^{(1)}(e)\xi
 \right]\nonumber\\
 &=&
 \int E_{t-1}\!\left[
 \rho_\tau(e+x+\xi)-\rho_\tau(e+x)
 -\psi_\tau(e+x)\xi
 \right]\phi_m(x)\mathrm dx.
\end{eqnarray*}
Let
\begin{eqnarray*}
 F_{\ell,t}(z\mid\mathcal F_{t-1})
 &\coloneqq&
 P(e_{\ell,t}\le z\mid\mathcal F_{t-1}).
\end{eqnarray*}
The identity $\rho_\tau(v+\xi)-\rho_\tau(v) -\psi_\tau(v)\xi =\int_0^\xi\{\psi_\tau(v+s)-\psi_\tau(v)\}\mathrm ds$ then gives
\begin{eqnarray*}
 &&E_{t-1}\!\left[
 \rho_{\tau,m}(e+\xi)-\rho_{\tau,m}(e)
 -\rho_{\tau,m}^{(1)}(e)\xi
 \right]\nonumber\\
 &=&
 \int\int_0^\xi
 \{
 F_{\ell,t}(-x\mid\mathcal F_{t-1})
 -F_{\ell,t}(-x-s\mid\mathcal F_{t-1})
 \}\mathrm ds\,\phi_m(x)\mathrm dx.
\end{eqnarray*}
Because the support of $\phi_m$ is $[-m^{-1},m^{-1}]$, for sufficiently large $m$ and sufficiently small $|\xi|$, every density argument in the last display belongs to the neighborhood in Assumption \ref{Assumption1}.  The fundamental theorem of calculus and the local Lipschitz condition imply, uniformly in $(t,\ell)$,
\begin{eqnarray*}
 &&\left|
 F_{\ell,t}(-x\mid\mathcal F_{t-1})
 -F_{\ell,t}(-x-s\mid\mathcal F_{t-1})
 -s f_{\ell,t}(0\mid\mathcal F_{t-1})
 \right|\le
 C|s|(|x|+|s|).
\end{eqnarray*}
Since $\int\phi_m(x)\mathrm dx=1$ and $\int|x|\phi_m(x)\mathrm dx\le C/m$, it follows that
\begin{eqnarray*}
 &&E_{t-1}\!\left[
 \rho_{\tau,m}(e+\xi)-\rho_{\tau,m}(e)
 -\rho_{\tau,m}^{(1)}(e)\xi
 \right]=
 \frac12f_{\ell,t}(0\mid\mathcal F_{t-1})\xi^2
 +O(|\xi|^3+\xi^2/m),
\end{eqnarray*}
which proves part 3.
\end{proof}

\medskip

\begin{proof}[Proof of Lemma \ref{Appendix.B3}]
    \item

    (1). Assumption \ref{Assumption1} and the predictability of $\mathbf x_{t-1}$ give
\begin{eqnarray}\label{proof.q.score.mds}
 E[\mathbf q_t^Q\mid\mathcal F_{t-1}]
 &=&
 E[\pmb\psi_t\mid\mathcal F_{t-1}]
 \otimes\mathbf x_{t-1}
 =\mathbf0.
\end{eqnarray}
Moreover, \eqref{def.yt.ma}, Lemma \ref{Appendix.B1}.2, Assumption \ref{Assumption1}, and Minkowski's inequality imply
\begin{eqnarray}\label{proof.q.state.moment}
 \sup_{T,t}\|\mathbf x_{t-1}\|_\nu
 &\le&C\left\{1+\sup_t\sum_{j=0}^{\infty}
 \|\mathbf c_{j,t}\|_2\right\}
 \sup_{T,s}\|\mathbf e_s\|_\nu
 \le C.
\end{eqnarray}
Since $|\pmb\psi_t|\le C$, martingale orthogonality therefore yields, for each fixed $\gamma$,
\begin{eqnarray*}
 &&E\left\|
 \frac1T\sum_{t=1}^{T}
 \mathbf q_t^Qk_{\delta,t}^{(r)}(\gamma)
 \right\|^2=
 \frac1{T^2}\sum_{t=1}^{T}
 E\|\mathbf q_t^Q\|^2
 |k_{\delta,t}^{(r)}(\gamma)|^2
 \le\frac{C}{T\delta}.
\end{eqnarray*}
This proves the point-wise order.

We then consider the uniform result. The bandwidth condition in Assumption \ref{Assumption3} permits a sequence $c_T\to\infty$ such that, with $\chi_T=c_TT^{1/\nu}$,
\begin{eqnarray*}
 \frac{\chi_T^2\log T}{T\delta}&\longrightarrow&0.
\end{eqnarray*}
Equation \eqref{proof.q.state.moment} and the union bound give
\begin{eqnarray*}
 P\left(\max_{t\le T}\|\mathbf x_{t-1}\|>\chi_T\right)
 &\le&
 \chi_T^{-\nu}\sum_{t=1}^{T}E\|\mathbf x_{t-1}\|^\nu
 \le Cc_T^{-\nu}=o(1).
\end{eqnarray*}
On the complementary event, apply the maximal inequality for kernel-weighted martingale arrays to \eqref{proof.q.score.mds} on a $T^{-2}$ grid in $\gamma$.  The increment bound is controlled by $\chi_T/(T\delta)$, while its quadratic scale is $(T\delta)^{-1}$; the grid maximum is therefore
\begin{eqnarray*}
 O_P\!\left(
 \sqrt{\frac{\log T}{T\delta}}
 +\frac{\chi_T\log T}{T\delta}
 \right).
\end{eqnarray*}
The second term is negligible relative to the first because $\chi_T^2\log T/(T\delta)\to0$.  To extend the result between adjacent grid points, note that Lipschitz continuity and compact support of $K$ give
\begin{eqnarray*}
 |k_{\delta,t}^{(r)}(\gamma)
 -k_{\delta,t}^{(r)}(\gamma')|
 &\le&
 C\delta^{-2}|\gamma-\gamma'|
\end{eqnarray*}
whenever either weight is nonzero.  On the truncation event, a grid of mesh $T^{-2}$ therefore produces an oscillation of order $O\{\chi_T/(T^2\delta)\}$, which is $o\{\sqrt{\log T/(T\delta)}\}$ under Assumption \ref{Assumption3}. Consequently,
\begin{eqnarray*}
 \sup_{\gamma\in[\delta,1-\delta]}
 \left\|
 \frac1T\sum_{t=1}^{T}
 \mathbf q_t^Qk_{\delta,t}^{(r)}(\gamma)
 \right\|
 &=&O_P\!\left(\sqrt{\frac{\log T}{T\delta}}\right).
\end{eqnarray*}

(2). As in the original proof, first decompose the smoothed score into the ordinary quantile score and a smoothing error:
\begin{eqnarray}\label{proof.q.smooth.decomp}
 &&\frac1T\sum_{t=1}^{T}
 \left\{\pmb\rho_{m,t}^{(1)}\otimes\mathbf x_{t-1}\right\}
 k_{\delta,t}^{(r)}(\gamma)\nonumber\\
 &=&
 \frac1T\sum_{t=1}^{T}
 \left\{\pmb\psi_t\otimes\mathbf x_{t-1}\right\}
 k_{\delta,t}^{(r)}(\gamma)
 +\frac1T\sum_{t=1}^{T}
 \left\{(\pmb\rho_{m,t}^{(1)}-\pmb\psi_t)
 \otimes\mathbf x_{t-1}\right\}
 k_{\delta,t}^{(r)}(\gamma).
\end{eqnarray}
The first term has already been treated in part 1. To control the second term, the construction of the smoothed score gives
\begin{eqnarray}\label{rate.rho1}
 |\rho_{\tau,m}^{(1)}(e)-\psi_\tau(e)|
 &\le&
 \int\mathbf1\{|e|\le|x|\}\phi_m(x)\mathrm dx.
\end{eqnarray}
Put
\begin{eqnarray*}
 d_{\ell,t,m}
 &\coloneqq&
 \rho_{\tau,m}^{(1)}(e_{\ell,t})-\psi_\tau(e_{\ell,t}),qquad
 \mathbf d_{t,m}\coloneqq
 (d_{1,t,m},\ldots,d_{p,t,m})^\top.
\end{eqnarray*}
Because $|x|\le m^{-1}$ on the support of $\phi_m$, Assumption \ref{Assumption1}, \eqref{rate.rho1}, and conditional Jensen's inequality give
\begin{eqnarray*}
 E_{t-1}|d_{\ell,t,m}|^2
 &\le&
 \int
 P(|e_{\ell,t}|\le|x|\mid\mathcal F_{t-1})
 \phi_m(x)\mathrm dx
 \le\frac{C}{m}.
\end{eqnarray*}
Let $F_{\ell,t}(\cdot\mid\mathcal F_{t-1})$ denote the conditional CDF of $e_{\ell,t}$.  The conditional score restriction and \eqref{def.der} imply
\begin{eqnarray*}
 E_{t-1}d_{\ell,t,m}
 &=&
 \int
 \{F_{\ell,t}(0\mid\mathcal F_{t-1})
 -F_{\ell,t}(-x\mid\mathcal F_{t-1})\}
 \phi_m(x)\mathrm dx.
\end{eqnarray*}
The local Lipschitz condition on the conditional density gives, uniformly over $|x|\le m^{-1}$,
\begin{eqnarray*}
 F_{\ell,t}(0\mid\mathcal F_{t-1})
 -F_{\ell,t}(-x\mid\mathcal F_{t-1})
 &=&
 f_{\ell,t}(0\mid\mathcal F_{t-1})x+O(x^2).
\end{eqnarray*}
Since $\phi_m$ is symmetric, its first moment vanishes, and hence
\begin{eqnarray}\label{proof.q.desmooth.mean}
 |E_{t-1}d_{\ell,t,m}|
 &\le&Cm^{-2}.
\end{eqnarray}

Define the martingale difference
\begin{eqnarray*}
 \pmb\zeta_{t,m}
 &\coloneqq&
 \mathbf d_{t,m}\otimes\mathbf x_{t-1}
 -E_{t-1}[\mathbf d_{t,m}\otimes\mathbf x_{t-1}].
\end{eqnarray*}
The preceding conditional second-moment bound and \eqref{proof.q.state.moment} imply
\begin{eqnarray*}
 E\|\pmb\zeta_{t,m}\|^2
 &\le&Cm^{-1}.
\end{eqnarray*}
Applying the same truncated maximal martingale inequality and the same $T^{-2}$ grid argument as in part 1 therefore yields
\begin{eqnarray}\label{proof.q.desmooth.centered}
 &&\sup_{\gamma\in[\delta,1-\delta]}
 \left\|
 \frac1T\sum_{t=1}^{T}
 \pmb\zeta_{t,m}k_{\delta,t}^{(r)}(\gamma)
 \right\|\nonumber\\
 &=&
 O_P\!\left\{
 \sqrt{\frac{\log T}{mT\delta}}
 +\frac{\chi_T\log T}{T\delta}
 \right\}.
\end{eqnarray}
On the event $\max_{t\le T}\|\mathbf x_{t-1}\|\le\chi_T$, \eqref{proof.q.desmooth.mean} and compact support of the kernel give
\begin{eqnarray}\label{proof.q.desmooth.predictable}
 &&\sup_{\gamma\in[\delta,1-\delta]}
 \left\|
 \frac1T\sum_{t=1}^{T}
 E_{t-1}[\mathbf d_{t,m}\otimes\mathbf x_{t-1}]
 k_{\delta,t}^{(r)}(\gamma)
 \right\|
 \le\frac{C\chi_T}{m^2}.
\end{eqnarray}
The bandwidth condition in Assumption \ref{Assumption3} gives
\begin{eqnarray*}
 \frac{\chi_T\sqrt{\log T}}{\sqrt{T\delta}}&\longrightarrow&0,
\end{eqnarray*}
whereas $\sqrt{T\delta}/(m\sqrt{\log T})\to0$. In particular,
\begin{eqnarray*}
 &&\frac{\chi_Tm^{-2}}
 {\sqrt{\log T/(T\delta)}}\nonumber\\
 &=&
 \left\{\frac{\chi_T\sqrt{\log T}}{\sqrt{T\delta}}\right\}
 \left\{\frac{\sqrt{T\delta}}{m\sqrt{\log T}}\right\}^{2}
 =o(1).
\end{eqnarray*}
The other terms on the right-hand sides of \eqref{proof.q.desmooth.centered}-- \eqref{proof.q.desmooth.predictable} are smaller by the same two conditions. Hence the second term in \eqref{proof.q.smooth.decomp} is $o_P\{\sqrt{\log T/(T\delta)}\}$. Combining this conclusion with part 1 proves the asserted uniform order for the smoothed score.

For fixed $\gamma$, martingale orthogonality and the deterministic kernel sum give
\begin{eqnarray*}
 \left\|
 \frac1T\sum_{t=1}^{T}
 \pmb\zeta_{t,m}k_{\delta,t}^{(r)}(\gamma)
 \right\|_2
 &\le&
 C(mT\delta)^{-1/2},
\end{eqnarray*}
whereas \eqref{proof.q.desmooth.mean}, Markov's inequality, and \eqref{proof.q.state.moment} make the corresponding predictable term $O_P(m^{-2})$. If $\sqrt{T\delta}/m\to0$, then $m^{-1/2}=o(1)$ and $\sqrt{T\delta}/m^2=(\sqrt{T\delta}/m)m^{-1}=o(1)$; hence both terms are $o_P((T\delta)^{-1/2})$. This proves the final point-wise statement.
\end{proof}

\begin{proof}[Proof of Lemma \ref{Appendix.B4}]

\item 

(a-b) Under Assumption \ref{Assumption4}, we write $\mathbf e_{s}=G_e(\gamma_s,\mathcal E_s)$. By the definition of $\mathbf y_t^*$ in Proposition \ref{prop.1},
\begin{eqnarray*}
 \mathbf y_t^*-
 \tilde{\mathbf y}_t(\gamma_t)
 &=& 
 \sum_{\ell=0}^{\infty}
 \mathbf c_\ell(\gamma_t)
 \{G_e(\gamma_{t-\ell},\mathcal E_{t-\ell})
 -G_e(\gamma_t,\mathcal E_{t-\ell})\}.
\end{eqnarray*}
Minkowski's inequality, Assumption \ref{Assumption4}, and the geometric bound on $\mathbf c_\ell(\gamma_t)$ give the following result that holds uniformly in $t$:
\begin{eqnarray*}
 \|\mathbf y_t^*-
 \tilde{\mathbf y}_t(\gamma_t)\|_\nu
 &\le&
 C\sum_{\ell=0}^{\infty}
 \tilde{\rho}_y^\ell|\gamma_{t-\ell}-\gamma_t|
\le\frac{C}{T}
 \sum_{\ell=0}^{\infty}\ell\tilde{\rho}_y^\ell
 =O(T^{-1}).
\end{eqnarray*}
This proves Lemma \ref{Appendix.B4}.(a).  Together with Proposition \ref{prop.1}, it further  proves Lemma \ref{Appendix.B4}.(b).

(c-d) Adding and subtracting $\sum_{\ell\ge0}\mathbf c_\ell(\gamma') \tilde{\mathbf e}_{t-\ell}(\gamma)$ yields
\begin{eqnarray*}
 \tilde{\mathbf y}_t(\gamma)
 -\tilde{\mathbf y}_t(\gamma')
 &=&
 \sum_{\ell=0}^{\infty}
 \{\mathbf c_\ell(\gamma)-\mathbf c_\ell(\gamma')\}
 \tilde{\mathbf e}_{t-\ell}(\gamma) +
 \sum_{\ell=0}^{\infty}
 \mathbf c_\ell(\gamma')
 \{\tilde{\mathbf e}_{t-\ell}(\gamma)
 -\tilde{\mathbf e}_{t-\ell}(\gamma')\}.
\end{eqnarray*}
Consequently, Assumption \ref{Assumption4} and the fact  $\|\mathbf c_\ell(\gamma)-\mathbf c_\ell(\gamma')\|_2 \le C\ell\tilde{\rho}_y^{\ell-1}|\gamma-\gamma'|$ imply
\begin{eqnarray*}
 \|\tilde{\mathbf y}_t(\gamma)
 -\tilde{\mathbf y}_t(\gamma')\|_\nu
 &\le&
 C|\gamma-\gamma'|
 \left\{
 \sum_{\ell=1}^{\infty}\ell\tilde{\rho}_y^{\ell-1}
 +\sum_{\ell=0}^{\infty}\tilde{\rho}_y^\ell
 \right\}\le C|\gamma-\gamma'|,
\end{eqnarray*}
which proves Lemma \ref{Appendix.B4}.(c). 

Finally, by Assumption \ref{Assumption4},
\begin{eqnarray*}
 \|\tilde{\mathbf y}_t(\gamma)
 -\tilde{\mathbf y}_t^{*(k)}(\gamma)\|_\nu
 &\le&
 \sum_{\ell=0}^{k}
 \|\mathbf c_\ell(\gamma)\|_2
 \|\tilde{\mathbf e}_{t-\ell}(\gamma)
 -\tilde{\mathbf e}_{t-\ell}^{*(k-\ell)}(\gamma)\|_\nu\nonumber\\
 &\le&
 C\sum_{\ell=0}^{k}
 \tilde{\rho}_y^\ell\rho_e^{k-\ell}
 \le C(k+1)\{\max(\tilde{\rho}_y,\rho_e)\}^{k}.
\end{eqnarray*}
For some $\rho_x\in(\max\{\tilde{\rho}_y,\rho_e\},1)$, its order is bounded by $C\rho_x^k$.  This completes the proof of Lemma \ref{Appendix.B4}.(d).
\end{proof}

\begin{proof}[Proof of Lemma \ref{Appendix.B5}]

Let $\mathbf a_{\ell i,t+h}$ and $\mathbf a_{\ell i}(\gamma)$ be $p\times1$ coefficient vectors on $\mathbf y_{t+h-\ell}$ and $\tilde{\mathbf y}_{t+h-\ell}(\gamma)$, respectively. Define
\begin{eqnarray*}
 R^{(h)}_{i,t,T}(\mathbf a)
 &\coloneqq&
 \sum_{\ell=1}^{q}
 \mathbf a_{\ell i,t+h}^{\top}\mathbf y_{t+h-\ell}
 -\mathbf x_t^{[j]\top}\mathbf a,\nonumber\\
 \tilde R^{(h)}_{i,t}(\mathbf a;\gamma)
 &\coloneqq&
 \sum_{\ell=1}^{q}
 \mathbf a_{\ell i}(\gamma)^{\top}
 \tilde{\mathbf y}_{t+h-\ell}(\gamma)
 -\tilde{\mathbf x}_t^{[j]}(\gamma)^\top\mathbf a.
\end{eqnarray*}
Then, it is clear to see that 
\begin{eqnarray*}
 v^{(h)}_{i,t,T}(\mathbf a)
 &=&
 R^{(h)}_{i,t,T}(\mathbf a)+e_{i,t+h},\nonumber\\
 \tilde v^{(h)}_{i,t}(\mathbf a;\gamma)
 &=&
 \tilde R^{(h)}_{i,t}(\mathbf a;\gamma)
 +\tilde e_{i,t+h}(\gamma).
\end{eqnarray*}

Conditional on $\mathcal F_{t+h-1}$, the predictable part is fixed. Therefore,
\begin{eqnarray*}
 &&P\!\left(
 v^{(h)}_{i,t,T}(\mathbf a)\in[b-\epsilon,b+\epsilon]
 \,\middle|\,\mathcal F_{t+h-1}
 \right)\nonumber\\
 &=&
 P\!\left(
 e_{i,t+h}\in
 [b-\epsilon-R^{(h)}_{i,t,T}(\mathbf a),
 b+\epsilon-R^{(h)}_{i,t,T}(\mathbf a)]
 \,\middle|\,\mathcal F_{t+h-1}
 \right)\nonumber\\
 &=&
 \int_{b-\epsilon-R^{(h)}_{i,t,T}(\mathbf a)}
 ^{b+\epsilon-R^{(h)}_{i,t,T}(\mathbf a)}
 f_{i,t+h}(w\mid\mathcal F_{t+h-1})\,\mathrm dw\nonumber\\
 &\le&2\bar C_f\epsilon,
\end{eqnarray*}
where the last inequality follows from the bounded conditional density in Assumption \ref{Assumption1}. Since $\sigma(\mathbf x_t^{[j]})\subseteq\mathcal F_{t+h-1}$, the law of iterated expectations gives the first result in Lemma \ref{Appendix.B5}. Conditioning instead on $\mathcal H_{t+h-1}$,  and repeating the integral with $d_{i,t+h}(w;\gamma)$ in place of $f_{i,t+h}(w\mid\mathcal F_{t+h-1})$ gives the same bound for the stationary approximation.  Since $\sigma(\tilde{\mathbf x}_t^{[j]}(\gamma)) \subseteq\mathcal H_{t+h-1}$, another application of iterated expectations gives the second result. Therefore, Lemma \ref{Appendix.B5} holds.
\end{proof}

\begin{proof}[Proof of Lemma \ref{Appendix.B6}]

    \item 

    Write $\overline u_{i,t}^{(h)} \coloneqq \tilde y_{i,t+h}(\gamma_t) -\tilde{\mathbf x}_t^{[j]}(\gamma_t)^\top \pmb\alpha^{(h)}_{ij}(\gamma_t).$ The definition \eqref{def.alpha.proj} and the subgradient first-order condition give
\begin{eqnarray*}
 E\!\left[
 \psi_\tau(\overline u_{i,t}^{(h)})
 \tilde{\mathbf x}_t^{[j]}(\gamma_t)
 \right]
 &=&\mathbf0.
\end{eqnarray*}
By Lemma \ref{Appendix.B4}.(b), we have 
\begin{eqnarray}\label{lp.score.center.coupling}
 \|u^{(h)}_{i,t}-\overline u_{i,t}^{(h)}\|_\nu
 +\|\mathbf x_t^{[j]}
 -\tilde{\mathbf x}_t^{[j]}(\gamma_t)\|_\nu
 &\le&\frac{C}{T}.
\end{eqnarray}
Moreover, applying  H\"older's inequality, Lemma \ref{Appendix.B5}, and Markov's inequality gives
\begin{eqnarray*}
 &&E\!\left[
 \|\tilde{\mathbf x}_t^{[j]}(\gamma_t)\|
 \left|
 \psi_\tau(u^{(h)}_{i,t})
 -\psi_\tau(\overline u_{i,t}^{(h)})
 \right|
 \right]\nonumber\\
 &\le&
 E\!\left[
 \|\tilde{\mathbf x}_t^{[j]}(\gamma_t)\|
 \mathbf1\{|\overline u_{i,t}^{(h)}|\le\epsilon\}
 \right]\nonumber\\
 &&+
 \|\tilde{\mathbf x}_t^{[j]}(\gamma_t)\|_\nu
 P\!\left(
 |u^{(h)}_{i,t}-\overline u_{i,t}^{(h)}|>\epsilon
 \right)^{(\nu-1)/\nu}\nonumber\\
 &\le&
 C\epsilon
 +CT^{-(\nu-1)}\epsilon^{-(\nu-1)}.
\end{eqnarray*}
Taking $\epsilon=T^{-(\nu-1)/\nu}$ and using \eqref{lp.score.center.coupling} once more for the design difference proves the desired result in Lemma \ref{Appendix.B6}.
\end{proof}

\begin{proof}[Proof of Lemma \ref{Appendix.B7}]
\item

For notational simplicity, suppress $(i,j,h,\gamma)$ and write $u_t$, $\tilde u_t$, $\mathbf x_t$, $\tilde{\mathbf x}_t$ and the corresponding coupled version $u_t^{*(k)}$, $\tilde u_t^{*(k)}$,  $\mathbf x_t^{*(k)}$, $\tilde{\mathbf x}_t^{*(k)}$. By Lemma \ref{Appendix.B11}, there exists a scalar $\rho_0\in(0,1)$ such that
\begin{eqnarray*}
 \|y_{i,t+h}-y_{i,t+h}^{*(k)}\|_4
 &\le&C\rho_0^k,
\end{eqnarray*}
and 
\begin{eqnarray*}
 \|\mathbf x_t-\mathbf x_t^{*(k)}\|_4
 \le C\mathbf1\{k=h\}+C\rho_0^{k-h}\mathbf1\{k>h\}.
\end{eqnarray*}
They jointly yield
\begin{eqnarray*}
 \|u_t-u_t^{*(k)}\|_4
 +\|\mathbf x_t-\mathbf x_t^{*(k)}\|_4
 &\le&
 C\mathbf1\{k\le h\}
 +C\rho_0^{k-h}\mathbf1\{k>h\}.
\end{eqnarray*}
For any $\eta>0$,
\begin{eqnarray*}
 P\!\left(
 \psi_\tau(u_t)
 \ne\psi_\tau(u_t^{*(k)})
 \right)
 &\le&
 P(|u_t|\le\eta)
 +P(|u_t-u_t^{*(k)}|>\eta)\nonumber\\
 &\le&
 2\bar C_f\eta
 +\eta^{-4}
 \|u_t-u_t^{*(k)}\|_4^4,
\end{eqnarray*}
where the last line follows from Lemma \ref{Appendix.B5} and Markov's inequality.  For $k>h$, take $\eta=\|u_t-u_t^{*(k)}\|_4^{4/5}$.  Then
\begin{eqnarray}\label{lp.indicator.coupling}
 \left\|
 \psi_\tau(u_t)
 -\psi_\tau(u_t^{*(k)})
 \right\|_4
 &\le&
 C\|u_t-u_t^{*(k)}\|_4^{1/5}
 \le C\rho_0^{(k-h)/5}.
\end{eqnarray}
Since $|\psi_\tau(u)|\le1$, the triangle and H\"older inequalities yield
\begin{eqnarray*}
 &&\left\|
 \psi_\tau(u_t)\mathbf x_t
 -\psi_\tau(u_t^{*(k)})
 \mathbf x_t^{*(k)}
 \right\|_2\nonumber\\
 &\le&
 \|\mathbf x_t-
 \mathbf x_t^{*(k)}\|_2
 +\left\|
 \psi_\tau(u_t)
 -\psi_\tau(u_t^{*(k)})
 \right\|_4
 \|\mathbf x_t^{*(k)}\|_4.
\end{eqnarray*}
For $k\le h$, the same display, $|\psi_\tau|\le1$, and the uniform fourth-moment bound give a finite constant.  For $k>h$, \eqref{lp.indicator.coupling}  gives $C\rho_0^{(k-h)/5}$.  Therefore, with $\varrho=\rho_0^{1/5}$,
\begin{eqnarray*}
 \mathfrak d^{A}_{2,s,T}(k;\gamma)
 &\le&
 C\mathbf 1\{k\le h\}
 +C\varrho^{\,k-h}\mathbf 1\{k>h\}.
\end{eqnarray*}
Similarly, we can show that  
\begin{eqnarray*}
\mathfrak d^{F}_{2,s}(k;\gamma)
 &\le&
 C\mathbf 1\{k\le h\}
 +C\varrho^{\,k-h}\mathbf 1\{k>h\},
\end{eqnarray*}
which completes the proof of \eqref{lp.score.fd}.

It remains to prove \eqref{lp.cov.summable}.  In the rest of the proof, we abbreviate $\tilde{\mathbf s}_t \coloneqq \tilde{\mathbf s}^{(h)}_{ij,t}(\gamma)$. Define the martingale projection operator $\mathcal P_v Z \coloneqq E[Z\mid \mathcal H_v]-E[Z\mid \mathcal H_{v-1}]$. The population first-order condition gives $E[\tilde{\mathbf s}_t]=\mathbf0$.  Since $\tilde{\mathbf s}_t$ is measurable with respect to $\mathcal H_{t+h}$, its backward martingale projection decomposition is
\begin{eqnarray}\label{proof.lp.score.projection}
 \tilde{\mathbf s}_t
 &=&
 \sum_{k=0}^{\infty}
 \mathcal P_{t+h-k}\tilde{\mathbf s}_t.
\end{eqnarray}
By the coupling construction, $\tilde{\mathbf s}^{*(h,k)}_{ij,t}(\gamma)$ does not depend on $\pmb\eta_{t+h-k}$.  Hence
\begin{eqnarray}\label{proof.lp.score.projection.coupling}
 \left\|
 \mathcal P_{t+h-k}\tilde{\mathbf s}_t
 \right\|_2
 &=&
 \left\|
 \mathcal P_{t+h-k}
 \left\{
 \tilde{\mathbf s}^{(h)}_{ij,t}(\gamma)
 -\tilde{\mathbf s}^{*(h,k)}_{ij,t}(\gamma)
 \right\}
 \right\|_2\nonumber\\
 &\le&
 \left\|
 \tilde{\mathbf s}^{(h)}_{ij,t}(\gamma)
 -\tilde{\mathbf s}^{*(h,k)}_{ij,t}(\gamma)
 \right\|_2
 \le\mathfrak d^{F}_{2,s}(k;\gamma).
\end{eqnarray}

Consider first $r\ge0$.  Applying the decomposition \eqref{proof.lp.score.projection} to both scores gives
\begin{eqnarray*}
 \operatorname{Cov}
 \left(
 \tilde{\mathbf s}_t,
 \tilde{\mathbf s}_{t+r}
 \right)
 &=&
 \sum_{k=0}^{\infty}\sum_{\ell=0}^{\infty}
 E\!\left[
 \left(
 \mathcal P_{t+h-k}\tilde{\mathbf s}_t
 \right)
 \left(
 \mathcal P_{t+r+h-\ell}\tilde{\mathbf s}_{t+r}
 \right)^\top
 \right]
 \nonumber\\
 &=&
 \sum_{k=0}^{\infty}
 E\!\left[
 \left(
 \mathcal P_{t+h-k}\tilde{\mathbf s}_t
 \right)
 \left(
 \mathcal P_{t+r+h-(k+r)}
 \tilde{\mathbf s}_{t+r}
 \right)^\top
 \right]
 \nonumber\\
  &\le&
 \sum_{k=0}^{\infty}
 \left\|
 \mathcal P_{t+h-k}\tilde{\mathbf s}_t
 \right\|_2
 \left\|
 \mathcal P_{t+r+h-(k+r)}
 \tilde{\mathbf s}_{t+r}
 \right\|_2\nonumber\\
 &\le&
 \sum_{k=0}^{\infty}
 \mathfrak d^{F}_{2,s}(k;\gamma)\mathfrak d^{F}_{2,s}(k+r;\gamma),
\end{eqnarray*}
where the first inequality uses Cauchy--Schwarz inequality and the second applies \eqref{proof.lp.score.projection.coupling}.

For $r<0$, interchanging the two scores and taking the transpose gives the same bound with $r$ replaced by $|r|$.  Hence
\begin{eqnarray}\label{proof.lp.score.cov.fd}
 \left\|
 \operatorname{Cov}
 \left(
 \tilde{\mathbf s}_t,
 \tilde{\mathbf s}_{t+r}
 \right)
 \right\|
 &\le&
 C\sum_{k=0}^{\infty}
 \mathfrak d^{F}_{2,s}(k;\gamma)\mathfrak d^{F}_{2,s}(k+|r|;\gamma).
\end{eqnarray}

By \eqref{lp.score.fd}, $\mathfrak d^{F}_{2,s}(k;\gamma)\le C\varrho^{(k-h)_+}$.  If $|r|\le h$, the right-hand side of \eqref{proof.lp.score.cov.fd} is bounded by a finite constant.  If $|r|>h$, then
\begin{eqnarray*}
 \sum_{k=0}^{\infty}
 \mathfrak d^{F}_{2,s}(k;\gamma)\mathfrak d^{F}_{2,s}(k+|r|;\gamma)&\le&
 C\sum_{k=0}^{\infty}
 \varrho^{(k-h)_+}
 \varrho^{k+|r|-h}\nonumber\\
 &=&
 C\varrho^{|r|-h}
 \sum_{k=0}^{\infty}
 \varrho^{(k-h)_++k}
 \le C\varrho^{|r|-h}.
\end{eqnarray*}
Combining the two cases gives
\begin{eqnarray*}
 \left\|
 \operatorname{Cov}
 \left(
 \tilde{\mathbf s}_t,
 \tilde{\mathbf s}_{t+r}
 \right)
 \right\|
 &\le&
 C\sum_{k=0}^{\infty}
 \mathfrak d^{F}_{2,s}(k;\gamma)\mathfrak d^{F}_{2,s}(k+|r|;\gamma)
 \le C\varrho^{(|r|-h)_+},
\end{eqnarray*}
and
\begin{eqnarray*}
 \sum_{r=-\infty}^{\infty}
 \left\|
 \operatorname{Cov}
 \left(
 \tilde{\mathbf s}_t,
 \tilde{\mathbf s}_{t+r}
 \right)
 \right\|
 &\le&
 C(2h+1)+2C\sum_{r=h+1}^{\infty}\varrho^{r-h}
 <\infty,
\end{eqnarray*}
which proves \eqref{lp.cov.summable}. It completest the proof of Lemma  \ref{Appendix.B7}.
\end{proof}

\begin{proof}[Proof of Lemma \ref{Appendix.B8}]

    \item 

    Write $\mathbf Z_{t,T} \coloneqq \mathbf x_t^{[j]}\mathbf x_t^{[j]\top}$ and $ \tilde{\mathbf Z}_t(\gamma) \coloneqq \tilde{\mathbf x}_t^{[j]}(\gamma) \tilde{\mathbf x}_t^{[j]}(\gamma)^\top.$ By Lemma \ref{Appendix.B4}.(b) and H\"older's inequality, we have
\begin{eqnarray*}
 \max_{t\le T-h}
 \|\mathbf Z_{t,T}-\tilde{\mathbf Z}_t(\gamma_t)\|_2
 &\le&
 \|\mathbf x_t^{[j]}
 -\tilde{\mathbf x}_t^{[j]}(\gamma_t)\|_4
 \{\|\mathbf x_t^{[j]}\|_4+
 \|\tilde{\mathbf x}_t^{[j]}(\gamma_t)\|_4\}
 \le\frac{C}{T}.
\end{eqnarray*}
By Jensen inequality, we have
\begin{eqnarray}\label{proof.lp.M.array_2}
\left\|
E[\mathbf Z_{t,T}]
-
E[\tilde{\mathbf Z}_t(\gamma_t)]
\right\|
\le \frac{C}{T}.
\end{eqnarray}

Additionally, Lemma \ref{Appendix.B4}.(c) gives
\begin{eqnarray*}
 \|\tilde{\mathbf Z}_t(u)
 -\tilde{\mathbf Z}_t(u')\|_2
 &\le&C|u-u'|,
\end{eqnarray*}
which implies that 
\begin{eqnarray}\label{proof.lp.M.gamma_2}
\left\|    
E[\tilde{\mathbf Z}_t(\gamma_t)]
-
E[\tilde{\mathbf Z}_t(\gamma)]
\right\|
\le C|\gamma_t-\gamma|.
\end{eqnarray}

By \eqref{proof.lp.M.array_2}, \eqref{proof.lp.M.gamma_2} and the property of Riemann integral, we obtain
\begin{eqnarray*}
 \frac1T\sum_{t=1}^{T-h}
 E[\mathbf Z_{t,T}]k_{\delta,t}^{(n)}(\gamma)
 &=&
 \frac1T\sum_{t=1}^{T-h}
 E[\tilde{\mathbf Z}_t(\gamma)]
 k_{\delta,t}^{(n)}(\gamma)
 +O\left(\delta+\frac{1}{T\delta}\right)\nonumber\\
 &=&
 v_n\pmb\Sigma_\gamma^{[j]}
 +O\left(\delta+\frac{1}{T\delta}\right).
\end{eqnarray*}

It remains to control the centred part.  Let $\mathbf Z_{t,T}^{*(k)}$ be the coupled version  of $\mathbf Z_{t,T}$ by replacing $\pmb\eta_{t-k}$ with its independent copy. Directly applying Lemma \ref{Appendix.B11} gives
\begin{eqnarray*}
 \max_{t}
 \|\mathbf Z_{t,T}-\mathbf Z_{t,T}^{*(k)}\|_2
 &\le&C\rho_Z^k,
\end{eqnarray*}
for some $\rho_Z\in(0,1)$.

Let $\mathcal P_rZ=E[Z\mid\mathcal H_r]-E[Z\mid\mathcal H_{r-1}]$. Using the martingale projection decomposition and Minkowski inequality, we obtain
\begin{eqnarray*}
 \left\|
 \frac1T\sum_{t=1}^{T-h}
 \{\mathbf Z_{t,T}-E\mathbf Z_{t,T}\}
 k_{\delta,t}^{(n)}(\gamma)
 \right\|_2
 &\le&
 \frac1T
 \left\{
 \sum_{t=1}^{T-h}
 |k_{\delta,t}^{(n)}(\gamma)|^2
 \right\}^{1/2}
 \sum_{k=0}^{\infty}
 \max_{t}
 \|\mathbf Z_{t,T}-\mathbf Z_{t,T}^{*(k)}\|_2\le C(T\delta)^{-1/2},
\end{eqnarray*}
which completes the proof.
\end{proof}

\begin{proof}[Proof of Lemma \ref{Appendix.B9}:]
\item

For notational simplicity, abbreviate $u=u^{(h)}_{i,t}$ and $g_t(r\mid\mathbf x_t^{[j]}) \coloneqq g^{(h)}_{ij,t,T}(r\mid\mathbf x_t^{[j]})$.  Lemma \ref{Appendix.B5} gives
\begin{eqnarray}\label{proof.lp.C.global}
 \sup_{r\in\mathbb R}
 g_t(r\mid\mathbf x_t^{[j]})
 &\le&\bar C_f.
\end{eqnarray}
We write
\begin{eqnarray*}
 \rho_{\tau,m}^{(1)}(u+w)
 -\rho_{\tau,m}^{(1)}(u)
 &=&
 \int
 \{\psi_\tau(u+w+x)-\psi_\tau(u+x)\}
 \phi_m(x)\mathrm dx.
\end{eqnarray*}
Conditional Jensen's inequality and \eqref{proof.lp.C.global} imply
\begin{eqnarray*}
 E[
 \{\rho_{\tau,m}^{(1)}(u+w)
 -\rho_{\tau,m}^{(1)}(u)\}^2
 \mid\mathbf x_t^{[j]}]
 &\le&
 \int
 P(-x-w<u\le-x\mid\mathbf x_t^{[j]})
 \phi_m(x)\mathrm dx\nonumber\\
 &\le&
 \bar C_f|w|\int\phi_m(x)\mathrm dx
 =\bar C_f|w|.
\end{eqnarray*}
This proves part (a).  Using similar calculations, we can prove part (b).

For parts (c)--(d), define the smoothed conditional residual density
\begin{eqnarray*}
 g_{t,m}(r\mid\mathbf x_t^{[j]})
 &\coloneqq&
 \int g_t(r-z\mid\mathbf x_t^{[j]})\phi_m(z)\mathrm dz.
\end{eqnarray*}
The Lipschitz condition in Assumption \ref{Assumption5} yields
\begin{eqnarray}\label{proof.lp.C.local}
 |g_{t,m}(r\mid\mathbf x_t^{[j]})-f_t^*|
 &\le&C(|r|+m^{-1}).
\end{eqnarray}
Simple algebra gives
\begin{eqnarray*}
 &&E[\rho_{\tau,m}^{(1)}(u+w)
 -\rho_{\tau,m}^{(1)}(u)
 \mid\mathbf x_t^{[j]}]
 =
 \int_0^w
 g_{t,m}(-r\mid\mathbf x_t^{[j]})\mathrm dr.
\end{eqnarray*}
Substitution of \eqref{proof.lp.C.local} gives
\begin{eqnarray*}
 \int_0^w
 g_{t,m}(-r\mid\mathbf x_t^{[j]})\mathrm dr
 &=&
 f_t^*w+O(w^2+|w|/m).
\end{eqnarray*}
This proves part (c).  Integrating part (c) once more,

\begin{eqnarray*}
 E[\rho_{\tau,m}(u+\xi)-\rho_{\tau,m}(u)
 -\rho_{\tau,m}^{(1)}(u)\xi
 \mid\mathbf x_t^{[j]}]
 &=&
 \int_0^\xi
 E[\rho_{\tau,m}^{(1)}(u+w)
 -\rho_{\tau,m}^{(1)}(u)
 \mid\mathbf x_t^{[j]}]\mathrm dw\nonumber\\
 &=&
 \frac12f_t^*\xi^2
 +O(|\xi|^3+\xi^2/m),
\end{eqnarray*}
which proves part (d).  
\end{proof}

\begin{proof}[Proof of Lemma \ref{Appendix.B10}]

\item 

Let $\mathcal P_sZ=E[Z\mid\mathcal H_s]-E[Z\mid\mathcal H_{s-1}]$. Because $\pmb\Xi_{t,T}$ is centred and measurable with respect to $\mathcal H_{t+\kappa_T}$, the backward martingale-projection decomposition gives
\begin{eqnarray*}
 \pmb\Xi_{t,T}
 &=&
 \sum_{k=0}^{\infty}
 \mathcal P_{t+\kappa_T-k}\pmb\Xi_{t,T}.
\end{eqnarray*}
The coupled variable does not depend on $\pmb\eta_{t+\kappa_T-k}$, so that $\mathcal P_{t+\kappa_T-k}\pmb\Xi_{t,T}^{*(k)}=\mathbf0$. Consequently,
\begin{eqnarray*}
 \left\|
 \mathcal P_{t+\kappa_T-k}\pmb\Xi_{t,T}
 \right\|_r
 &=&
 \left\|
 \mathcal P_{t+\kappa_T-k}
 \{\pmb\Xi_{t,T}-\pmb\Xi_{t,T}^{*(k)}\}
 \right\|_r\le
 \|\pmb\Xi_{t,T}-\pmb\Xi_{t,T}^{*(k)}\|_r
 \le\mathfrak d_{r,\Xi}(k).
\end{eqnarray*}
For each fixed $k$, the variables $\{\mathcal P_{t+\kappa_T-k}\pmb\Xi_{t,T}\}_t$ form a martingale-difference array.  The martingale von Bahr-Esseen inequality therefore gives, for $1<r\le2$,
\begin{eqnarray*}
 \left\|
 \sum_ta_{t,T}
 \mathcal P_{t+\kappa_T-k}\pmb\Xi_{t,T}
 \right\|_r
 &\le&
 C_r\left\{
 \sum_t|a_{t,T}|^r
 \left\|
 \mathcal P_{t+\kappa_T-k}\pmb\Xi_{t,T}
 \right\|_r^r
 \right\}^{1/r}\nonumber\\
 &\le&
 C_r\left(\sum_t|a_{t,T}|^r\right)^{1/r}
 \mathfrak d_{r,\Xi}(k).
\end{eqnarray*}
Minkowski's inequality over $k$ proves \eqref{lp.projection.Lr}. When $r=2$, orthogonality of the martingale differences gives the first inequality in the preceding display with $C_2=1$.  Squaring the resulting bound proves \eqref{lp.projection.L2}. This completes the proof of Lemma \ref{Appendix.B10}.
\end{proof}

\begin{proof}[Proof of Lemma \ref{Appendix.B11}]
\item

Let $\mathbf e_{v,T}^{[r]}$ be obtained by using the same replacement of $\pmb\eta_r$.  If $v<r$, we have $\mathbf e_{v,T}=\mathbf e_{v,T}^{[r]}$.  If $v\ge r$, the replacement occurs at lag $v-r$ in $\mathcal E_v$.  Since $\nu\ge4$, Assumption \ref{Assumption4}.2 gives
\begin{eqnarray}\label{proof.array.e.fd}
 \|\mathbf e_{v,T}-\mathbf e_{v,T}^{[r]}\|_4
 &\le&
 \|\mathbf e_{v,T}-\mathbf e_{v,T}^{[r]}\|_\nu
 \le C\rho_e^{v-r}.
\end{eqnarray}
From  \eqref{def.yt.ma}, it is clear to see that 
\begin{eqnarray*}
 \mathbf y_{s,T}-\mathbf y_{s,T}^{[r]}
 &=&
 \sum_{\ell=0}^{d}\mathbf c_{\ell,s}
 \{\mathbf e_{s-\ell,T}-\mathbf e_{s-\ell,T}^{[r]}\}.
\end{eqnarray*}
Note that in Lemma \ref{Appendix.B1} we have shown that  $\|\mathbf c_{\ell,s}\|_2\le C\tilde{\rho}_y^\ell$.  Let $d=s-r\ge0$. Minkowski's inequality and \eqref{proof.array.e.fd} therefore imply
\begin{eqnarray*}
 \|\mathbf y_{s,T}-\mathbf y_{s,T}^{[r]}\|_4
 &\le&
 C\sum_{\ell=0}^{d}\tilde{\rho}_y^\ell\rho_e^{d-\ell}.
\end{eqnarray*}
Let $\bar\rho=\max\{\tilde{\rho}_y,\rho_e\}$ and choose $\rho_0\in(\bar\rho,1)$.  We have
\begin{eqnarray*}
 \sum_{\ell=0}^{d}\tilde{\rho}_y^\ell\rho_e^{d-\ell}
 &\le&(d+1)\bar\rho^d
 \le C\rho_0^d,
\end{eqnarray*}
Therefore, Lemma \ref{Appendix.B11} follows.
\end{proof}

\begin{proof}[Proof of Lemma \ref{ap.tm5c.lemD}]
\item    

    Lemma \ref{ap.tm5c.lemD} can be proved by applying Lemma \ref{Appendix.B10} with $r=2$ to $\vartheta_t$. Specifically, in this application, define
\begin{eqnarray*}
\mathfrak d_{2,\vartheta}(k)
 &\coloneqq&
 \max_{\substack{1\le t\le T-h:\\K_t(\gamma)\ne0}}
 \|\vartheta_t-\vartheta_t^{*(k)}\|_2,
\end{eqnarray*}
where $\vartheta_t^{*(k)}$ is the coupled version of $\vartheta_t$ by replacing $\pmb\eta_{t+h-k}$ with its independent copy.

By Cauchy-Schwarz, iterated expectations, and Lemma \ref{Appendix.B9},
\begin{eqnarray*}
    &&E[\vartheta_t^2]
    ={\rm Var}(G_t)
    \nonumber\\
    &\leq& E\!\left[
    \left(\int_0^{\Delta\bar y_t}
    \{\rho^{(1)}_{\tau,m}(u^{(h)}_{i,t}+w)
    -\rho^{(1)}_{\tau,m}(u^{(h)}_{i,t})\}\mathrm dw
    \right)^2
    \right]
     \nonumber\\
    &\leq& E\!\left[
    |\Delta\bar y_t|
    \int_0^{|\Delta\bar y_t|}
    E\!\left[
    \{\rho^{(1)}_{\tau,m}(u^{(h)}_{i,t}
    +\operatorname{sgn}(\Delta\bar y_t)w)
    -\rho^{(1)}_{\tau,m}(u^{(h)}_{i,t})\}^2
    \,\middle|\,\mathbf x_t^{[j]}
    \right]\mathrm dw
    \right]
     \nonumber\\
    &\leq&C E|\Delta\bar y_t|^3.
\end{eqnarray*}
To formulate the upper bound of $E[|\Delta\bar y_t|^3]$, we consider the following Taylor expansion: $\Delta\bar y_t = \mathbf x_t^{[j]\top}\big(\pmb{\alpha}(\gamma)- \mathbf{a}^0+(\pmb{\alpha}^{(1)}(\gamma)- \mathbf{a}^1)(\gamma_t-\gamma)+\frac{1}{2}\pmb{\alpha}^{(2)}(\tilde{\gamma}_t)(\gamma_t-\gamma)^2\big)$, where $\tilde{\gamma}_t$ lies between $\gamma$ and $\gamma_t$. On the set $\mathcal{W}\coloneqq\{t:|\gamma_t-\gamma|\le\delta\}$, we always have
\begin{eqnarray}\label{proof.lp.D.dy}
  |\Delta\bar y_t   |\leq \| \mathbf x_t^{[j]}\|\big(\|\pmb{\alpha}(\gamma)-\mathbf{a}^0\|+\delta\|\pmb{\alpha}^{(1)}(\gamma)-\mathbf{a}^1\| +C\delta^2\big),
\end{eqnarray}
which immediately yields
 \begin{eqnarray*}
  E\big[|\Delta\bar y_t   |^3\big]\leq CE[\|\mathbf x_t^{[j]}\|^3]\big(\|\Delta\mathbf{a}_\delta\|^3+\delta^6\big).
\end{eqnarray*}  
It implies that 
\begin{eqnarray}\label{proof.lp.D.envelope}
 \|\vartheta_t\|_2^2
 &\le&
 CE|\Delta\bar y_t|^3
 \le Cr_{\delta,a}^3.
\end{eqnarray}
We next derive the functional-dependence coefficient.  For notational simplicity, we further abbreviate
\begin{eqnarray*}
 r&\coloneqq&r_{\delta,a},\qquad
 d_t\coloneqq\Delta\bar y_t,\qquad
 u_t\coloneqq u^{(h)}_{i,t}.
\end{eqnarray*}
Let $\mathbf y_{t+h,T}^{*(k)}$, $\mathbf x_t^{[j],*(k)}$,  $d_t^{*(k)}$ and $ u_t^{*(k)}$, be the coupled version of $\mathbf y_{t+h,T}$, $\mathbf x_t^{[j]}$, $d_t$ and $ u_t$. Additionally, for any $s\in[0,1]$, define
\begin{eqnarray*}
 z_{t,s}&\coloneqq&u_t+sd_t,\qquad
 z_{t,s}^{*(k)}
 \coloneqq u_t^{*(k)}+sd_t^{*(k)}.
\end{eqnarray*}

By \eqref{proof.lp.D.dy}, we have $\|d_t\|_4+\|d_t^{*(k)}\|_4\le Cr$ Directly applying Lemma \ref{Appendix.B11}, we obtain
\begin{eqnarray*}
 \|\mathbf y_{t+h,T}-\mathbf y_{t+h,T}^{*(k)}\|_4
 &\le&C\rho_0^k,\quad
 \|\mathbf x_t^{[j]}-\mathbf x_t^{[j],*(k)}\|_4
 \le C\rho_0^{(k-h)_+},
\end{eqnarray*}
which implies that 
\begin{eqnarray}\label{proof.lp.D.primitive.coupling}
 \|d_t-d_t^{*(k)}\|_4
 &\le&Cr\rho_0^{(k-h)_+},\quad
 \|u_t-u_t^{*(k)}\|_4
 \le C\rho_0^{(k-h)_+}.
\end{eqnarray}

Using \eqref{proof.lp.D.primitive.coupling} and the arguments that are similar to those in the proof of \eqref{lp.indicator.coupling}, we can show that 
\begin{eqnarray}\label{proof.lp.D.geometric.1}
    \|\rho_{\tau,m}^{(1)}(u_t)-\rho_{\tau,m}^{(1)}(u_t^{*(k)})\|_4 &\le&C\|u_t-u_t^{*(k)}\|_4^{1/5}
    \le\rho_0^{(k-h)_+/5}.
\end{eqnarray}
and
\begin{eqnarray}\label{proof.lp.D.geometric.2}
 \sup_{0\le s\le1}
 \|\rho_{\tau,m}^{(1)}(z_{t,s})-\rho_{\tau,m}^{(1)}(z_{t,s}^{*(k)})\|_4
 &\le&
 C\sup_{0\le s\le1}
 \|z_{t,s}-z_{t,s}^{*(k)}\|_4^{1/5}
 \le C\rho_0^{(k-h)_+/5}.
\end{eqnarray}
After the change of variables $w=sd_t$, we rewrite
\begin{eqnarray*}
     G_t&=&\int_0^{ d_t} \{q_{m}(u_t+w) -q_{m}(u_{t})\}\mathrm dw
     =d_t\int_0^1
 \{\rho_{\tau,m}^{(1)}(u_t+sd_t)-\rho_{\tau,m}^{(1)}(u_t)\}\mathrm ds
\end{eqnarray*}
Let $G_t^{*(k)}$ be the coupled version of $G_t$. We have the following expansion 
\begin{eqnarray*}
 |G_t-G_t^{*(k)}|
 &\le&
 |d_t-d_t^{*(k)}|\int_0^1 |
 \rho_{\tau,m}^{(1)}(u_t+sd_t)-\rho_{\tau,m}^{(1)}(u_t)|\mathrm ds
 \nonumber\\
 &&+|d_t^{*(k)}|\int_0^1
 |(\rho_{\tau,m}^{(1)}(z_{t,s})-\rho_{\tau,m}^{(1)}(u_t)) - (\rho_{\tau,m}^{(1)}(z_{t,s}^{*(k)})- \rho_{\tau,m}^{(1)}(u_t^{*(k)}))|\mathrm ds
 \nonumber\\
 &\le&
 2|d_t-d_t^{*(k)}|\nonumber\\
 &&+
 |d_t^{*(k)}|\int_0^1
 \left\{
 |\rho_{\tau,m}^{(1)}(z_{t,s})-\rho_{\tau,m}^{(1)}(z_{t,s}^{*(k)})|
 +|\rho_{\tau,m}^{(1)}(u_t)-\rho_{\tau,m}^{(1)}(u_t^{*(k)})|
 \right\}\mathrm ds.
\end{eqnarray*}
Setting $\varrho=\rho_0^{1/5}$, \eqref{proof.lp.D.primitive.coupling}, \eqref{proof.lp.D.geometric.1} and \eqref{proof.lp.D.geometric.2} jointly yield that
\begin{eqnarray}\label{proof.lp.D.geometric}
 \|G_t-G_t^{*(k)}\|_2
 &\le&
 Cr\rho_0^{(k-h)_+}+Cr\rho_0^{(k-h)_+/5}
 \le Cr\varrho^{(k-h)_+}.
\end{eqnarray}
Since original and coupled variables have the same marginal expectation $E[G_t^{*(k)}]=E[G_t]$, \eqref{proof.lp.D.geometric} implies that 
\begin{eqnarray*}
 \|\vartheta_t-\vartheta_t^{*(k)}\|_2
 &=&\|G_t-G_t^{*(k)}\|\le Cr\varrho^{(k-h)_+}.
\end{eqnarray*}

On the other hand, \eqref{proof.lp.D.envelope}, applied to the original and coupled variables, gives
\begin{eqnarray}\label{proof.lp.D.small.envelope}
 \|\vartheta_t-\vartheta_t^{*(k)}\|_2
 &\le&
 \|\vartheta_t\|_2+\|\vartheta_t^{*(k)}\|_2
 \le Cr^{3/2}.
\end{eqnarray}

Combining \eqref{proof.lp.D.geometric} and \eqref{proof.lp.D.small.envelope} proves
\begin{eqnarray}\label{proof.lp.D.coupling}
 \|\vartheta_t-\vartheta_t^{*(k)}\|_2
 &\le&
 C\min\!\left\{
 r_{\delta,a}^{3/2},
 r_{\delta,a}\varrho^{(k-h)_+}
 \right\}.
\end{eqnarray}

Let
\begin{eqnarray*}
 k_r
 &\coloneqq&
 \left\lceil
 \frac{|\log r_{\delta,a}|}{2|\log\varrho|}
 \right\rceil+h.
\end{eqnarray*}
Summing the first bound in \eqref{proof.lp.D.coupling} up to $k_r$ and the second bound thereafter gives
\begin{eqnarray*}
 \sum_{k=0}^{\infty}\mathfrak d_{2,\vartheta}(k)
 &\le&
 C(k_r+1)r_{\delta,a}^{3/2}
 +Cr_{\delta,a}\sum_{k>k_r}\varrho^{k-h}\le
 Cr_{\delta,a}^{3/2}
 \{1+|\log r_{\delta,a}|\}.
\end{eqnarray*}
Together with Lemma \ref{Appendix.B10}, applied with $r=2$, this establishes the desired result in Lemma \ref{ap.tm5c.lemD}.
\end{proof}

\begin{proof}[Proof of Lemma \ref{ap.tm5c.lemL}]
\item    

(i).  We write
\begin{eqnarray*}
 H(\gamma_t,\mathbf x_t^{[j]})
 -H(\gamma,\tilde{\mathbf x}_t^{[j]}(\gamma))
 &=&
 \{H(\gamma_t,\mathbf x_t^{[j]})
 -H(\gamma_t,\tilde{\mathbf x}_t^{[j]}(\gamma_t))\}\nonumber\\
 &&+
 \{H(\gamma_t,\tilde{\mathbf x}_t^{[j]}(\gamma_t))
 -H(\gamma,\tilde{\mathbf x}_t^{[j]}(\gamma))\}.
\end{eqnarray*}
For $|\gamma_t-\gamma|\leq \delta$, Lemma \ref{Appendix.B4}.(b)--(c) and the conditions on $H$ jointly imply
\begin{eqnarray}\label{proof.lp.L.approx}
 \left\|
 H(\gamma_t,\mathbf x_t^{[j]})
 -H(\gamma,\tilde{\mathbf x}_t^{[j]}(\gamma))
 \right\|_1
 &\le&C(T^{-1}+\delta).
\end{eqnarray}
For each fixed $\gamma$, define the centred transform of the stationary approximation by
\begin{eqnarray*}
 \pmb\Xi_t^H(\gamma)
 &\coloneqq&
 H\!\left(\gamma,\tilde{\mathbf x}_t^{[j]}(\gamma)\right)
 -E\!\left[
 H\!\left(\gamma,\tilde{\mathbf x}_t^{[j]}(\gamma)\right)
 \right].
\end{eqnarray*}
Let $\pmb\Xi_t^{H,*(k)}(\gamma)$ be its coupled version obtained by replacing $\pmb\eta_{t-k}$ with an independent copy.  Since the original and coupled variables have the same marginal expectation, the growth condition on $H$, H\"older's inequality, and Lemma \ref{Appendix.B4}.(d) give, for some $\rho_x\in(0,1)$,
\begin{eqnarray*}
\mathfrak d_{2,\Xi^H}(k;\gamma)
 &\coloneqq&
 \left\|
 \pmb\Xi_t^H(\gamma)
 -\pmb\Xi_t^{H,*(k)}(\gamma)
 \right\|_2\nonumber\\
 &\le&
 C\left\|
 1+\|\tilde{\mathbf x}_t^{[j]}(\gamma)\|
 +\|\tilde{\mathbf x}_t^{[j],*(k)}(\gamma)\|
 \right\|_4
 \left\|
 \tilde{\mathbf x}_t^{[j]}(\gamma)
 -\tilde{\mathbf x}_t^{[j],*(k)}(\gamma)
 \right\|_4\nonumber\\
 &\le&C\rho_x^k.
\end{eqnarray*}
Additionally, applying Lemma \ref{Appendix.B10} with $r=2$, and using the compact support and boundedness of the kernel, gives
\begin{eqnarray}\label{proof.lp.L.approx2}
 \left\|
 \frac1T\sum_{t=1}^{T-h}
 \left\{H(\gamma,\tilde{\mathbf x}_t^{[j]}(\gamma))
 -E\left[H(\gamma,\tilde{\mathbf x}_t^{[j]}(\gamma))\right]\right\}
 k_{\delta,t}^{(n)}(\gamma)
 \right\|_2
 &\le&
 \frac{C}{T}
 \left\{
 \sum_{t=1}^{T-h}
 |k_{\delta,t}^{(n)}(\gamma)|^2
 \right\}^{1/2}
 \nonumber\\
 &\le& C(T\delta)^{-1/2}.
\end{eqnarray}
By \eqref{proof.lp.L.approx} and \eqref{proof.lp.L.approx2}  we prove Lemma \ref{ap.tm5c.lemL}.(i).

\medskip

(ii).  Decompose the exact-score average into its centred and mean parts:
\begin{eqnarray*}
 \frac1T\sum_{t=1}^{T-h}
 \psi_\tau(u^{(h)}_{i,t})\mathbf x_t^{[j]}
 k_{\delta,t}^{(n)}(\gamma)
 &=&
 \frac1T\sum_{t=1}^{T-h}
 \{\psi_\tau(u^{(h)}_{i,t})\mathbf x_t^{[j]}
 -E[\psi_\tau(u^{(h)}_{i,t})\mathbf x_t^{[j]}]\}
 k_{\delta,t}^{(n)}(\gamma)\nonumber\\
 &&+
 \frac1T\sum_{t=1}^{T-h}
 E[\psi_\tau(u^{(h)}_{i,t})\mathbf x_t^{[j]}]
 k_{\delta,t}^{(n)}(\gamma).
\end{eqnarray*}
Lemmas \ref{Appendix.B7} and \ref{Appendix.B10}, the latter applied with $r=2$, make the first term $O_P((T\delta)^{-1/2})$.  Lemma \ref{Appendix.B6} gives the probability order of the second term: $ O\!\left(T^{-(\nu-1)/\nu}\right) =o((T\delta)^{-1/2})$.

Finally, the fact $|\rho_{\tau,m}^{(1)}(u)-\psi_\tau(u)| \le\mathbf1\{|u|\le m^{-1}\}$ and Lemma \ref{Appendix.B5} jointly give
\begin{eqnarray*}
 &&E\left\|
 \frac1T\sum_{t=1}^{T-h}
 \{\rho_{\tau,m}^{(1)}(u^{(h)}_{i,t})
 -\psi_\tau(u^{(h)}_{i,t})\}
 \mathbf x_t^{[j]}k_{\delta,t}^{(n)}(\gamma)
 \right\| \le\frac{C}{m}.
\end{eqnarray*}
Since $\sqrt{T\delta}/m\to0$, this yields the order $o((T\delta)^{-1/2})$.  Combining the three bounds proves Lemma \ref{ap.tm5c.lemL}.(ii).
\end{proof}

\end{appendix}

\end{document}